\documentclass[a4paper,12pt]{article}
\pdfoutput=1
\usepackage{epsfig,graphicx,amsbsy,amssymb,latexsym,amsfonts,amsmath}
\usepackage{pstricks,setspace}
\usepackage{hyperref,soul}
\usepackage{xcolor,color,tcolorbox}
\usepackage{eurosym}
\usepackage[a4paper]{geometry}
\makeatletter
\newsavebox{\@brx}
\newcommand{\llangle}[1][]{\savebox{\@brx}{\(\m@th{#1\langle}\)}%
  \mathopen{\copy\@brx\kern-0.5\wd\@brx\usebox{\@brx}}}
\newcommand{\rrangle}[1][]{\savebox{\@brx}{\(\m@th{#1\rangle}\)}%
  \mathclose{\copy\@brx\kern-0.5\wd\@brx\usebox{\@brx}}}
\makeatother

\newcommand {\beq}{\begin{equation}}
\newcommand {\eeq}{\end{equation}}
\newcommand {\beqa}{\begin{eqnarray}}
\newcommand {\eeqa}{\end{eqnarray}}
\renewcommand{\d}{{\rm d}}
\def\pd{\partial}

\def\al{\alpha}

\def\nn{\nonumber}

\title{\bf Massless and Massive Higher Spins
\\
from
\\
Anti-de Sitter Space Waveguide
\\
{}
}

\author{ \\
 \textsc{Seungho Gwak}$^{[1]}$, \ \textsc{Jaewon Kim}$^{[1]}$, \ \textsc{Soo-Jong Rey}$^{[1],[2],[3]}$
 \\
 }
\date{}
\begin{document}
\maketitle

\begin{center}
\renewcommand{\thefootnote}{\fnsymbol{footnote}} 
${}^{\thefootnote[1]}$ \sl School of Physics and Astronomy \& Center for Theoretical Physics\\
          Seoul National University, Seoul 08826 \rm KOREA\\[0.4cm]
${}^{\thefootnote[2]}$ \sl Center for Theoretical Physics, College of Physical Sciences \\
Sichuan University, Chengdu 610064 \rm PR CHINA \\[0.4cm]
${}^{\thefootnote[3]}$ \sl Fields, Gravity \& Strings, Center for Theoretical Physics of the Universe \\
Institute for Basic Sciences, Daejeon 34047 \rm KOREA\\[1cm]
\end{center}

\begin{abstract}
Understanding Higgs mechanism for higher-spin gauge fields is an outstanding open problem. We investigate this problem in the context of Kaluza-Klein compactification. Starting from a free massless higher-spin field in $(d+2)$-dimensional anti-de Sitter space and compactifying over a finite angular wedge, we obtain an infinite tower of heavy, light and massless higher-spin fields in $(d+1)$-dimensional anti-de Sitter space. All massive higher-spin fields are described gauge invariantly in terms of  Stueckelberg fields. The spectrum depends on the boundary conditions imposed at both ends of the wedges. We observed that higher-derivative boundary condition is inevitable for spin greater than three. 
For some higher-derivative boundary conditions, equivalently, spectrum-dependent boundary conditions, we get a non-unitary representation of partially-massless higher-spin fields of varying depth. We present intuitive picture which higher-derivative boundary conditions yield non-unitary system in terms of boundary action. We argue that isotropic Lifshitz interfaces in $O(N)$ Heisenberg magnet or $O(N)$ Gross-Neveu model provides the holographic dual conformal field theory and propose experimental test of (inverse) Higgs mechanism for massive and partially massless higher-spin fields.  

\end{abstract}
\newpage
\vskip-1cm
\tableofcontents

\vskip2cm
\rightline{\sl People used to think that when a thing changes, it must be in a state of change,}
\rightline{\sl and that when a thing moves, it is in a state of motion.}
\rightline{\sl This is now known to be a mistake.}
\rightline{\textsc{bertrand russell}}

\section{Introduction}
Massive particles of spin higher than two are not only a possibility  --- both theoretically as in string theory and experimentally as in hadronic resonances (See, for example, \cite{Chow:1997sg}) -- but also a necessity for consistent dynamics of lower spin gauge fields they interact to (See, for example, \cite{Camanho:2014apa}). As for their lower spin counterpart, one expects that their masses were generated by a sort of Higgs mechanism, combining higher-spin Goldstone fields \cite{Porrati:2001db, Porrati:2003sa} to massless higher spin gauge fields. Conversely, one expects that, in the massless limit, massive higher-spin fields undergo inverse Higgs mechanism and split its polarization states irreducibly into massless higher-spin gauge fields and higher-spin Goldstone fields. On the other hand, it is well-known that the higher-spin gauge invariance is consistent only in curved background such as (anti)-de Sitter space ((A)dS). As such, gauge invariant description of massive higher-spin fields and their Higgs mechanism would necessitate any dynamical description of the (inverse) Higgs mechanism formulated in (A)dS background. 

A novel feature in (A)dS background, which opens up a wealth of the Higgs mechanism, is that a massive higher-spin field may have different number of possible polarizations as its mass is varied. The (A)dS extension of massive higher-spin field in flat space can be in all possible polarizations. They have arbitrary values of mass and are called massive higher-spin fields. In (A)dS background, there are also massive higher-spin fields for which part of possible polarizations is eliminated by partial gauge invariance. They have special values of mass-squared and are called partially massless higher-spin fields.  Just as the Higgs mechanism of massive higher-spin fields are not yet fully understood, the Higgs mechanism (if any) of partially massless higher-spin fields remains mysterious. For both situations, what are origins and patterns of massive higher-spin fields?

In this work, we lay down a concrete framework for addressing this question and, using it, to analyze the pattern of the massive higher-spin fields as well as Higgs mechanism that underlies the mass spectrum. The idea is to utilize the Kaluza-Klein approach \cite{Kaluza:1921 Klein:1926} for compactifying higher-dimensional (A)dS space to lower-dimensional (A)dS space and to systematically study mass spectrum of compactified higher-spin field in gauge invariant manner. This Kaluza-Klein setup also permits a concrete realization of holographic dual conformal field theory from which the above Higgs mechanism can also be understood in terms of conventional global symmetry breaking.

The Kaluza-Klein compactification provides an elegant geometric approach for dynamically generating  masses. Compactifying a massless field in higher dimensions on a compact internal space, one obtains in lower dimensions not only a massless field but also a tower of massive fields. In (A)dS space, a version of compactification of spin-zero, spin-one and spin-two field theories were studied in \cite{Metsaev, Dimensional Degression}. In flat space, compactification of higher-spin field theories was studied in \cite{DD Stueckel}.
The Kaluza-Klein mass spectrum depends on specifics of the compact internal space. Here, the idea is that we start from unitary, massless higher-spin field in a higher-dimensional (A)dS space and Kaluza-Klein compactify to a lower-dimensional (A)dS space. 
One of our main results is that, to produce not only massive but also partially massless higher-spin fields upon compactification, differently from the above situations, we must choose the internal space to have boundaries and specify suitable boundary conditions at each boundaries. We shall refer to the compactification of higher-dimensional (A)dS space over an internal space with boundaries to lower-dimensional (A)dS space as ``(A)dS waveguide" compactification \footnote{Formally, the compactifications \cite{DD Stueckel} and its (A)dS counterparts \cite{Metsaev, Dimensional Degression} may be viewed as projectively reducing on a conformal hypersurface. This viewpoint was further studied for partially massless spin-two system in \cite{Gover:2014vxa}. Starting from non-unitary conformal gravity, this work showed that projective reduction yields partially massless spin-two system, which is also non-unitary. Here, we stress we are taking an entirely different route of physics. We start from unitary, Einstein gravity and compactify on a suitable internal space with boundaries to obtain non-unitary, partially massless spin-two system. Our setup has another added benefit of physics that Higgs mechanism can be triggered by dialing choices of boundary conditions.}.

Compactifying a unitary, massless spin-$s$ field on an (A)dS waveguide whose internal space is a one-dimensional angular wedge of size $[-\alpha, \alpha]$, we show that presence of boundaries and rich choices of boundary condition permit a variety of mass spectra of higher-spin fields in lower-dimensional (A)dS space  (as summarized at the end of Section \ref{mode s}  and in Fig. \ref{spectrumspins}). These spectra reveal several interesting features:
\begin{itemize}
\item
The spectra contain not only massless and massive fields but also partially massless fields \cite{PaMa}. The partially massless fields arises only if the internal compact space has boundaries and specific boundary conditions. The (non)unitarity of partially massless fields in (A)dS space can be intuitively understood by the presence of boundary degrees of freedom and (non)unitarity of their dynamics (see Section \ref{sec7}).  

\item The spectra split into two classes, distinguished by dependence on the waveguide size, $\alpha$. The modes that depend on the size is the counterpart of massive Kaluza-Klein states in flat space compactification, so they all become infinitely heavy as the size $\alpha$ is reduced to zero. We call them {\sl Kaluza-Klein modes}. The modes independent of the wedge angle is the counterpart of zero-mode states in flat space compactification. We call them {\sl ground modes}. 

\item    
The (A)dS compactification features {\sl two} independent scales: the scale of waveguide size and the scale of (A)dS curvature radius. This entails an interesting pattern of the resulting mass spectra. While the Kaluza-Klein modes are all fully massive higher-spin fields (thus the same as for the flat space compactification), the ground modes comprise of full variety of mass spectra: massless, partially massless and fully massive higher-spin fields. 
 \label{groundmode1}
 
 \item All massive higher-spin fields, both fully massive and partially massless, are structured by the Stueckelberg mechanism, in which the Goldstone modes are provided by a tower of higher-spin fields of varying spins. The Higgs mechanism can be understood in terms of branching rules of Verma $\mathfrak{so}(d,2)$ modules. It turns out that massless spin-$s$ gauge symmetry on (A)dS$_{d+2}$ is equivalent to the Stueckelberg gauge symmetries \cite{Stue} for spin-$s$ on (A)dS$_{d+1}$ \cite{Zino1}.
 
\item The ground mode spectra and (inverse) Higgs mechanism therein match perfectly with the critical behavior we expect from $d$-dimensional isotropic Lifshitz interfaces in $(d+1)$-dimensional conformal field theories such as $O(N)$ Heisenberg system or Gross-Neveu fermion system in the large $N$ limit. Both are realizable in heavy fermion magnetic materials and in multi-stack graphene sheets at Dirac point, respectively. This suggests an exciting possibility for condensed-matter experimental realizations / tests of (inverse) Higgs mechanism for higher-spin gauge fields. 
\end{itemize}

In obtaining these results, we utilized several technicalities that are worth of highlighting.  
\begin{itemize}

\item As our focus is on mass spectra and their Higgs mechanism, we limit our analysis to non-interacting higher-spin fields. Moreover, we analyze linearized field equations  instead of quadratic action. It is known that the two approaches are equivalent in so far as the gauge transformations are also kept to the linear order. 

\item For spectral analysis, we further bypass working on the linearized field equations. Instead, we extract mass spectra and Stueckelberg structure from the linearized gauge transformations. This is because the linearized field equation and hence the quadratic part of action for massless spin-$s$ field are uniquely determined by the spin-$s$ gauge symmetries.

\item We recast the spectral analysis in terms of a pair of first-order differential operators. They play the role of raising and lowering operators. The associated Sturm-Liouville problem is factorized into quadratic product of these operators, akin to the supersymmetric quantum mechanics Hamiltonian. 

\item We associate the origin of partially massless higher-spin fields to Sturm-Liouville problems with non-unitary boundary conditions involving higher-order derivatives. We obtain self-adjointness of the spectral analysis by extending the Hilbert space. Physically, we interpret the newly introduced Hilbert space as {\sl boundary degrees of freedom}. 

\end{itemize}

To highlight novelty and originality of our approach, we compare it with previous works. There have been various approaches for higher-dimensional origin of higher-spin fields. {The work \cite{RaRe} proposed so-called `radial reduction' that reduces a higher-spin theory in $(d+1)$-dimensional Minkowski spacetime to that  in $d$-dimensional (A)dS spacetime. This approach describes one-to-one correspondence between flat and (A)dS interaction vertices but does not guarantee consistency of reduced theory as interacting higher-spin theory is not known in flat space. Our approach starts from higher-spin gauge theory in 
(A)dS$_{d+2}$ space and compactifies it to (A)dS$_{d+1}$ space. Both theories are well-defined.} 
The work \cite{Dimensional Degression} proposes to decompose the higher-spin representations of $\mathfrak{so}(d+1, 2)$ in terms of higher-spin representations of $\mathfrak{so}(d,2)$, viz. decomposing (A)dS$_{d+2}$ space to foliation leaves of (A)dS$_{d+1}$. While it takes an advantage of the discrete spectrum of these unitary representations, this approach is rather limited for not having a tunable Kaluza-Klein parameter (such as $\alpha$ in our approach) that specifies compactification size or with a set of boundary conditions that yields the requisite mass spectrum. In particular, it does not give rise to massless or partially massless higher-spin fields in (A)dS$_{d+1}$ space. In our approach, we have both of them. We summarize more specifics of these comparisons in Section \ref{alpha two pi limit}.

The rest of this paper is organized as follows. In section \ref{sec2}, we start with the spin-one waveguide in flat space. We emphasize that the Kaluza-Klein compactification manifests the Stueckelberg structure and consequent Higgs mechanism by combining various polarization components. We also show that the consistency of equations of motion or of gauge transformations restricts possible set of boundary conditions among various components of spin-one field. In section \ref{sec3}, we explain how the Kaluza-Klein compactification works for (A)dS space. We demonstrate that the so-called Janus geometry provides conformal compactification of (A)dS$_{d+2}$ space down to (A)dS$_{d+1}$ space, and refer to it as  AdS waveguides. On this geometry, we study mass spectra for spin-one, spin-two and spin-three fields
in section \ref{sec4}, \ref{sec5} and \ref{sec6}, respectively.
We explain in detail how the spectral analysis of equations of motion and of gauge transformations fit consistently each other,
and confirm that, in lower dimensions and at linearized level, the gauge transformations are sufficient to uniquely fix the equations of motion.
We show that a variety of boundary conditions are possible and rich pattern of Higgs mechanism and mass spectra are obtained from them.  
In particular, we show that, in addition to fully massive higher-spin fields,  massless and partially massless(PM) fields on (A)dS can be realized. For the latter, we show that they arise from higher-derivative boundary conditions (HDBCs) and that such boundary conditions can arise for spin two or higher.
A simple example is considered in section \ref{sec7} 
to provide the physical  meaning of the higher derivative boundary conditions and 
we provide the intuitive picture why non-unitary representation appears by reduction.
All procedure extend to spin-$s$ in section \ref{sec8}.
In Section 9, we argue that isotropic Lifshitz interface of $O(N)$ Heisenberg magnet or Gross-Neveu model is the simplest dual conformal field theory which exhibits the (inverse) breaking of global higher-spin symmetries.  
Section \ref{sec10} discusses various open issues for further investigation.
Appendix \ref{adsC} summarizes our conventions for the AdS space. The $\mathfrak{so}(d,2)$-modules is briefly reviewed in Appendix \ref{Verma M}. The non-abelian AdS waveguide method via the Kaluza-Klein compactification from (A)dS$_{d+k}$ to (A)dS$_d$ for $k \ge 2$ is demonstrated in Appendix \ref{DR}. 

\section{Flat Space Waveguide and Boundary Conditions} \label{sec1example}\label{sec2}

The salient feature of our approach is to compactify AdS$_{d+2}$ space to AdS$_{d+1}$ space times an open internal manifold with boundaries.  A complete specification of the compactification requires to impose a suitable set of boundary condition at each boundary, which in turn uniquely determine the mass spectrum in AdS$_{d+1}$.  The choice of boundary condition provides a new, tunable parameter in addition to the size of internal manifold that features the conventional compactification, triggering the (inverse) Higgs mechanism. 

\subsection{Kalauza-Klein mode expansion}
To gain physics intuition, we first warm up ourselves with the electromagnetic -- massless spin-one -- waveguide in $(d+2)$-dimensional flat spacetime with two boundaries, paying particular attention to relations between boundary conditions and spectra for fields of different spins. The flat spacetime is $\mathbb{R}^{1,\,d} \times I_L$, where interval $I_L\equiv \{0\leq z\leq L\}$. 
We decompose the $(d+2)$-dimensional coordinates into parallel and perpendicular directions, $x^M=(x^\mu, z)$, and the $(d+2)$-dimensional spin-one field to a spin-one field and a spin-zero field  in $(d+1)$ dimensions, $A_M = (A_\mu, \phi)$.
The equations of motions are decomposed as 
\begin{align}
\label{exeom1} \partial^M\, F_{M \nu} &= \partial^{\mu}\, F_{\mu \nu} - \partial_z{} (\partial_\nu \phi - \partial_z \, A_{\nu} ) = 0\,, \\
\label{exeom2} \partial^M\, F_{M z} &= \partial^{\mu}\,(\partial_{\mu}\, \phi - \partial_{z}\,A_{\mu}) =0\,,
\end{align}
while the gauge transformations are decomposed as
\beq
\delta\,A_\mu=\partial_\mu\,\Lambda\,,\qquad \qquad 
\delta\,\phi=\partial_z\,\Lambda\,.\label{flat gauge}
\eeq
We note that both the equations of motion and the gauge transformations manifest the structure of Stueckelberg system \cite{Stue}. Recall that the Stueckelberg Lagrangian of massive spin-one vector field is given by 
\beq
{\cal L} = -\frac{1}{4}\,F_{\mu\nu}F^{\mu\nu} + \frac{1}{2}\,\partial_\mu\,\phi\,\partial^\mu\,\phi
+m\,A_\mu\left(\frac{m}{2}\, A^\mu-\partial^\mu\,\phi\right)\,,
\eeq
which is invariant under the Stueckelberg gauge transformations
\beq
\delta\, A_\mu = \partial_\mu\, \lambda\,\qquad \mbox{and} \qquad \delta\, \phi= m\,\lambda\,.\label{StueckelbergG}
\eeq
The field $\phi$ is referred to as the Stueckelberg spin-zero field. This field is redundant for $m \ne 0$ because it can be eliminated by a suitable gauge transformation. In the massless limit, $m \rightarrow 0$, the Stueckelberg system dissociate into a spin-one gauge system and a massless spin-zero system.

Inside the waveguide, the $(d+2)$-dimensional spin-one field $A_M$ is excited along the $z$-direction. The field can be mode-expanded, and expansion coefficients are $(d+1)$-dimensional spin-one and spin-zero fields of various masses. Importantly, mode functions can be chosen from any complete set of basis functions. 
 It is natural to choose them by the eigenfunctions of $\Delta := - \left(\partial_z\right)^2$ 
 with a prescribed boundary condition that ensures the self-adjointness. 
 
Inside the waveguide, the mode functions of the gauge parameter $\Lambda$ should be chosen compatible with the mode functions of the spin-one field $A_M$. Combining the two gauge variations Eq.\eqref{flat gauge}, we learn that the mode functions ought to be related to each other as 
\beq
\partial_z\,(\mbox{ mode function of spin-one field  $A_\mu (x, z)$}) \propto
(\mbox{ mode function of spin-zero field $\phi(x, z)$})\,.\label{proportionality}
\eeq
Being a local expression, this relation must hold at each boundaries as well. 

It would be instructive to understand what might go wrong if, instead of the required Eq.(\ref{proportionality}), one imposes the same boundary conditions for both $A_{\mu}$ and $\phi$, such as zero-derivative (Dirichlet) or one-derivative (Neumann) boundary conditions. Suppose one adopts the zero-derivative (Dirichlet) boundary condition for both fields. From $A_{\mu}(z)|_{z=0,\,L}=0$, $\phi(z)|_{z=0, L}= 0$ and from the field equation of $\phi$, Eq. \eqref{exeom2}, it follows that
\begin{align}
\Big( \partial^{\mu}\, \partial_{\mu}\,\phi (z) - \partial^{\mu}\, \partial_z\, A_{\mu} (z) \Big)\Big|_{z=0,\,L} = - \partial^{\mu}\, \partial_z\, A_{\mu} (z) |_{z=0,\,L} = 0\,,\label{over determinent}
\end{align}
and hence $\partial_z\, A_{\mu} (z) |_{z=0,L} = 0$. But $A_{\mu}$ satisfies second-order partial differential equation, so these two sets of boundary conditions
--- $A_{\mu}(z)|_{z=0,\,L} = 0$ and $\partial_z\, A_{\mu} (z) |_{z=0,L} = 0$ --- imply that $A_{\mu}(z)$ must vanish everywhere. Likewise, $\phi$ satisfies a first-order differential equation Eq.(\ref{exeom1}), so the two sets of boundary conditions imply that $\phi (z)$ vanishes everywhere as well. One concludes that there is no nontrivial field excitations satisfying such boundary conditions. We remind that this conclusion follows from the fact that these boundary conditions do not preserve the relation Eq.\eqref{proportionality}. 

The most general boundary conditions compatible with the relation Eq.\eqref{proportionality} restrict the form of boundary conditions for spin-one and spin-zero fields. For example, if we impose the Robin boundary condition for the spin-zero field, ${\cal M}(\partial_z) \phi |_{z=0, L} := (a \partial_z + b) \phi |_{z=0, L}= 0$ where $a, b$ are arbitrary constants, the relation Eq.\eqref{proportionality} imposes the boundary condition for the spin-one field as ${\cal M}\,\partial_z\,A_\mu|_{z=0,\,L} =0$. Modulo higher-derivative generalizations, we have two possible boundary conditions: $a=0, b \ne 0$ corresponding to the vector boundary condition and $a \ne 0, b = 0$ corresponding to the scalar boundary condition. Hereafter, we analyze each of them explicitly. 

\subsection{Vector boundary condition}
We may impose one-derivative (Neumann) boundary condition on the spin-one field $A_{\mu}(x,z)$ 
and zero-derivative (Dirichlet) boundary condition on the spin-zero field $\phi(x,z)$ at $z=0,\, L$. The corresponding mode expansion for $A_{\mu}$ and $\phi$ reads
\begin{align}
A_{\mu}(z) = \sum^{\infty}_{n=0}\, A^{(n)}_{\mu}\, \hbox{cos}\,\left(\frac{n\,\pi}{L}\,z\right)
 \qquad \mbox{and} \qquad 
\phi (z) = \sum^{\infty}_{n=1}\, \phi^{(n)}\, \hbox{sin}\,\left(\frac{n\,\pi}{L}\,z\right),
\end{align}
so the field equations Eq.(\ref{exeom1}) and Eq.(\ref{exeom2}) are also expanded 
in a suggestive form
\begin{align}
&\sum_{n=0}^{\infty} \hbox{cos}\left( \frac{n\,\pi}{L}\,z\right) \left[ \partial^{\mu}\, F^{(n)}{}_{\mu \nu} - \left(\frac{n\,\pi}{L}\right) \left( \, \frac{n\pi}{L}\, A^{(n)}_{\nu} +  \partial_{\nu}\,\phi^{(n)} 
\right) \right] =0\,,\\
&\sum_{n=1}^{\infty} \hbox{sin}\left(\frac{n \,\pi}{L}\,z\right) \partial^\mu \left( \, \frac{n\,\pi}{L}\, \, A^{(n)}_{\mu} + \partial_{\mu}\, \phi^{(n)} \right) =0\,.
\end{align}
The mode functions $\sin (n \pi z / L)$ for $n=0,1,\dots$ form a complete set of the orthogonal basis for square-integrable functions over $I_L$,  
so individual coefficient in the above equations ought to vanish. 
The zero-mode $n=0$ is special, as only the first equation is nonempty  
and gives the equation of motion for massless spin-one field.  
All Kaluza-Klein modes, $n \ge 1$, satisfies the Stueckelberg equation of motion for massive spin-one field \footnote{
For Kaluza-Klein compactification of flat spacetime, the Stueckelberg structure of higher-spin fields was first noted in \cite{DD Stueckel}.}
 with mass $m_n = {n\,\pi}/{L}$.  
 The second equation follows from divergence of the first equation,  
 so just confirms consistency of the prescribed boundary conditions.  
 In the limit $L\rightarrow 0$, all Stueckelberg fields become infinitely massive.  
 As such, there only remains the massless spin-one field $A^{(0)}_{\mu}$ 
 with associated gauge invariance.  Also, there is no spin-zero field $\phi^{(0)}$, 
 an important result that follows from the prescribed boundary conditions.  
 Intuitively, $A_\mu^{(0)}$ remains massless and gauge invariant,  
 so Stueckelberg spin-zero field $\phi^{(0)}$ is not needed.  
 Moreover, the spectrum is consistent with the fact 
that this boundary condition ensures no energy flow across the boundary $z = 0, L$. 

The key observation crucial for foregoing discussion is that the same result is obtainable from Kaluza-Klein compactification of gauge transformations Eq.\eqref{flat gauge}. The gauge transformations that preserve the vector boundary conditions can be expanded by the Fourier modes:
\beq
\Lambda=\sum_{n=0}^\infty\,
\Lambda^{(n)}\,\hbox{cos}\,\left(\frac{n\,\pi}{L}\,z\right)\,.
\eeq
The gauge transformations of $(d+1)$-dimensional fields read
\beqa
\delta\, A_{\mu}^{(n)}
=\partial_\mu\,\Lambda^{(n)}\,
\quad\mbox{($n\geq0$)}\,
\qquad \mbox{and} \qquad
\delta\, \phi^{(n)}
=-\frac{n\,\pi}{L}\,\Lambda^{(n)}\,
\quad\mbox{($n \ge 1$)}\,.
\eeqa
We note that the $n=0$ mode is present only for the gauge transformation of spin-one field. This is the gauge transformation of a massless gauge vector field.  We also note that gauge transformations of all higher $n = 1, 2, \cdots$ modes take precisely the form of Stueckelberg gauge transformations. Importantly, the Stueckelberg gauge invariance fixes quadratic part of action as the Stueckelberg action for a tower of Proca fields with masses $m_n = {n\,\pi}/{L}$, $(n = 1, 2, \cdots)$. 

The fact that normal modes and their mass spectra are extractible equally well from the linearized equations of motion and from the linearized gauge transformations is an elementary consequence of Fourier analysis. At the risk of being pedantic, here we recall this trivial fact. Consider fields $A_M(z)$ belonging to the Hilbert space of $I_L$. Denote the normal modes of the Sturm-Liouville operator $- \partial_z^2$ as $\langle z | n \rangle$ and their completeness relation as $ \sum_n \langle z | n \rangle \langle n | z' \rangle = \delta (z - z') $. Varying the quadratic part of action with respect to the gauge variation and integrating over $z \in I_L$, we have
\beqa
0 = \langle {\delta L^{(2)} \over \delta A_M } \vert  \delta A_M \rangle  = \sum_n \langle {\delta L^{(2)} \over \delta A_M} \vert n \rangle \langle n \vert \delta A_M \rangle
\eeqa
It is elementary to conclude from this equation that, projecting the gauge transformation onto $n$-th mode, the equations of motion is projected to the same $n$-th mode. It follows that the spectrum of gauge transformation $\delta A_M$ dictates the spectrum of equations of motion $\delta L^{(2)} / \delta A_M$. The converse also follows straightforwardly.  Note that this argument is universal in the sense that it holds for any linear Sturm-Liouville system which is derivable from action or energy functional. In particular, it holds for linearized higher-spins and for curved internal space for which the operator $-\partial_z^2$ is replaced by the most general Sturm-Liouville operator $-\nabla_z^2$ and the measure ${\rm d} z$ is replaced by the covariant counterpart ${\rm d} z \sqrt{g_{zz}}$. We will practice this elementary fact repeatedly throughout this paper. 

We can turn the argument around. Suppose we want to retain massless spin-one field $A^{(0)}_\mu$ in $(d+1)$-dimensions, along with associated gauge invariance. This requirement then singles out one-derivative (Neumann) boundary condition for $A_\mu$. This and the divergence for $A_\mu$, in turn, single out zero-derivative (Dirichlet) boundary condition for $\phi$. Clearly, the massless fields are associated with gauge or global symmetries (namely invariances under inhomogeneous local or rigid transformations). So, this argument shows that proper boundary conditions for linearized field equations can be extracted just from linearized gauge transformations. 

\subsection{Scalar boundary condition}
Alternatively, one might impose no-derivative (Dirichlet) boundary condition to the spin-one field $A_{\mu}$ and one-derivative (Neumann) boundary condition to the spin-zero $\phi$. In this case, the equations of motion, when mode-expanded, take exactly the same form as the above except that the mode functions are interchanged:
\begin{align}
&\sum_{n=1}^{\infty} \hbox{sin}\left( \frac{n\,\pi}{L}\,z\right) \left[ \partial^{\mu}\, F^{(n)}{}_{\mu \nu} - \left(\frac{n\,\pi}{L}\right) \left( \, \frac{n\pi}{L}\, A^{(n)}_{\nu} -  \partial_{\nu}\,\phi^{(n)} 
\right) \right] =0\,,\\
&\sum_{n=0}^{\infty} \hbox{cos}\left(\frac{n\,\pi}{L}\,z\right) \partial^\mu \left( \, \frac{n \,\pi}{L}\, \, A^{(n)}_{\mu} - \partial_{\mu}\, \phi^{(n)} \right) =0\,.
\end{align}
Consequently, the zero-mode $n=0$ consists of massless spin-zero field $\phi^{(0)}$ only ($A^{(0)}_\mu$ is absent from the outset). All Kaluza-Klein modes $n \ne 0$ are again Stueckelberg massive spin-one fields with mass $m_n = n \pi/L$.
 In the limit $L \rightarrow 0$, these Stueckelberg field becomes infinitely massive. Below the Kaluza-Klein scale $1/L$, there only remains the massless spin-zero field $\phi^{(0)}$. Once again, this is consistent with the fact that this boundary condition ensures no energy flow across the boundary. 

Once again, the above results are also obtainable from the Kaluza-Klein compactification of gauge transformations. For the gauge transformation that preserves the scalar boundary condition, the gauge function can be expanded as
\beq
\Lambda(x, z) =\sum_{n=1}^\infty\,
\Lambda^{(n)} (x) \,\hbox{sin}\,\left(\frac{n\,\pi}{L}\,z\right)\,.
\eeq
With these modes, the gauge transformations of fields are
\beqa
\delta\, A_{\mu}^{(n)}
=\partial_\mu\,\Lambda^{(n)} \quad\mbox{($n \ge 1$)}
 \qquad \mbox{and} \qquad
\delta\, \phi^{(n)}
=\frac{n\,\pi}{L}\,\Lambda^{(n)}\,
\quad\mbox{($n \ge 0$)}\,. 
\eeqa
There is no $n=0$ zero-mode for the gauge transformation, and so no massless spin-one gauge field. The spin-zero zero-mode $\phi^{(0)}$ is invariant under the gauge transformations. We also note that the gauge transformations take the form of the Stueckelberg gauge symmetries with masses $m_n = {n\,\pi}/{L}$.

Once again, we can turn the argument around. Suppose we want to retain massless spin-zero field $\phi^{(0)}$ in $(d+1)$ dimensions. This then singles out one-derivative (Neumann) boundary condition for $\phi$. This and the divergence of $A_\mu$ equation of motion, in turn, put the spin-one field $A_\mu$ to zero-derivative (Dirichlet) boundary  condition. 

Summarizing, 
\vskip0.1cm
\begin{tcolorbox}
\begin{itemize}
\item Kaluza-Klein spectrum is obtainable either from linearized field equations or from linearized gauge transformations.   
\item 
Stueckelberg structure naturally arises  from Kaluza-Klein compactification not only for flat space but also for (A)dS space. 
\item Boundary conditions of lower-dimensional component fields (for example, $A_{\mu}$ and $\phi$ from $A_M$) are correlated each other (for example as in Eq.\eqref{proportionality}).
\end{itemize}
\end{tcolorbox}
Before concluding this section, we comment how these features are realized in string theory in terms of the brane configurations and S-duality for $d=3$ case.

\subsection{D3-branes ending on five-branes and S-duality}
The two possible boundary conditions discussed above are universal for all dimensions $d$.   When $d=3$ and adjoined with maximal supersymmetry, the two boundary conditions are related each other by the electromagnetic duality. This feature can be neatly seen in the context of brane configurations in Type IIB string theory, studied most recently in \cite{Gaiotto:2008sa}.

Consider a D3 brane ending on parallel five-branes (D5 or NS5 brane). From the viewpoint of world-volume dynamics, the stack of five-branes provides boundary conditions to flat space waveguide. The original low-energy degree of freedom of D3 brane is four-dimensional $\mathcal{N}=4$ vector multiplet. In the presence of five-branes, half of sixteen supersymmetries is broken. At the boundary, the four-dimensional $\mathcal{N}=4$ vector multiplet is split into three-dimensional $\mathcal{N}=4$ vector multiplet and ${\cal N}=4$ hypermultiplet. If the five-brane were D5 brane, the zero-mode is the three-dimensional  $\mathcal{N}=4$ hypermultiplet. If the five-branes were NS5 brane, the zero-mode is the three-dimensional $\mathcal{N}=4$ vector multiplet. In terms of D3 brane world-volume theory, D5 brane sets ``D5-type" boundary condition: Dirichlet boundary condition on three dimensional vector multiplet and Neumann boundary condition on three-dimensional hypermultiplet, while NS5 brane sets ``NS5-type" boundary condition: Neumann boundary condition on three dimensional vector multiplet and Dirichlet boundary condition on three-dimensional hypermultiplet. 

The Type IIB string theory has $SL(2, \mathbb{Z})$ duality symmetry, under which the two brane configurations are rotated each other. In terms of D3-brane world-volume dynamics, the three-dimensional ${\cal N}=4$ vector multiplet and hypermultiplet are interchanged with each other. This is yet another way of demonstrating the well-known mirror symmetry in three-dimensional gauge theory, which exchanges two hyperK\"ahler manifolds provided by the vector multiplet moduli space ${\cal M}_{\rm V}$ and the hypermultiplet moduli space ${\cal M}_{\rm H}$. Here, following our approach, we see that they can also be derived entirely from the viewpoint of gauge and global symmetries of component fields. 

\section{Waveguide in Anti-de Sitter Space}\label{sec3}
We now move to waveguide in (A)dS space. Here, we first explain how, starting from AdS$_{d+2}$ space, we can construct a ``tunable" AdS$_{d+1}$ waveguide -- a waveguide which retains $\mathfrak{so}(d, 2)$ sub-isometry within $\mathfrak{so}(d+1,2)$ isometry and which has a tunable size of internal space. 

Consider the AdS$_{d+2}$ space in the Poincar\'e patch with coordinates $(t,\, {\bf x}_{d-1},\,y,\,z) \in \mathbb{R}^{1,\, d} \times \mathbb{R}^+$:
\begin{align}
\d s(\mbox{AdS}_{d+2}){}^2 
= \frac{\ell^2}{z^2}\left(-\d t^2 + \d {\bf x}_{d-1}^{\,2} + \d z^2 \right)+\frac{\ell^2}{z^2}\d y^2 
= \d s(\mbox{AdS}_{d+1}){}^2{} + g_{yy}\d y^2\,.
\label{try1}
\end{align}
The $(d+2)$-dimensional Poincar\'e metric is independent of $y$, and remaining $(d+1)$-dimensional space is again Poincar\'e patch. Therefore, it appears that this foliation of AdS metric would work well for the AdS compactification we look for. Actually, it is not. The reason is as follows. Locally at each $y$, the
isometry $\mathfrak{so}(d, 2)$ is part of the original isometry $\mathfrak{so}(d+1, 2)$. 
However, globally, this does not hold in the Poincar\'e patch. 
The reason is that $\mathfrak{so}(d, 2)$ isometry transformation does not commute with translation along $y$ direction at the Poincar\'e horizon, $z = \infty$. Moreover, when compactifying along the $y$-direction, the $(d+2)$-dimensional tensor 
does  \emph{not} give rise to $(d+1)$-dimensional tensors.  Consider, for example, a small fluctuation of the metric. The tensor $\nabla_{\mu}\,h_{\nu y}$ is dimensionally reduced to $\nabla_{\mu} \,A_{\nu} + \delta_{\mu z}\,\frac{1}{z}\,A_{\nu}$, where $A_{\mu} \equiv h_{\mu y}$. The second term is a manifestation of non-tensorial transformation in $(d+1)$ dimensions. 

In fact, any attempt of compactifying along an isometry direction faces the same difficulties. As such, we shall instead foliate AdS$_{d+2}$ into a semi-direct product of AdS$_{d+1}$ hypersurface and an angular coordinate $\theta$ and Kaluza-Klein compactify along the $\theta$-direction over a finite interval:
\vskip0.1cm
\begin{tcolorbox}
\beq
\d s(\mbox{AdS}_{d+2}){}^2=\frac{1}{\hbox{cos}^2 \theta}\left[\d s(\mbox{AdS}_{d+1}){}^2 + \ell^2\, \d\theta^2\right]. \label{metricre}
\eeq\end{tcolorbox}
\vskip0.1cm
\noindent  
Here, the conformal factor arises because we compactified the internal space along a direction which is globally non-isometric.
This compactification bypasses the issues that arose in the compactification Eq.(\ref{try1}). In particular, $(d+2)$-dimensional tensors continue to be $(d+1)$-dimensional tensors. For instance, $\nabla_{\mu}\, h_{\nu \theta}$ becomes $\nabla_{\mu}\, A_{\nu} - \hbox{tan}\theta\, h_{\mu \nu} + \hbox{tan}\theta\, \frac{1}{\ell^2}\,g_{\mu \nu}\, \phi$. In Appendix \ref{DR}, under mild assumptions, we prove that the semi-direct product waveguide Eq.(\ref{metricre}) is the unique compactification that preserves covariance of tensors. 

\begin{figure}[!h] 
 \label{fig1}
\centering
{\includegraphics[scale=0.46]{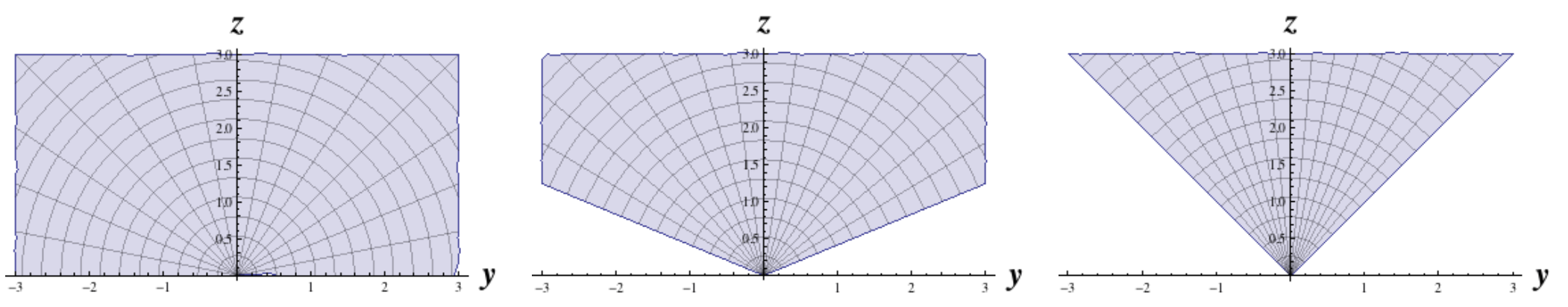}} \quad \quad \quad \quad
\caption{ \sl Anti-de Sitter waveguide: The left depicts a slice of AdS space in Poincar\'e coordinates $(y,z)$. In polar coordinates $(\rho, \theta)$, the AdS boundary is located at $\theta = \pm \pi/2 (z=0)$. The waveguide is constructed by taking the angular domain $ - \alpha \le \theta \le + \alpha$ for $\alpha < \pi/2$, as in the middle figure. For example, the waveguide for $\alpha=\pi/4$ is given in the right figure. }\label{fol}
\end{figure}

We can explicitly construct the semi-direct product metric from appropriate foliations of AdS$_{d+2}$ space. 
We start from the Poincar\'e patch of AdS$_{d+2}$ space and change bulk radial coordinate $z$ and another spatial coordinate  $y$ to polar coordinates, $z =\rho \, \hbox{cos}\theta$, $y = \rho \, \hbox{sin}\theta$ \footnote{The choice of spatial Poincar\'e direction ``$y$" does not play a special role. It can be chosen from any of the $SO(d)/SO(d-1)$ coset space. The semi-direct product structure can be straightforwardly generalized to other descriptions of the (A)dS space. See Appendix \ref{DR}}. With this parametrization, the AdS$_{d+2}$ space can be represented as a fibration of AdS$_{d+1}$ space over the interval, $\theta \in [-\frac{\pi}{2},\frac{\pi}{2}]$:
\begin{align}
\d s_{d+2}{}^2&=\frac{\ell^2}{z^2}\left(-\d t^2 + \d \vec{x}^{\,2}+\d y^2+\d z^2\right) \nonumber =\frac{\ell^2}{\rho^2 \, \hbox{cos}^2 \theta}\left(-\d t^2+\d \vec{x}^{\,2}+\d \rho^2 + \rho^2\, \d \theta^2 \right) \\
&=\frac{1}{\hbox{cos}^2 \theta}(\d s_{d+1}{}^2 + \ell^2\,\d \theta^2) \, . 
\end{align}
The boundary of AdS$_{d+2}$ space is at $\theta = \pm \frac{\pi}{2}$. From this foliation, we can construct the AdS waveguide by taking the wedge $- \alpha \le \theta \le \alpha$ where $\alpha < {\pi \over 2}$. See Fig.\ref{fol}. Note that this waveguide is embeddable to string theory: such geometry arises as a solution of Type IIB supergravity for nontrivial dilaton and axion field configurations and is known as the Janus geometry \cite{Bak:2003jk}.

An important consequence of  compactifying along non-isometry direction is the appearance of the conformal factor $\frac{1}{\hbox{cos}^2 \theta}$. 
One might try an alternative compactification scheme of AdS tube by putting periodic boundary condition that identifies the two boundaries at $\theta  = \pm \alpha$. This is not possible. The vector $\partial_\theta$ is not a Killing vector, so although the metric at  hyper-surfaces $\theta = \pm \alpha$ are equal, their first derivatives differ each other. We reiterate that the AdS waveguide is the unique choice for tunable compactification. Another consequence is that the integration measure of the waveguide is nontrivial
\beqa
\d \mbox{Vol}(AdS_{d+2}) = \d \mbox{Vol}(AdS_{d+1}) \ {\d \mu}_{d+2} [\theta] \quad \mbox{where} \quad \d \mu_{d+2} [\theta] := {\d \theta \over (\cos \theta)^{d+2}} \quad ( - \alpha \le \theta \le \alpha). \ \ 
\label{measure}
\eeqa

Before concluding this section, we introduce the notations that will be extensively used in later sections. We introduce the mode functions of AdS waveguide as follows:
\vskip0.3cm
\begin{tcolorbox}
\vskip-0.8cm
\begin{eqnarray}
\Theta^{s\vert S}_n(\theta)  &=& \mbox{$n$-th mode function for $(d+1)$-dimensional spin-$s$ component that arise from}
\nonumber \\
&&\mbox{ $(d+2)$-dimensional spin-$S$ field upon waveguide compactification  }.
\label{theta}
\end{eqnarray}
\vskip0.2cm
\end{tcolorbox}
\vskip0.3cm
\noindent    
Evidently, $ s = 0, 1, \cdots, S$. We also introduce the first-order differential operators $\L_n$ ($n \in \mathbb{Z}$) of Weyl scaling weight $n$ in the Hilbert space ${\cal L}^2[-\alpha, \alpha]$ spanned by the above mode functions:
\vskip0.3cm
\begin{tcolorbox}
\vskip-0.1cm
\beq 
\, \, \L_n= {1 \over \ell} \big( \pd_\theta+n\,\tan\theta \big) = {1\over \ell} (\cos \theta)^n \partial_\theta (\cos \theta)^{-n}\,.
\label{Lop}
\eeq
\end{tcolorbox}
\vskip0.3cm
\noindent    
From the general covariance, it follows that all of $\partial_\theta$ derivatives in the (A)dS wavegude are in the combination of these operators. So we will use $\L_n$'s to express Kaluza-Klein normal-mode equations, gauge transformations and boundary conditions. As we will see, the Sturm-Liouville differential operator acting on spin-$s$ field will carry the Weyl scaling weight $(d - 2s)$. The differential operator is quadratic in $\partial_\theta$, so acting on inner product of spin-$s$ fields defined by the integration measure in Eq.(\ref{measure}),  $\L_n$ and $L_{d-2s-n}$ are adjoint each other.  

A comment is in order. In this paper, we mainly concentrate on AdS waveguide.
However, we can straightforwardly convert the results to the dS waveguide 
which we introduce in Appendix \ref{DR}.
The only technical difference is to replace $(\tan\theta)$ to $(-\tanh\theta)$, as can be seen in Table \ref{AdStoAdS}.
The spectral differential operators in dS are obtainable by changing the tangent functions in Eq.\eqref{Lop}.

\section{AdS Waveguide Spectrum of Spin-One Field}\label{spin-1 wave}\label{sec4}
In this section, we focus on the lowest spin field,  spin-one, in AdS space and systematically work out Kaluza-Klein compactification on the Janus waveguide. In section \ref{sec1example},  we learned that boundary conditions of different polarization fields are combined one another to facilitate the Stueckelberg mechanism. In our AdS compactification, where the semi-direct product structure is the key feature, the choice of boundary conditions were left unspecified a priori. 
Here, we develop methods for identifying the proper boundary conditions. As this will be extended to higher-spin fields in later sections, modulo some technical complications (some of which actually open new physics), we will explain in detail
how the boundary conditions are identifiable. 

Through lower-spin examples, spin-one in this section and spin-two and spin-three in later sections, we shall compare two alternative but equivalent methods at the level of quadratic part of the action. One method is using the equations of motions, while the other method is using the gauge transformations. As the equation of motion contains increasingly complex structure for higher spin field (even at the linearized level), the first method is less practical for adopting to general higher-spin fields.  The second method is relatively easier to deal with and can be applied to general higher spin fields. The second method has one more important advantage: the dimensionally reduced equations of motion can be derived by the second method. After compactification, the gauge transformations become Stueckelberg transformation. The point is that the Stueckelberg symmetries are as restrictive as the gauge symmetries (because the latter follows from the former, as we will show below), so it completely fixes the equations of motion for all massive higher-spin fields \cite{Zino1} to the same extent that the higher-spin gauge symmetries fix the equations of motion of higher-spin gauge fields. Therefore, it suffices to use the gauge transformations for obtaining information about the mass spectra of dimensionally reduced higher-spin fields. 

For foregoing analysis, we use the following notations and conventions.   
The capital letters $M, N, \cdots$ will be used to represent the indices of AdS$_{d+2}$:
they run from 0 to $d+1$. The greek letters $\mu, \nu, \cdots$ are the indices of AdS$_{d+1}$ space: they run from 0 to $d$. For the waveguide, index for the internal direction is $\theta$. Therefore, $M = \{\mu, \theta\}$.
The barred quantities represent tensors in AdS$_{d+2}$ space, while unbarred quantities are tensors of AdS$_{d+1}$. The AdS radius is denoted by $\ell$.

\subsection{Mode functions of spin-one waveguide}\label{spin1B}

We first consider the method using the equation of motion. 
The spin-one field equation in AdS$_{d+2}$ space decomposes into two polarization components:
\begin{align}
\sec^2\theta \; \bar{g}^{M N}\, \overline{\nabla}_M\, \bar{F}_{ \mu N}&
=\nabla^\nu\, {F}_{\mu\nu}
-\L_{d-2}\left(\L_0\, {A}_\mu - \pd_\mu\,\phi\right)=0, \label{spin11}\\
\sec^2\theta \; \bar{g}^{M N}\, \overline{\nabla}_M\, \bar{F}_{ \theta N}&
=\nabla^\mu \left( \L_0\, {A}_\mu - 
\partial_\mu \,\phi\right) =0\,,\label{spin10}
\end{align} 
where $\bar A_M = (\bar A_\mu, \bar A_\theta) := (A_\mu,  \phi)$. 
The $(d+1)$-dimensional fields $ A_\mu$, $ \phi$ can be mode-expanded
in terms of a complete set of mode functions $\Theta_n^{s|1}(\theta)$, labelled by the mode harmonics $n= 0, 1, 2, \cdots$,  on the interval $\theta\in[-\al,\al]$: 
\beqa
{A}_{\mu}=\sum_{n=0}^\infty\, A^{(n)}_{\mu}\,\Theta_n^{1|1}(\theta) \qquad \mbox{and} 
\qquad
{\phi} = \sum_{n=0}^\infty\, \phi^{(n)}\, \Theta_n^{0|1}(\theta). \label{procedure}
\eeqa
Mode functions are determined once proper boundary conditions are prescribed. 
As stated above, our key strategy is not to specify boundary conditions at the outset. Rather, we first require gauge invariance of various higher-spin fields 
and then classify all possible boundary conditions that are compatible with such gauge invariances. 

What we learn from Section \ref{sec2} is that boundary conditions, equivalently, mode functions for ${A}_{\mu}$ and for ${\phi}$ must be related each other such that each term of Eqs. \eqref{spin11} and \eqref{spin10} obey the same boundary condition. Otherwise, as we learned in Section \ref{sec2}, equations of motion are accompanied with independent boundary conditions for each field and there would be no degree of freedom left after dimensional reduction. Therefore, each term of Eq. \eqref{spin11} and Eq. \eqref{spin10} must to be expanded by the same set of mode functions. We find that this consistency condition leads to the relations 
\vskip0.2cm
\begin{tcolorbox}
\vskip-0.2cm
\begin{align}
\begin{pmatrix} 0 & \L_{d-2} \\ \L_0 & 0 \end{pmatrix} 
\begin{pmatrix} \Theta^{1|1} \\ \Theta^{0|1} \end{pmatrix} 
= 
\begin{pmatrix} c^{01} \Theta^{1|1} \\ c^{10} \Theta^{0|1} \end{pmatrix}\, 
\label{spin1spec}
\end{align}
\vskip-0.3cm
\end{tcolorbox}
\noindent  
among the spin-one modes and the spin-zero modes. 
Here, $c^{01}, c^{10}$'s are in general complex-valued coefficients. 
These equations reveals that the Sturm-Liouville (SL) operator $-\Delta_{(s)}$ in Eq.\eqref{spin11} that determines the mass spectra of spin-one field in $(d+1)$ dimensions is factorized to a product of two first-order elliptic differential operators, 
\beqa
\L_{d-2} \Theta^{0|1} = c^{01} \Theta^{1|1} \qquad \mbox{and} \qquad
\L_0 \Theta^{1|1} = c^{10} \Theta^{0|1}_n. 
\label{more}
\eeqa
We first note that $\L_0$ and $\L_{d-2}$ are adjoint each other with respect to the measure Eq.\eqref{measure} for the field strength of $(d+2)$-dimensional spin-one field strength:
\beqa
\int {\rm d} \mu_{d+2} [\theta] (\cos^2 \theta)^2 A(\theta) (\L_{d-2} B(\theta)) = 
\int {\rm d} \mu_{d+2} [\theta] (\cos^2 \theta)^2 (\L_0 A(\theta)) B(\theta)
\label{conjugate}
\eeqa
{\sl provided} we impose Dirichlet boundary conditions at $\theta = \pm \alpha$. 
Acting $\L_0$ and $\L_{d-2}$ to each equations of Eq.\eqref{more}, respectively, one obtains two Sturm-Liouville systems for spin-one and spin-zero modes,
\begin{align}
& \mbox{spin-one Sturm-Liouville}: \qquad \qquad  \Delta_{(1)} \Theta^{1|1} :=  - (\L_{d-2}\, \L_{0}) \, \Theta^{1|1}  = - c^{11}\, \Theta^{1|1} \label{MSE11} \\
& \mbox{spin-zero Sturm-Liouville:} \qquad \qquad \, \Delta_{(0)} \Theta^{0|1} :=  - (\L_0 \, \L_{d-2}) \, \Theta^{0|1} = - c^{00} \, \Theta^{0|1} \, , \label{MSE10}
\end{align}
with the property that the eigenvalues of respective spins are paired up
\beqa
- c^{11} = - c^{10} c^{01} = - c^{00} := \lambda^2 \, . 
\label{eigenvalue}
\eeqa
Here, we took into account that $\L_0$ and $\L_{d-2}$ are conjugate to each other and hence the eigenvalue $\lambda^2$ is positive semi-definite. This also puts  the coefficients $c^{01}, c^{10}$ pure imaginary, and the eigenvalues $c^{00}, c^{11}$ pure real. Hereafter, we label the eigenmodes in the ascending order of their eigenvalues and label them by $n=0, 1, 2, \cdots$, viz. $0 \le \lambda_0^2 \le \lambda_1^2 \le \lambda_2^2 \le \cdots$.
The relations Eq.\eqref{spin1spec} are then the statement that the SL spectrum is doubly degenerate: for a spin-zero mode $\Theta^{0|1}_m$ for some $m$ there ought to be present a spin-one mode $\Theta^{1|1}_n$ for some $n$ proportional to $\L_{d-2} \Theta^{0|1}_m$, and for a spin-one mode $\Theta^{1|1}_m$ for some $m$ there ought to be present a spin-zero mode 
$\Theta^{0|1}_n$ for some $n$ proportional to $\L_0 \Theta^{1|1}_m$. By the aforementioned ordering of eigenmodes, we labeled the paired spin-zero and spin-zero modes by the same index $m = n = 0, 1, 2, \cdots$. 

Stated differently, the two first-order elliptic operators $\L_{d-2}$, $\L_0$ are not only conjugate each other but also act as raising and lowering operators between spin-one and spin-zero modes with doubly degenerate spectra
\vskip0.2cm
\begin{tcolorbox}
\beq
\begin{array}{rcll}
 & \Theta^{1|1}_n & &\\
\L_{d-2} & \upharpoonleft \downharpoonright & \L_{0}& \qquad  \mbox{with} \qquad
-\lambda_n^2  =c_n^{11} = c_n^{00} = c_n^{10} c_n^{01}  \, .  \\
 & \Theta^{0|1}_n & &\\
\end{array}
\label{spin1}
\eeq
\end{tcolorbox}
\vskip0.1cm
\noindent 
As such, we refer to Eq.\eqref{spin1spec}, equivalently, Eq.\eqref{spin1} as ``spectrum generating complex" for spin-one field in AdS$_{d+2}$ space. 

In fact, we can attribute such double-degeneracy to a hidden supersymmetry of the complex Eq.\eqref{spin1} \footnote{Note that the hidden supersymmetry is unrelated to the ${\cal N}=2$ spacetime supersymmetry of ten-dimensional Type IIB supergravity in which the Janus geometry is a classical solution that preserves half of the supersymmetry.}.
To see this, let us combine the two Sturm-Liouville problems for spin-one and spin-zero modes into one Sturm-Liouville problem acting on two-component modes
\beqa
\mathbf{H} \left[ \begin{matrix} \Theta^{1|1}_n \\ \Theta^{0|1}_n \end{matrix} \right] = \left[ \begin{matrix} c^{11}_n & 0 \\ 0 & c^{00}_n \end{matrix} \right] \left[ \begin{matrix} \Theta^{1|1}_n \\ \Theta^{0|1}_n \end{matrix} \right] \, ,  \qquad
\mbox{where} \qquad \mathbf{H} = \left[ \begin{matrix} - \L_{d-2} \L_0 & 0 \\
0 & - \L_0 \L_{d-2}   \end{matrix} \right].
\label{susyform}
\eeqa
Let us also introduce two supercharges
\beqa
\mathfrak{Q} = \left[ \begin{matrix} 0 & 0 \\ i \L_0 & 0 \end{matrix} \right] 
\qquad \mbox{and} \qquad
\mathfrak{Q}^\dagger = \left[ \begin{matrix} 0 & i \L_{d-2} \\ 0 & 0 \end{matrix}
\right].
\eeqa
Then, the two-component SL operator $\mathbf{H}$ in Eq.\eqref{susyform} is nothing but
\beqa
\mathbf{H} = \{ \mathfrak{Q}, \mathfrak{Q}^\dagger \} \, , 
\qquad 
\{ \mathfrak{Q}, \mathfrak{Q} \} = \{ \mathfrak{Q}^\dagger, \mathfrak{Q}^\dagger \} = 0
\eeqa
and the spectral relation Eq. \eqref{spin1spec} is the statement that
\beqa
\mathfrak{Q} \left[ \begin{matrix} \Theta^{1|1}_n \\ 0 \end{matrix} \right]
= i c^{10}_n \left[ \begin{matrix} 0 \\ \Theta^{1|1}_n \end{matrix} \right]
\qquad
\mbox{and}
\qquad
\mathfrak{Q}^\dagger 
\left[ \begin{matrix} 0 \\ \Theta^{0|1}_n \end{matrix} \right]
= i c^{01}_n 
\left[ \begin{matrix} \Theta^{0|1}_n \\ 0 \end{matrix} \right] \, , 
\eeqa
reinforcing the fact in Eq.\eqref{conjugate} that $\L_{d-2}$ and $\L_0$ are conjugate to each other with respect to the inner product defined by the measure Eq.\eqref{measure}.  Moreover, the double degeneracy $c^{11}_n = c^{00}_n$ is a consequence of the fact that the supercharges $\mathfrak{Q}$ and $\mathfrak{Q}^\dagger$ commute with the SL operator $\mathbf{H}$. 

Returning to the Kaluza-Klein compactification, the relations Eq.\eqref{spin1spec}
allow to decompose the the $(d+1)$-dimensional field equations into spin-one and spin-zero modes as
\begin{align}
& \sum_{{n}}\left[\nabla^\mu\, F^{(n)}_{\mu\nu}+c_n^{01} (c^{10}_n \,A^{(n)}_\nu - \, \pd_\nu\,\phi^{(n)}) \right]\Theta_n^{1|1}=0\, \nonumber \\
& \sum_{{n}}\nabla^\mu \left[ c_n^{10}\, A^{(n)}_\mu - \partial_\mu \phi^{(n)} \right] \Theta_n^{0|1}=0\,.\label{Tem spin1eq}
\end{align}
We see that these equations take precisely the form of Stueckelberg coupling, 
triggering the Higgs mechanism for massive spin-one field in AdS$_{d+1}$ space with mass
\begin{equation}
 M_n^2 = - \lambda_n^2\,. 
 \label{eigenvalue}
\end{equation}

Recalling the flat space counterpart in Sec.2,  it may so happen that there exist massless -- thus unHiggsed --  spin-one or spin-zero fields in AdS$_{d+1}$ space. This is actually more interesting situation, so we would like to understand when and how this comes about. Recalling that the SL eigenvalue is product of $c^{10}_n$ and $c^{01}_n$ and that the eigenvalue $\lambda_n^2$ is positive semidefinite, there are three possible situations for the lowest eigenvalue:
\begin{itemize}
\item (1) Doubly Degenerate Kaluza-Klein Modes: This case is when both of $c^{01}_0, c^{10}_0$ are nonzero. This implies that the spin-zero eigenvalue $c^{00}_n$ and spin-one eigenvalue $c^{11}_n$ are nonzero for {\sl all} $n=0, 1, 2, \cdots$. By the spectrum generating relations Eq.\eqref{more}, none of the corresponding modes $\Theta^{0|1}_n$ and $\Theta^{1|1}_n$ are annihilated by $\L_0$ and $\L_{d-2}$, respectively. By the double degeneracy, the eigenvalue for spin-zero $-c^{00}_n$ and the eigenvalue for spin-one $-c^{11}_n$ are positive definite, and are paired up. The spectrum consists of doubly degenerate Kaluza-Klein modes. A special case is when both $c^{01}_0$ and $c^{10}_0$ become zero. In this case, the spectrum includes doubly degenerate ground modes.  Nevertheless, we shall distinguish these two cases.
\vskip0.2cm
\begin{tcolorbox}
\vskip-0.7cm
\beqa
\begin{matrix}
\mbox{double Kaluza-Klein modes:} &  \qquad \L_0 \Theta^{1|1}_0 \ne 0, & \qquad \L_{d-2} \Theta^{0|1}_0 \ne 0 \\
\mbox{double ground modes:} & \qquad \L_0 \Theta^{1|1}_0 = 0, & \qquad \L_{d-2} \Theta^{0|1}_0 = 0
\end{matrix}
\label{kkmode}
\eeqa
\end{tcolorbox}

\item (2) Spin-One Ground Mode: This case is when $c^{10}_0 \rightarrow 0$ from the situation (1), leading to $\L_{0} \Theta^{1|1}_0 = 0$.  This means that $\Theta^{1|1}_0$ is the ground mode with vanishing eigenvalue $c^{11}_0 =0$ of the spin-one SL operator $\L_{d-2} \L_0$. All higher modes, $\Theta^{1|1}_n$ for $n =1, 2, \cdots$ are necessarily massive. On the other hand, spin-zero eigenmodes $\Theta^{0|1}_n$ for $n = 0, 1, 2, \cdots$ have positive eigenvalue $c^{00}_n >0$. By the double degeneracy property, they are paired up with $\Theta^{1|1}_n$ for $n= 1, 2, \cdots$. 
There is one spin-one ground mode of zero eigenvalue. 
\begin{tcolorbox}
\vskip-0.7cm
\beqa
\mbox{spin-one ground mode}: \qquad \L_0 \Theta^{1|1}_0 = 0, \qquad \Theta^{0|1}_0 = 0.
\label{spin-one-ground-mode}
\eeqa
\end{tcolorbox}
\item (3) Spin-Zero Ground Mode: This case is when $c^{01} \rightarrow 0$ from the above situation (1), leading to $\L_{d-2} \Theta^{0|1}_0 = 0$. This means that $\Theta^{0|1}_0$ is the ground mode with vanishing eigenvalue $c^{00}_0$ of the spin-zero SL operator $\L_0 \L_{d-2}$.  All higher modes, $\Theta^{0|1}_n$ for $n=1, 2, \cdots$ are necessarily massive. On the other hand, spin-one eigenmodes $\Theta^{1|1}_n$ for $n=0, 1, 2, \cdots$ have positive eigenvalue $c^{11}_n >0$. By the double degeneracy property, they are paired up with $\Theta^{0|1}_n$ for $n= 1, 2, \cdots$.  There is one spin-zero ground mode of zero eigenvalue. 
\begin{tcolorbox}
\vskip-0.7cm
\beqa
\mbox{spin-zero ground mode}: \qquad
\Theta_0^{1|1} = 0, \qquad \L_{d-2} \Theta^{0|1}_0 = 0.
\label{spin-zero-ground-mode}
\eeqa
\end{tcolorbox}
\end{itemize}
These situations are depicted in Fig.2. 
\begin{figure}[!h] 
\vskip-3cm
\centering
\hskip-0.75cm
{\includegraphics[scale=0.8]{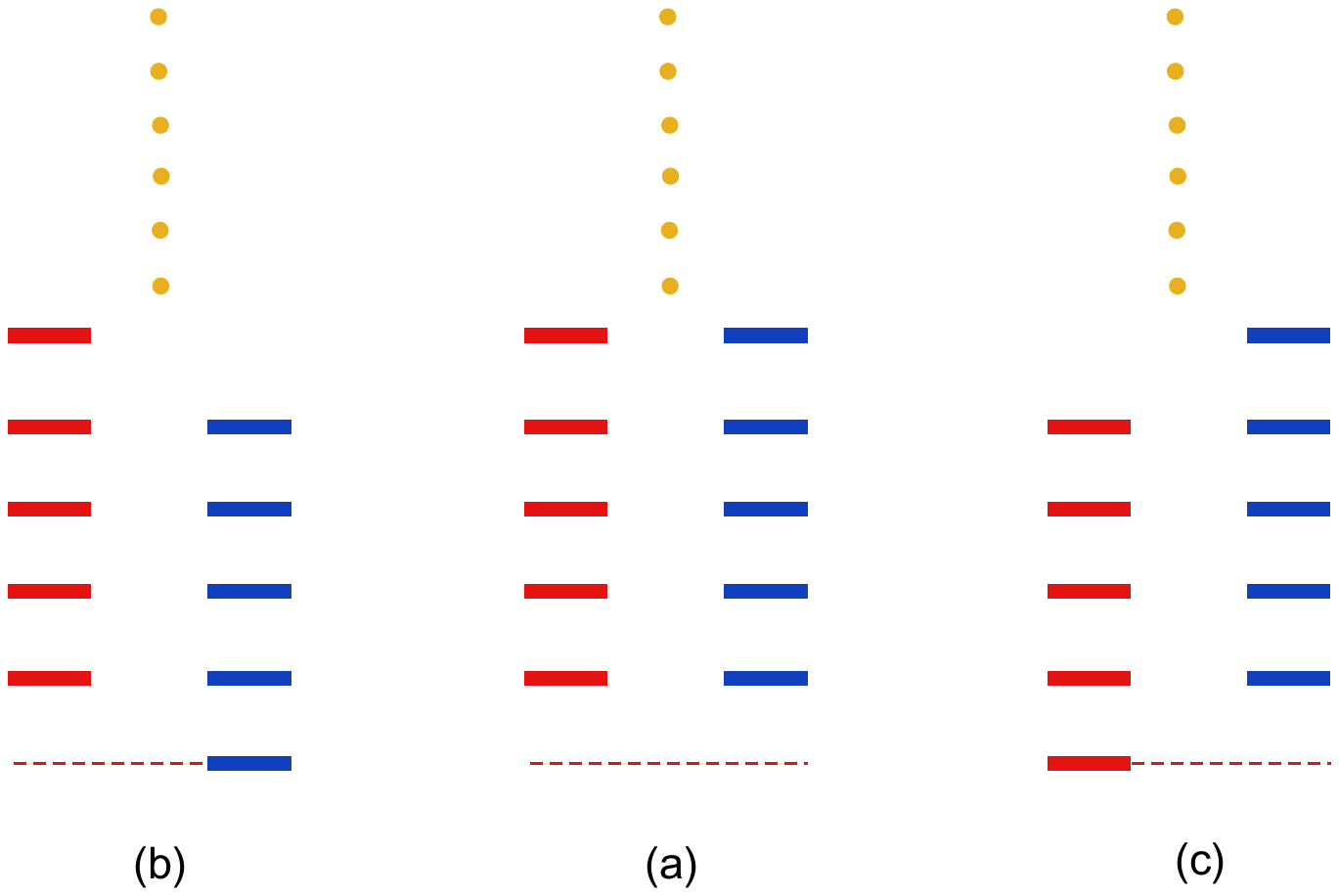}} \quad \quad \quad \quad
\vskip-11.5cm
\caption{\sl Various situations of double degeneracy between spin-zero (red color) and spin-one (blue color) modes. The middle spectrum (a) depicts the situation (1) that both $c^{01}_0$ and $c^{10}_0$ coefficients are nonzero. The spin-zero and spin-one modes are degenerate and have nonzero eigenvalues and so have no ground mode. The left spectrum (b) depicts the situation (2) that $c^{10} \rightarrow 0$. The spin-zero modes have no ground mode, while the spin-one modes have ground mode. The right spectrum (c) depicts the situation (3) that $c^{01} \rightarrow 0$. The spin-one zero modes have no ground mode, while the spin-one modes have ground mode. If both $c^{01}_0$ and $c^{10}_0$ are taken to zero, the spectrum is again doubly degenerate, but now starting from the ground mode. 
}\label{spectra}
\end{figure} 

Remarks are in order. First, for the doubly degenerate modes of nonzero eigenvalues, if one of them is normalizable then the other is normalizable automatically. Thus, not only their eigenvalues but also their multiplicities also
pair up. Second, the zero modes are solutions of the first-order differential equations $\L_0 \Theta^{1|1}_0 = 0$ and $\L_{d-2}\Theta^{0|1}_0 = 0$ subject to the Dirichlet boundary condition that render the two differential operators adjoint each other. As the measure $\d \mu [\cos \theta]$ is nonsingular and as the interval $[-\alpha, +\alpha]$ is finite, the existence of normalizable zero modes is always guaranteed. 

We can classify the pattern of spectral pair-up by  the elliptic index defined by  
\beqa
{\cal I}_{(s) = 1} &\equiv& {\sum_n} \mbox{Multiplicity}(\Theta^{0|1}_n) - {\sum_n}
\mbox{Multiplicity}(\Theta^{1|1}_n)  \nonumber \\
&=& \mbox{dim Ker} \ \L_{d-2} - \mbox{dim Ker} \ \L_0
\eeqa
Here, we used the fact that all Kaluza-Klein modes $n >0$ are paired up and so do not contribute to the index.  In the situation (1), we have 
\beqa
\begin{matrix}
c_0^{10} \ne 0, \ c_0^{01} \ne 0: & \qquad \mbox{dim Ker} \L_0 = 0 & \qquad \mbox{dim Ker} \ \L_{d-2} = 0 \\
c_0^{10}=0, \ c_0^{01} = 0 : & \qquad \mbox{dim Ker} \L_0 =1 & \qquad \mbox{dim Ker} \ \L_{d-2} = 1 
\end{matrix}
\eeqa
In the situations (2) and (3), we have
\begin{eqnarray}
\begin{matrix}
c_0^{10} = 0: & \qquad \mbox{dim Ker }\L_{0} \ = 1&   \qquad \mbox{dim Ker} \ \L_{d-2} = 0  
\\ 
c_0^{01} = 0: & \qquad \mbox{dim Ker } \L_{0} \ = 0 &   \qquad \mbox{dim Ker} \ \L_{d-2} \ =  1 
\end{matrix}
\label{susy-like}
\end{eqnarray}
So we see that spectral asymmetry is present whenever the elliptic index 
${\cal I}_{(s)=1}$ is nonzero. For the situation (1), the index is zero. For the situations (2) and (3), the index is nonzero.  

Once again, the above results have close parallels to the supersymmetric quantum mechanics. A vacuum $\vert \mbox{vac} \rangle$ of supersymmetric system preserves the supersymmetry if $\mathfrak{Q} \vert \mbox{vac} \rangle = 0$ or if $\mathfrak{Q}^\dagger \vert \mbox{vac} \rangle = 0$. We see that the situations (2) and (3) preserves the supersymmetry, while the situation (1) preserves the supersymmetry only if the ground modes are present. 

We can also determine the spectrum from the gauge invariances. In the waveguide, only those gauge transformations that do not change the boundary condition would make sense, viz. gauge fields and gauge transformation parameters ought to obey the same boundary conditions and hence the same mode functions. So, we have
\begin{align}
\delta A_\mu = \sum_n \delta A^{(n)}_{\mu}\, \Theta_n^{1|1} (\theta) 
&= \sum_n\, 
\partial_{\mu}\, \Lambda^{(n)}\, \Theta_n^{1|1} (\theta)\,, \\
\delta \phi \ \ = \sum_n \delta \phi^{(n)}\, \Theta_n^{0|1} (\theta) 
&= \sum_n\, 
\L_0\, \Lambda^{(n)}\, \Theta_n^{1|1} (\theta) 
= \sum_n c^{10}_n\, \Lambda^{(n)}\, \Theta_n^{0|1} (\theta).
\label{gaugemode}
\end{align}
We see that the relations Eq.\eqref{spin1}, which was obtained by the method using the equation of motion, can now be derived by the variations Eq.\eqref{gaugemode} and the Sturm-Liouville equations, Eqs.(\ref{MSE11},\ref{MSE10}).

Putting together the field equations and the gauge transformations of $n$-th Kaluza-Klein modes, we have
\begin{align}
&\nabla^{\mu}\,F^{(n)}_{\mu \nu} + c^{01}_n [c^{10}_n\, A^{(n)}_{\nu} - \partial_{\nu}\, \phi^{(n)} ]=0 \nonumber\\
&\nabla^\mu [\, c^{10}_n\, A^{(n)}_{\mu} - \partial_\mu \phi^{(n)} ] = 0 \label{stsystem} \\
&\, \delta A^{(n)}_{\mu} = \partial_{\mu}\, \Lambda^{(n)} \nonumber \\ 
&\, \delta \phi^{(n)} = c^{10}_n\, \Lambda^{(n)} \nonumber
\end{align}
We recognize these equations as precisely the Stueckelberg equations of motion 
and Stueckelberg gauge transformations that  describe a massive spin-one gauge field (Proca field) in AdS$_{d+1}$ space. Comparing them with the standard form of Stueckelberg system, we also identify the coefficients $c_n$'s with the Stueckelberg coupling, viz. the mass of the Proca field, $c^{10}_n = - c^{01}_n = M_n$ for all $n > 0$. 

The double degeneracy, as seen above, between spin-one and spin-zero fields is at the core of the Higgs mechanism, much as in the flat space counterpart in Sec. 2. The Stueckelberg coupling that realizes the Higgs mechanism follows from two ingredients. First, Kaluza-Klein modes of spin-one and spin-zero are coupled together, such that the second equation in Eq.\eqref{Tem spin1eq} follows from the first equation by consistency condition. Second, the factorization property that the SL operator $\Delta_{(s)}$ is a product of two first-order elliptic differential operators $\L_0$ and $\L_{d-2}$ implies that the spectrum of spin-one mode is equal to the spectrum of spin-zero mode. The spin-zero mode provides the Goldstone mode to the massive spin-one (Proca) field when the mass of spin-one is zero, and this picture continues to hold in AdS space. A novelty for the AdS space is that the scalar field, despite being a Goldstone mode, is massive.

The double degeneracy and hence the Higgs mechanism breaks down for the ground mode. From Eq.\eqref{Tem spin1eq}, we see that $c_n^{01}, c_n^{10}$ are the Stueckelberg coupling parameters. For the ground modes, either $c^{01}_0$, $c^{10}_0$ or both is set to zero and so the corresponding Stueckelberg couplings vanish. In terms of the hidden supersymmetry, we see that inverse of the Higgs mechanism takes place whenever the supersymmetry is unbroken.

\subsection{Waveguide boundary conditions for spin-one field}
Having identified the mode functions as well as raising and lowering operators relating them, we are now ready to examine boundary conditions these mode functions must satisfy. To simplify and systematize the analysis, we shall first concentrate on boundary conditions which do not contain derivatives higher than first-order \footnote{For $s \ge 2$, as we shall show in next section, boundary conditions necessarily involve higher derivative terms in order to accommodate all possible mass spectra of higher-spin fields. In Section \ref{sec7}, we discuss in detail origin and physical interpretation of higher-derivative boundary conditions (HDBC). }.  In this case, all possible boundary conditions can be related to all possible choice of mode functions with nontrivial ground modes. This is because the ground modes Eqs.(\ref{spin-one-ground-mode}, \ref{spin-zero-ground-mode}), which are valid everywhere in the waveguide $\theta = [-\alpha, \alpha]$, trivially satisfy the zero-derivative (Dirichlet) boundary condition and the one-derivative (Neumann) boundary condition, respectively, at $\theta = \pm \alpha$. Moreover, by an argument similar to the reasoning around Eq.\eqref{over determinent} in flat space, we see that the situation in Eq.\eqref{kkmode} does not give rise to massless fields and that the situations in Eqs.(\ref{spin-one-ground-mode}, \ref{spin-zero-ground-mode}) do give rise to massless spin-one and spin-zero fields, respectively. 

So, to have massless fields in AdS$_{d+1}$ space, we can choose the boundary conditions as 
 \beq
 \left\{ \begin{array}{lll}
\Theta^{1|1}\,|_{\theta=\pm\alpha}=0\,,\quad& \L_{d-2}\,\Theta^{0|1}\,|_{\theta=\pm\alpha}=0\,&\qquad\text{Dirichlet}\\
 \L_{0}\,\Theta^{1|1}\,|_{\theta=\pm\alpha}=0\,,\quad&  \Theta^{0|1}\,|_{\theta=\pm\alpha}=0\,&\qquad\text{Neumann}
 \end{array} \right.
 \label{spin1bc}
 \eeq
 that give rise to massless spin-zero field and spin-one field, respectively, in AdS$_{d+1}$ space. The first corresponds to the situation that $\Theta^{0|1}$ is a zero mode belonging to Ker $\L_{d-2}$ and the second case corresponds to the situation that $\Theta^{1|1}$ is a zero mode belonging to Ker $\L_0$. 
 
\begin{figure}[!h] 
\centering
{\includegraphics[scale=0.55]{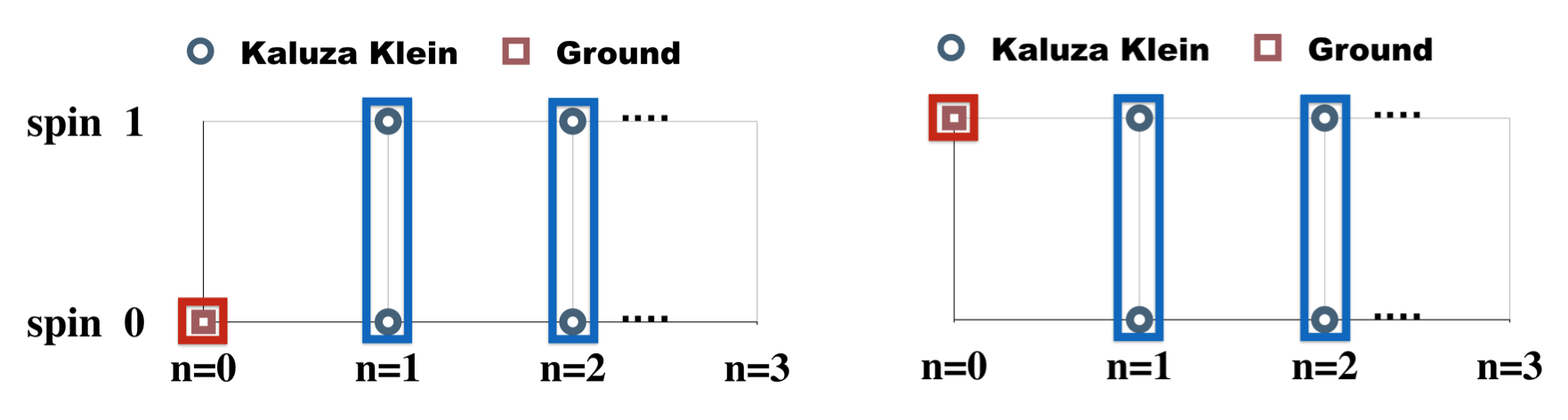}} \quad \quad \quad \quad
\caption{\sl 
Mass spectrum $(M_n, s)$ for spin-one field in AdS waveguide. The left is for Dirichlet boundary condition to spin-one component, and the right is for Dirichlet boundary condition to spin-zero component.  The red squares are the ground modes, while the blue circles are the Kaluza-Klein modes. Circles inside the same rectangle have the same eigenvalue and form a Stueckelberg system of massive spin-one field. 
}\label{spectra1}
\end{figure}

For each of the above two boundary conditions, the mass spectrum is determined by the Sturm-Liouville problem Eq.\eqref{MSE10}. We emphasize again that the above choice of boundary conditions put all Kaluza-Klein modes to Stueckelberg coupling, leading to Higgsed spin-one fields. The ground mode of Dirichlet boundary condition, $ \Theta^{1|1}_0 (\pm \alpha)=0 $ has vanishing mass  for spin-zero field and there is no massless spin-one field. The ground mode of Neumann boundary condition, $\L_0 \Theta^{1|1}_0 (\pm \alpha) =0$  has vanishing mass for spin-one field and there is no massless spin-zero field. 
So, we see that the two possible boundary conditions Eq.(\ref{spin1bc}) are precisely the AdS counterparts of ``vector" and ``scalar" boundary conditions for flat space waveguide studied in Sec. \ref{sec1example}.
We summarize the spectrum of each boundary conditions in Fig. \ref{spectra1}.

Our result for the ground modes, which comprises of massless spin-one or spin-zero fields, fits perfectly to the $\mathfrak{so}(d,2)$ representation theory of AdS$_{d+1}$ space. The set of normalizable solutions to the free field equation form a $\mathfrak{so}(d,2)$-module~\footnote{We summarize our convention of the $\mathfrak{so}(d,2)$-module and representations in Appendix \ref{Verma M}.
}.  Consider an irreducible  representation $\mathcal{D}(\Delta, s)$ of $\mathfrak{so}(d,2)$. 
The conformal weight $\Delta$ (the Casimir of $\mathfrak{so}(2)$ subalgebra) is related to the mass-squared of spin-$s$ field by
\beq
m_{\rm{\text spin}-0}^2\,\ell^2=\Delta\left(\Delta-d\right)\,,\qquad
m_{\rm{\text spin}-1}^2\,\ell^2=\Delta\left(\Delta-d\right)+\left(d-1\right)\, . 
\label{massrelation}
\eeq
The ground modes of Dirichlet and Neumann boundary conditions are massless spin-zero  and spin-one fields, respectively. We see from Eq.(\ref{massrelation}) that each of them corresponds to the $\mathfrak{so}(d,2)$ representations, $\mathcal{D}\left(d,\,0\right)$ and $\mathcal{D}\left(d-1,\,1\right)$, respectively.
Moreover, both are irreducible parts of the reducible Verma $\mathfrak{so}(d,2)$-module $\mathcal{V}\left(d-1,\,1\right)$, viz. 
\beq
\mathcal{V}\left(d-1,\,1\right)= 
\underbrace{\mathcal{D}\left(d-1,\,1\right)}_{\rm Neumann} \oplus
\underbrace{\mathcal{D}\left(d,\,0\right)}_{\rm Dirichlet}\,.
\eeq
The pattern that ground mode comes from the irreducible representations of reducible Verma module
continues to hold for higher-spin fields as well, and is an integral part of our main results in this paper. 

Summarizing, from Kaluza-Klein compactification of spin-one field in AdS$_{d+2}$ space, we take following lessons.
\vskip0.2cm
\begin{tcolorbox}
\vskip-0.2cm
\begin{itemize}
\item The mode functions of different spins in AdS$_{d+1}$ space are related to each other, which permits Stueckelberg structure. For spin-one, this relation is shown in Eq.\eqref{spin1}. 
\item It is known that free part of Stueckelberg equation and action are uniquely determined by Stueckelberg gauge transformations.  Therefore, we could derive the lower-dimensional equations of motion just from consideration of the lower-dimensional gauge transformations.
\end{itemize}
\end{tcolorbox}

\section{Waveguide Spectrum of Spin-Two Field}\label{spin-2 wave}\label{sec5}
In this section, we extend the analysis to spin-two field in the AdS waveguide. 
The idea is basically the same as the spin-one case, but the result turns out more interesting for the ground modes. We shall present the analysis as closely parallel as possible to the spin-one case and highlight salient differences that begin to show up for spin two and higher. 

\subsection{Mode functions of spin-two waveguide}
We begin with the method based on the equation of motion. 
The Pauli-Fierz equation of motion for a massive spin-two field in AdS$_{d+2}$ is given by
\beqa
{\cal K}_{MN}(\bar{h})-(d+1)\,(2\,\bar{h}_{MN}-\bar{g}_{MN}\, \bar{h} ) 
- M^2\, (\bar{h}_{MN}-\bar{g}_{MN}\, \bar{h})=0\,, 
\eeqa
where $M^2$ is the mass-squared, $\bar{g}_{MN}$ is the metric of AdS$_{d+2}$ space, and ${\cal K}_{MN}(\bar{h})$ is the spin-two Lichnerowicz operator:
\begin{align}
{\cal K}_{MN}(\bar{h}) =& \square\, \bar{h}_{MN}
- (\overline\nabla^L\, \overline\nabla_N\, \bar{h}_{ML}
+ \overline\nabla^L\, \overline\nabla_M\, \bar{h}_{NL} )
+\bar{g}_{MN}\,\overline\nabla_K\, \overline\nabla_L\, \bar{h}^{KL}
+\overline\nabla_M\, \overline\nabla_N\, \bar{h}-\bar{g}_{MN}\, \overline\square\, \bar{h} \, ,  
\end{align}
where $\bar{h}$ denotes for the trace part, $\bar{g}^{MN} \bar{h}_{MN}$.  After the compactification, the $(d+2)$-dimensional spin-two field is decomposed 
to $(d+1)$-dimensional spin-two, spin-one, and spin-zero component fields, respectively:
\beq
h_{\mu\nu}=\bar{h}_{\mu\nu}+\frac{1}{d-1}\,g_{\mu\nu}\,\bar{h}_{\theta\theta}\,,\qquad  
\bar{h}_{\mu\theta}=A_{\mu}\,,\qquad    \bar{h}_{\theta\theta}=\phi\,.\label{lin spin2}
\eeq
Note that the spin-two field $h_{\mu\nu}$ is defined by the linear combination of $\bar{h}_{\mu\nu}$ and $\bar{h}_{\theta\theta}$\footnote{The equations of motion have cross terms between $\bar{h}$ and $\nabla^2\phi$. This linear combination removes such cross terms.
This specific combination is also the linear part of canonical metric in the original Kaluza-Klein compactification, $\bar{g}_{\mu\nu}=e^{\phi/(d-1)}\,g_{\mu\nu}$.}.

The massless spin-two equation of motion in AdS$_{d+2}$ space decomposes into equations of motion for the 
component fields $(h_{\mu \nu},\, A_\mu,\, \phi)$ in AdS$_{d+1}$ space:
\begin{align}
&{\cal K}_{\mu\nu}(h)
-d\left(2\,h_{\mu\nu}-g_{\mu\nu}\, h \right) 
+\L_{d-2}\,\L_{-2}\left( {h}_{\mu\nu}-g_{\mu\nu}\,  {h}\right)\nn\\
&
\hskip1.0cm -\L_{d-2}\left(\nabla_\mu\, {A}_\nu+\nabla_\nu\, {A}_\mu-2\, g_{\mu \nu}\, \nabla^\rho\, {A}_\rho \right)
+\frac{d}{d-1}\, g_{\mu\nu}\, \L_{d-2}\,\L_{d-3}\, \phi =0\,,\label{spin22}\\
&\nabla^\mu\, {F}_{\mu\nu} -2\,d\, {A}_\nu
-\L_{-2}\,\left(\nabla^\mu\, {h}_{\mu\nu}-\nabla_\nu\, {h}\right)
-\frac{d}{d-1}\,\L_{d-3}\,\nabla_\nu\, \phi=0\,,\label{spin21}\\
&\square\, \phi-\left(\frac{d+1}{d-1}\,\L_{-1}\,\L_{d-3}+d+1\right) \phi-2\, \L_{-1}\,\nabla^\mu\, {A}_\mu+\L_{-1}\,\L_{-2}\, {h}=0\, , 
\label{spin20}
\end{align}
where $h$ is the trace part, $g^{\mu \nu} h_{\mu \nu}$, in AdS$_{d+1}$ space. The mode expansion of $(d+1)$-dimensional spin-two, spin-one and spin-zero component fields reads
\begin{align}
&&{h}_{\mu\nu} = \sum_n h^{(n)}{}_{\mu\nu}\, \Theta^{2|2}_n (\theta)\,, 
\qquad
{A}_{\mu} =\sum_n  A^{(n)}{}_{\mu}\, \Theta^{1|2}_n (\theta)\,, 
\qquad
{\phi} =\sum_n  \phi^{(n)}\, \Theta^{0|2}_n (\theta)\label{spin 2 mode}\,.
\end{align}
From Eqs.(\ref{spin22}, \ref{spin21}, \ref{spin20}), 
we again expect relations among mode-functions
which can be summarized by the following two first-order coupled differential equations
\vskip0.2cm
\begin{tcolorbox}
\vskip-0.5cm
\begin{eqnarray}
&& \begin{pmatrix} 0 & \L_{d-2}  \\ \L_{-2}  & 0 \end{pmatrix} 
\begin{pmatrix} \Theta^{2|2}_n \\ \Theta^{1|2}_n \end{pmatrix}
=
\begin{pmatrix} c^{12}_n \Theta^{2|2}_n \\ c^{21}_n \Theta^{1|2}_n \end{pmatrix}
\label{21ladder} \\
\nonumber\\
&& \begin{pmatrix}  0 & \L_{d-3} \\  \L_{-1}  & 0 \end{pmatrix}
\begin{pmatrix} \Theta^{1|2}_n \\ \Theta^{0|2}_n \end{pmatrix}
=
\begin{pmatrix} c^{01}_n \Theta^{1|2}_n \\ c^{10}_n \Theta^{0|2}_n  \end{pmatrix} \, . 
\label{10ladder}
\end{eqnarray}
\end{tcolorbox}
\noindent
Here, $c_n$'s are pure imaginary coefficients. Compared to the previous section, we now have two sets of raising and lowering operators, one for connecting spin-zero and spin-one and another for connecting spin-one and spin-two, respectively. Accordingly, we have two pairs of Sturm-Liouville problems.  
The Eq. \eqref{21ladder} leads to the first set of Sturm-Liouville problems for spin-two and spin-one, respectively:
\begin{align}
& \L_{d-2}\,\L_{-2}\, \Theta^{2|2}_n 
= c^{21}_n\, c^{12}_n\, \Theta^{2|2}_n
=-M_n^2\, \Theta^{1|2}_n\,,
\nonumber \\
& \L_{-2}\,\L_{d-2}\, \Theta^{1|2}_n 
= c^{12}_n\, c^{21}_n\, \Theta^{1|2}_n
=-M_n^2\, \Theta^{1|2}_n\,.\label{spin2 MSE}
\end{align}
The Eq.\eqref{10ladder} leads to the second set of Sturm-Liouville problems for spin-one and spin-zero, respectively:
\begin{align}
& \L_{d-3}\,\L_{-1}\, \Theta^{1|2}_n 
= c^{10}_n\, c^{01}_n\, \Theta^{1|2}_n\,,
\nonumber \\
& \L_{-1}\,\L_{d-3}\, \Theta^{0|2}_n 
= c^{01}_n\, c^{10}_n\, \Theta^{0|2}_n\,.
\end{align}
The two sets of equations appear overdetermined, as the spin-one mode function $\Theta^{1|2}_n$ is the eigenmode that participate in two separate Sturm-Liouville problems. However, it can be shown that the two Sturm-Liouville problems are actually related each other upon using the relations 
\beq
{\L_m\, \L_n - \L_{n-1}\, \L_{m+1}=(n-m-1)} \, . 
\eeq
This also leads to relations to the two sets of eigenvalues that appear in the two separate sets of Sturm-Liouville problems. 
\beq
c^{10}_n c^{01}_n =c^{21}_n c^{12}_n-(d-1).
\label{relation}
\eeq
It should be noted that the difference between spin-two eigenvalues and spin-one eigenvalues is linearly proportional to the spacetime dimensions (measued in unit of the AdS$_{d+1}$ curvature scale).
 
We can summarize the coupled Sturm-Liouville problems by the following spectrum generating complexes
\vskip0.3cm
\begin{tcolorbox}
\beq
\begin{array}{rcll}
 & \Theta^{2|2}_n & &\\
\L_{d-2} & \upharpoonleft \downharpoonright & \L_{-2}& \quad : \quad  -M_{n,2|2}^2=-\lambda_n^2 = c^{21}_n c^{12}_n \\
 & \Theta^{1|2}_n & &\\
\L_{d-3} & \upharpoonleft \downharpoonright & \L_{-1}& \quad : \quad -M_{n,1|2}^2=-(\lambda_n^2 +d-1) =c^{10}_n c^{01}_n \\
& \Theta^{0|2}_n & &\\
\end{array}
\label{spin2}
\eeq
\end{tcolorbox}
\vskip0.1cm
\noindent
As in the spin-one counterpart, these complexes, defined by raising and lowering operators between $(d+1)$-dimensional fields of adjacent spins, are precisely the structure required by the Stueckelberg mechanism of spin-two field~\footnote{
Note, however, $M_{n,1|2}$ is not related to the mass-like term of spin-one field in Eq.\eqref{spin21}. It is only that $M_n=M_{n,2|2}$ is the mass of spin-two field in Eq. \eqref{spin22}.
}.
If $M_{n,\,2|2}$ and $M_{n,\,1|2}$ were nonzero, the corresponding modes among different spin fields combine and become the Stueckelberg spin-two system. From these complexes, we can draw two pieces of information. First, as anticipated from the flat space intuition, the Stueckelberg mechanism would work between two adjacent spin fields, one for spin-two and spin-one and another for spin-one and spin-zero. Second, the relation Eq.\eqref{relation}, which is already reflected in Eq.\eqref{spin2}, indicates that the Stueckelberg mechanism actually involve the whole tower of component spin fields. In the present case, this means that the spin-two, spin-one and spin-zero fields are all involved in the Higgs mechanism. 

Again, there are two special cases, vanishing $M_{n,\,2|2}$ or vanishing $M_{n,\,1|2}$. As these are important exceptional situations, leading to a new phenomenon involving so-called partially massless spin-two fields, we will analyze them separately in Sec. \ref{sec spin2 BC} with examples. There is also a special case of these two, namely, simultaneously vanishing $M_{n, \, 2|2}$ and $M_{n, \, 1|2}$. This case will lead to massless limit of all component spin fields.  

We can also obtain Eq.\eqref{spin2} from the method based on gauge invariances. 
The gauge transformations in AdS$_{d+1}$ space, with the gauge parameter $\bar{\xi}_M=\{\xi_\mu,\,\xi_\theta\}$, are decomposed into components
\begin{align}
\delta  {h}_{\mu \nu} &= \nabla_{(\mu}\, {\xi}_{\nu)} +\frac{1}{d-1} \, g_{\mu \nu} \,\L_{d-2}\,  {\xi}_{\theta}\,,
\nonumber \\
\delta { A}_\mu & =\frac{1}{2}\,\pd_\mu\, {\xi}_\theta+\frac{1}{2}\,\L_{-2}\,{ {\xi}}_\mu\,,
\label{spin2gauge}\\
\delta  {\phi} \, \, &=\L_{-1}\, {\xi}_\theta\nonumber\,.
\end{align}
We should stress that these component fields are in the basis of AdS$_{d+1}$ fields that diagonalize them at linearized level. The gauge transformations given above are those in this basis. 
  
To retain the gauge invariances, the mode functions of gauge parameter must be set the same as the mode functions of fields:
\beq
{\xi}_{\mu} = \sum_n \xi^{(n)}_{\mu}\, \Theta^{2|2}_n (\theta)\,,
\qquad \quad
{\xi}_\theta = \sum_n \xi^{(n)}_{\theta}\, \Theta^{1|2}_n (\theta)\,.
\eeq
By substituting these to Eq.\eqref{spin2gauge} and comparing mode expansion in the gauge variations, we see we can recover precisely the same raising and lowering operators as in Eq.\eqref{spin2}, which was previously derived from the field equations Eqs.(\ref{spin22}, \ref{spin21}, \ref{spin20}).

After the mode expansion, the component field equations read
\begin{align}
&{\cal{K}}_{\mu \nu} (h^{(n)}) - d\left[2\,h^{(n)}{}_{\mu \nu}-g_{\mu \nu}\,h^{(n)}\right] 
+c^{21}_n\,c^{12}_n\left[h^{(n)}{}_{\mu\nu}-g_{\mu\nu}h^{(n)}\right]
\nonumber \\ 
& \qquad  \qquad \quad -c^{12}_n\left[\nabla_{\mu}\,A^{(n)}{}_{\nu} + \nabla_{\nu}\, A^{(n)}{}_{\mu} - 2\,g_{\mu \nu}\, \nabla^{\rho}\, A^{(n)}{}_{\rho}\right]
 + c^{01}_n\, c^{12}_n\, \frac{d}{d-1}\,g_{\mu \nu}\,\phi^{(n)} = 0\,, \label{eqspin22} \\
&\nabla^\mu\, F^{(n)}{}_{\mu\nu}-2\,d\, A^{(n)}{}_\nu-c^{21}_n\, \nabla^\mu \left[ \, h^{(n)}{}_{\mu\nu}- g_{\mu \nu}\, h^{(n)}\right] -c^{01}_n\, \frac{d}{d-1}\,\nabla_\nu\, \phi^{(n)} =0\,, \label{eqspin21}\\
&\square\, \phi^{(n)} -\left[\frac{d+1}{d-1}\,c^{01}_n\, c^{10}_n +d+1\right] \phi^{(n)} -2\, c^{10}_n\, \nabla^\mu\, A^{(n)}{}_\mu+ c^{21}_n\, c^{10}_n\, h^{(n)} =0\,.
\label{eqspin20}
\end{align}
Their gauge transformations are
\begin{align}
\delta h^{(n)}_{\mu \nu} = \nabla_{( \mu}\, \xi^{(n)}_{\nu )} + \frac{c^{12}_n}{d-1}\, g_{\mu \nu}\, \xi^{(n)}\,, \qquad
\delta A^{(n)}_{\mu} = \frac{1}{2}\, \partial_{\mu}\, \xi^{(n)} + \frac{c^{21}_n}{2}\, \xi^{(n)}_{\mu}\,, \label{gauge2}
\qquad
\delta \phi^{(n)} = c^{10}_n \,\xi^{(n)}\,.
\end{align}
We see that this system, Eqs.(\ref{eqspin22},\ref{eqspin21},\ref{eqspin20},\ref{gauge2}), coincides precisely with the spin-two Stueckelberg system in AdS$_{d+1}$ space, once we redefine $c_n$'s as
\beqa
\, c^{12}_n=-\sqrt{2}M_n,\,\,\,
c^{21}_n=\frac{M_n}{\sqrt{2}}, \,\,\,
c^{01}_n=-\sqrt{\frac{d}{2\,(d-1)}(M_n{}^2+d-1)}, \,\,
c^{10}_n=\sqrt{\frac{2\,(d-1)}{d}(M_n{}^2+d-1)} . 
\nonumber
\eeqa

Time and again, the linearized gauge invariances uniquely fix the linearized field equations or equivalently the quadratic part of action. Therefore, from the knowledge of linearized gauge transformations Eq.\eqref{gauge2}, we can fully reconstruct the linearized field equations Eqs.(\ref{eqspin22},\ref{eqspin21},\ref{eqspin20}). In practice, the gauge transformations are much simpler to handle than the field equations. From now on, we shall analyze the spectrum  primarily using the linearized gauge invariances. Note that the modes with nonempty image of raising operators or nonempty image of lowering operators always combine together and undergo the Higgs mechanism for massive spin-two fields.

Before classifying possible boundary conditions, we summarize the Stueckelberg spin-two system and the Goldstone mode decomposition pattern of it. For general values of the masses, the Stueckelberg spin-two system describes the same physical degree of freedom as a massive spin-two field, having the maximal number of longitudinal polarizations. This is because the spin-one and spin-zero fields can be algebraically removed by the gauge symmetries Eq.\eqref{gauge2}, corresponding to the unitary gauge fixing. However, such gauge fixing is not possible were if the masses take special values:
\beq
\lambda^2_n=0\qquad \text{and}\qquad \lambda_n^2=-\frac{(d-1)}{\ell^2}\,.\label{spin2 special}
\eeq
At these special mass values, the Stueckelberg system breaks into two subsystems which can be deduced just from the gauge transformations. We now elaborate on this. 

For the situation that $\lambda_n=0$, the gauge transformations are
\beq
\delta\,h_{\mu\nu}=\nabla_{(\mu}\,\xi_{\nu)}\,,\qquad
\delta\,A_\mu=\frac12\,\partial_\mu\,\xi\,,\qquad
\delta\,\phi={1 \over \ell} \sqrt{\frac{2}{d}}(d-1)\,\xi\,.\label{StueckelbergG2}
\eeq
We see from the first transformation that the spin-two field ought to be massless gauge field as it has the spin-two diffeomorphism invariance. We also see that the remaining two equations precisely constitute the spin-one Stueckelberg system with $m^2=2\,d/\ell^2$. This implies that the Goldstone field of the massive spin-two is given by the massive spin-one system, which in turn was formed by the Stueckelberg system of massless spin-one and massless spin-zero fields~\footnote{Note that the normalization of each field is not the standard form.}. 

For the situation that $\lambda_n^2=-(d-1)/\ell^2$, a subtlety arises as the coefficients $c^{12}_n$ and $c^{21}_n$ are pure imaginary. Specifically, the relation Eq.(\ref{21ladder}) implies that one of the two mode functions $\Theta^{1|2}_n$, $\Theta^{2|2}_n$ and corresponding field become pure imaginary. We are thus led to redefine the mode functions $\tilde{\Theta}^{1|2}_n = \pm i \, \Theta^{1|2}_n$ and the fields $\tilde{A}_{\mu} = \pm i \, A_{\mu}$ \footnote{In the path integral formulation, this amounts to choosing that the integration contour purely imaginary.}. The gauge transformations now become
\begin{align}
\delta h_{\mu \nu} = \nabla_{( \mu}\, \xi_{\nu )} + \sqrt{\frac{2}{d-1}} \frac{1}{\ell}\, g_{\mu \nu}\, \xi\,, \qquad
\delta A_{\mu} = \frac{1}{2}\, \partial_{\mu}\, \xi + \sqrt{\frac{d-1}{2}}\frac{1}{2\,\ell}\, \xi_{\mu}\,, \qquad
\delta \phi = 0\,.\label{spin2PMgauge}
\end{align}
We see that the above redefinition does not alter the fact that the spin-two gauge transformations and spin-one gauge transformations are coupled each other. 
In fact, we recognize that these are precisely the gauge transformations for the Stueckelberg system of partially massless (PM) spin-two field \cite{Zino1}. 
We can always gauge-fix the spin-one field to zero, and the remanent gauge symmetry coincides with the partially-massless (PM) spin-two gauge symmetry \cite{PaMa}:
\beq
\delta h_{\mu\nu}=\nabla_\mu\,\nabla_\nu\,\lambda-\frac{1}{\ell^2}\,g_{\mu\nu}\,\lambda\,, 
\qquad  \mbox{where} \qquad
\lambda = \ell \sqrt{\frac{2}{d-1}}\, \xi \, .
\eeq
Therefore, as the mass-squared hits the special value $M_n^2 = - (d-1) / \ell^2$, the Stueckelberg system breaks into a spin-two partially-massless (PM) Stueckelberg system and a massive spin-zero field of mass-squared $m^2=(d+1)/\ell^2$, as  given in Eq.\eqref{eqspin20}.

This spectral decomposition pattern perfectly fits to the reducibility structure of the Verma $\mathfrak{so}(d,2)$-module $\mathcal{V}(\Delta,2)$ for spin-two field. For the special values of conformal weights, $\Delta=d$ and $\Delta=d-1$, the Verma module becomes reducible and break into
 \beqa
\mathcal{V}\left(d,\,2\right) \, \, &=& \, \,  
\underbrace{\mathcal{D}\left(d,\,2\right)}_{{\rm massless} \, s=2} \, \, \oplus \, \, 
\underbrace{\mathcal{D}\left(d+1,\,1\right)}_{{\rm massive} \, s=1}\,,\nonumber \\
\mathcal{V}\left(d-1,\,2\right) &=& 
\underbrace{\mathcal{D}\left(d-1,\,2\right)}_{{\rm partially \, massless}\,  s=2}\oplus
\underbrace{\mathcal{D}\left(d+1,\,0\right)}_{{\rm massive} \, s=0}\,.
\label{breaking verma2}
\eeqa
Here, $\mathcal{D}\left(d,\,2\right)$ and $\mathcal{D}\left(d-1,\,2\right)$ are the irreducible representations of massless and partially massless states, respectively.
Using the relation between the mass-squared and the conformal weights~\footnote{Here, we define the mass-squared equal to the mass-squared in flat space limit. Therefore, it differs from the mass-squared dictated by  the Fierz-Pauli equations. See Appendix \ref{Verma M}.}
\beq
m_{\rm{\text spin}-1}^2\,\ell^2=\Delta\left(\Delta-d\right)+\left(d-1\right) 
\qquad \mbox{and} \qquad 
m_{\rm{\text spin}-0,\,2}^2\,\ell^2=\Delta\left(\Delta-d\right)\,,\label{mass Delta}
\eeq
one finds that $\mathcal{D}\left(d+1,\,1\right)$ corresponds to the spin-one field with $m^2=2\,d/\ell^2$, 
and $\mathcal{D}\left(d+1,\,0\right)$ corresponds to the spin-zero field with $m^2=\left(d+1\right)/\ell^2$.
This result exactly matches with the spectral decomposition patterns we analyzed above.

Here, we tabulate the four types of fields that appear at four special values of masses, corresponding to the four irreducible representations that appear in Eqs.(\eqref{breaking verma2}). They will be shown to arise as the ground modes of the Sturm-Liouville problems with appropriate boundary conditions in 
section \ref{sec spin2 BC}.
\vskip1cm 
\begin{table}[h]
\centering
\begin{tabular}{|c|c|c|c|}
  \hline
\textbf{type}	&	$\mathcal{D}(\Delta,s)_{ \mathfrak{so}(d,2)}$	& field	& mass-squared \\
  \hline
\hline
\textbf{type \,\,  I}	&	$\mathcal{D}(d+1,1)$	& massive Stueckelberg spin-one	& $m^2=2\,d/\ell^2$ \\
\hline
\textbf{type \, II}	&	$\mathcal{D}(d+1,0)$	& massive spin-zero field	& $m^2=\left(d+1\right)/\ell^2$ \\
\hline
\textbf{type III}	&	$\mathcal{D}(\, d, \, 2 \, )$	& massless spin-two	& $m^2=0$ \\
\hline
\textbf{type IV}	&	$\mathcal{D}(d-1,2)$	& partially-massless Stueckelberg spin-two	& $m^2=-\left(d-1\right)/\ell^2$ \\
\hline
\end{tabular}
\vskip0.5cm
\caption{
\sl The types of field involved in the inverse Higgs mechanism when a spin-two Stueckelberg systems decompose into spin-two gauge field and Goldstone field. Type I and II are Goldstone fields of spin-zero and spin-one. In AdS space, these Goldstone fields are massive. Type III is massless, spin-two gauge field. Type IV is partially massless, spin-two gauge field.  }
 \label{spin2 actors}
\end{table}
\vskip0.5cm
\subsection{Waveguide boundary conditions for spin-two field}\label{sec spin2 BC}
With the spectrum generating complex at hand, we now move to classification  of possible boundary conditions.
In the spin-one counterpart, boundary conditions for different component fields (spin-one and spin-zero in that case) were related one another. This feature continues to hold for the spin-two situation.
For instance, suppose we impose Dirichlet boundary condition for the spin-one component field in AdS$_{d+1}$, $\Theta^{1|2}|_{\theta=\pm\alpha}=0$. Then, the spectrum generating complex Eq.\eqref{spin2} immediately imposes  unique boundary conditions for other component fields:
\beqa
\L_{-2}\,\Theta_n^{2|2}\sim\,\Theta_n^{1|2}\,,
&\quad &
\L_{-2}\,\Theta^{2|2}|_{\theta=\pm\alpha}=0\,,\nonumber\\
\L_{d-3}\,\Theta_n^{0|2}\sim\,\Theta_n^{1|2}\,,
&\quad &
\L_{d-3}\,\Theta^{0|2}|_{\theta=\pm\alpha}=0\,.\label{BC caused by}
\eeqa
Likewise, if we impose a boundary condition to one of the component fields,
the spectrum generating complex Eq.\eqref{spin2} uniquely fixes the boundary conditions of all other component fields. The simplest choice is to impose the Dirichlet boundary condition to one of the component fields. As there are $s+1 = 3$ component fields (spin-two, spin-one and spin-zero), there are then three possible boundary conditions~\footnote{
Note that the first and the third conditions are higher-derivative boundary conditions(HD BC). HD BC is not self-adjoint in the functional space ${\cal L}^2$, but can be made self-adjoint in a suitably extended functional space. We will explain this in Sec.\ref{sec7}.
}:
\beqa
\begin{array}{ccccccccc}
\textbf{B.C. 1:}&\quad
 	\{& \Theta^{2|2}|=0\,,&
\quad&   \L_{d-2}\,\Theta^{1|2}|=0\,,&
\quad&   \L_{d-2}\,\L_{d-3}\,\Theta^{0|2}| =0&\}\\
\textbf{B.C. 2:}&\quad
	\{& \L_{-2}\,\Theta^{2|2}| =0\,,&
\quad&  \Theta^{1|2}| =0\,,&
\quad&  \L_{d-3}\,\Theta^{0|2}| =0&\}\\
\textbf{B.C. 3:}&\quad
 	\{& \L_{-1}\,\L_{-2}\,\Theta^{2|2}| =0\,,&
\quad&   \L_{-1}\,\Theta^{1|2}| =0\,,&
\quad&  \Theta^{0|2}| =0&\}
 \end{array}\label{spin2 BC} \, , 
\eeqa
where $\Theta|$ is a shorthand notation for the boundary values of mode functions, $\Theta|_{\theta=\pm\alpha}$.
We reiterate that the boundary conditions on each set are automatically fixed by the spectrum generating complex Eq.\eqref{spin2}. 
We now examine mass spectra and mode functions for each of the three types of boundary conditions, Eq.\eqref{spin2 BC}.

To deliver our exposition clear and explicit, we shall analyze in detail for $d=2$, viz. compactification of AdS$_4$ to AdS$_3$ times the Janus wedge, where the mode solutions of the Sturm-Liouville problem, Eq.(\ref{spin2}), are elementary functions:
\beqa
\Theta^{2|2}&=&\left\{ \begin{array}{ll}
\sec\theta\left(\tan\theta\,\cos (z_n\theta)-z_n\,\sin(z_n\theta)\right)\,,\quad&\text{odd parity}\\ 
\sec\theta\left(\tan\theta\,\sin (z_n\theta)+z_n\,\cos(z_n\theta)\right)\,,&\text{even parity}\label{mode1}
\end{array} \right.\\
\Theta^{1|2}&=&\left\{ \begin{array}{ll}
\sec\theta\,\sin(z_n\theta)\,,\quad  &\text{odd parity}\\
\sec\theta\,\cos(z_n\theta)\,, & \text{even parity}\label{mode2}
\end{array}\right.\\
\Theta^{0|2}&=&\left\{ \begin{array}{ll}
\sec\theta\,\sin(z_n\theta)\,,\quad &\text{odd parity}\\
\sec\theta\,\cos(z_n\theta)\,, & \text{even parity}\label{mode3}
\end{array}\right.
\eeqa
with  $z_n^2=\lambda_n^2+1$. Note that the Sturm-Liouville equation and the boundary condition are symmetric under the parity $\theta \rightarrow -\theta$, so the modes are also classifiable as either odd or even under the parity. 

We begin our analysis with \textbf{B.C. 1}. Substituting  the above mode functions to the \textbf{B.C. 1}, we get the same expression for spin-two and spin-one component fields except the condition that the parity of mode functions must take opposite values:
\beq
\left\{ \begin{array}{ll}
\sec\theta\left(\tan\theta\,\cos (z_n\theta)-z_n\,\sin(z_n\theta)\right)|_{\theta=\pm\alpha}\,,\quad&\text{odd $\Theta^{2|2}$ and even $\Theta^{1|2}$}\\ 
\sec\theta\left(\tan\theta\,\sin (z_n\theta)+z_n\,\cos(z_n\theta)\right)|_{\theta=\pm\alpha}\,,\quad&\text{even $\Theta^{2|2}$ and odd $\Theta^{1|2}$}\label{BC11}
\end{array} \right. \, . 
\eeq
We also get the boundary condition for spin-zero component $\Theta^{0|2}$ as
\beq
\left\{ \begin{array}{ll}
z_n\,\sec\theta\left(\tan\theta\,\cos (z_n\theta)-z_n\,\sin(z_n\theta)\right)|_{\theta=\pm\alpha},\quad&\text{odd $\Theta^{(0|2)}$}\\ 
z_n\,\sec\theta\left(\tan\theta\,\sin (z_n\theta)+z_n\,\cos(z_n\theta)\right)|_{\theta=\pm\alpha},\quad&\text{even $\Theta^{(0|2)}$}
\end{array} \right. \, .  
\eeq
We note that, modulo the overall spectral factor $z_n$, this spin-zero boundary condition is the same as the  boundary condition Eq.\eqref{BC11}. This is not accidental but a consequence of  the spectrum generating complex Eq.\eqref{spin2} and the boundary condition Eq.\eqref{BC caused by}.

In general, solutions for each boundary condition, $z_n$, depend on the waveguide size parameter, $\alpha$. They are the AdS-counterpart of flat space compactification volume, and so $z_n$ and $\lambda_n$ would blow up as $\alpha$ is sent to zero. They also correspond to the ``Kaluza-Klein modes". For these modes, mode functions of different spin component fields couple together and form spin-two Stueckelberg system with mass-squared, $M^2_n=z^2_n-1$.
 
There are, however, two special limits that are independent of $\alpha$, $z_n=1$ and $z_n=0$. They correspond to ``ground modes" and have interesting features that are not shared by the Kaluza-Klein modes. First, masses of the ground modes are equal to the special masses Eq.\eqref{spin2 special} at which the unitary gauge-fixing ceases to work and the Stueckelberg system decomposes into subsystems. Second, mode function of some spin components are absent.
For $z_n=1$, the spin-two field is absent as $\Theta^{2|2}=0$ in this case. The spin-one and spin-zero fields combine and form the Stueckelberg spin-one system of \textbf{type I}. For $z_n=0$, only massive spin-zero field is present because $z_{n}=0$ is not a solution of boundary conditions Eq.\eqref{BC11} or corresponding mode function is 0. This spin-zero field is of \textbf{type II}.

\begin{figure}[!h] 
\centering
{\includegraphics[scale=0.52]{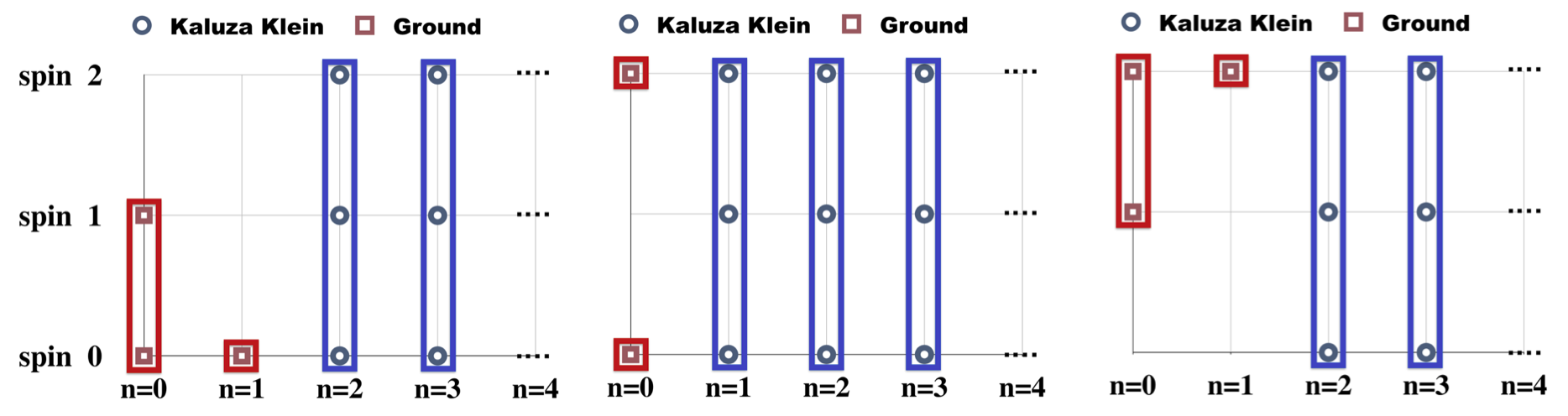}} \quad \quad \quad \quad
\caption{\sl
Spectral pattern for three types of Dirichlet conditions, \textbf{B.C.1}, \textbf{B.C.2} and \textbf{B.C.3} from left to right. The spin contents of each excitation level $n=0, 1, 2, \cdots$ are depicted.  Each point represents one mode: squares are from ground modes, while circles are from Kaluza-Klein modes. Points inside the same rectangle have the same eigenvalues and form Stueckelberg system. It is worth comparing this with the pattern for arbitrary higher-spin in Fig. \ref{spectrumspins}.
}\label{spectra2}
\end{figure}

Completing the analysis for all possible boundary conditions, we find the following spectrums of ground modes:
\beq
\begin{array}{ll}
\textbf{B.C. 1:}&\textbf{type I} \quad\text{and}\quad \textbf{type II}\\
\textbf{B.C. 2:}&\textbf{type II} \quad\text{and}\quad \textbf{type III}\\
\textbf{B.C. 3:}&\textbf{type III} \quad\text{and}\quad \textbf{type IV}
\end{array}
\eeq
We see that B.C.1 keeps mostly spin-zero, B.C.3 keeps mostly spin-two, while B.C.2 keeps spin-zero and spin-two. The complete spectrum of each set of boundary conditions is summarized in Fig. \ref{spectra2}.

\label{spin2 nonunitary}

The ground mode spectra associated with \textbf{B.C. 3} deserve further elaboration, as they in fact describe a non-unitary system. First, it is non-unitary because the mass-squared is below the Breitenlohner-Freedman bound of spin-two field in AdS$_{d+1}$ space. In section \ref{sec7}, we will explain Kaluza-Klein origin of this non-unitarity. Second, norms of some mode functions are negative-definite,
implying that the Hilbert space has the structure of indefinite metric, leading classically to unbound energy and instability and quantum mechanically to negative probability. Explicitly, for the mode functions
\beqa
\left\{ \begin{array}{ll}
\Theta_{1}^{2|2}=N_5\,\sec^2\theta\,&\qquad \textbf{type III} \\
\Theta_{0}^{2|2}=N_3\,\sec\theta\,\tan\theta\,, \qquad
\Theta_{0}^{1|2}=N_4\,\tan\theta\,&\qquad \textbf{type IV}\, , 
\end{array}\right.
\eeqa
the norms of $\Theta^{2|2}_0$ and $\Theta^{2|2}_1$ are $-2\alpha \, N_3{}^2$ and $-\frac{2}{\tan\alpha}N_5{}^2$ and hence negative-definite for all choices of $\alpha$. Such negative norms indicate that the higher-spin fields associated with these ground modes in {\bf B.C.3} have a wrong sign for their kinetic term. We will further confirm this more directly in subsection \ref{HD BC higher spin}, and offer intuitive explanations of them in terms of extra boundary degrees of freedom needed for extending the  indefinite Hilbert space to a definite Hilbert space. 

\label{alpha two pi limit1} 
As the waveguide size $\alpha$ tends to $\pi/2$,  the boundary surface of the waveguide approaches the timelike asymptotic boundary of the AdS$_{d+2}$ space. In other words, the Janus wedge decompactifies to the AdS$_{d + 2}$ space. In this limit, though,  the mass spectrum for each boundary conditions does not necessarily gets to the spectrum of massless spin-two field in AdS$_{d+2}$ space. The reason is that some of the boundary conditions we choose are singular in this limit as the associated mode function becomes ill-defined. Take for instance the mass spectrum for \textbf{B.C. 2}. It contains the massless spin-two ground mode as well as spin-zero ground mode whose normalized mode functions are
\beqa
\left\{ \begin{array}{ll}
\Theta_0^{0|2}=N_1\,\sec\theta\,,&\quad \textbf{type II}\,, \\
\Theta_0^{2|2}=N_2\, \sec^2\theta\,,&\quad \textbf{type III}\, , 
\end{array}\right.\quad
N_1=\frac{1}{\sqrt{2\alpha}}, \quad 
N_2=\sqrt{\frac{1}{2\tan\alpha}}\quad\label{massless}
\,.
\label{bc2spin2}
\eeqa
These ground-mode functions are not normalizable in AdS$_{d+2}$ space:
the normalized mode functions disappears as $N_2$ vanishes in the decompactification limit. This explains why there is no massless spin-two field in the ``dimensional degression" studied in Ref. \cite{Dimensional Degression}.
In section \ref{alpha two pi limit}, we will show that, for arbitrary spacetime dimension $d$ and spin $s$ of higher-spin field,  the mass spectrum of ``dimensional degression" spectrum is the spectrum of \textbf{B.C.1} in the decompactification limit.

Summarizing,
\begin{tcolorbox}
\vskip-0.2cm
\begin{itemize}
\item The mode functions of different spins in AdS$_{d+1}$ are related to each other by the spectrum generating complex Eq.\eqref{spin2}, 
whose structure is uniquely fixed by the consideration of Kaluza-Klein compactification of higher-spin gauge transformations.
\item 
At special values of masses, the Stueckelberg spin-two system decomposes into irreducible representations of massless or partially massless spin-two fields and massive Goldstone fields. The ground modes of Dirichlet boundary conditions Eq.\eqref{spin2 BC} are precisely these irreducible representations in Table \ref{spin2 actors} at the special mass values.  
\end{itemize}
\end{tcolorbox}

\section{Waveguide Spectrum of Spin-Three Field}\label{spin-3 wave}\label{sec6}
In this section, we further extend our analysis to spin-three field. This is not a mechanical extrapolation of lower spins. Compared to the situation of lower spins, for spin-three or higher, a new technical complication begins to show up: the spin-three gauge parameter needs to be traceless, $\bar{g}^{\mu\nu}\,\overline{\zeta}_{\mu\nu}=0$.
As we will demonstrate below, this complication forces us to consider appropriate linear combinations of gauge parameters. 

The field equation for a massless spin-three field $w_{\mu_1 \mu_2 \mu_3}$ in AdS$_{d+1}$ space is
\beq
{\cal K}_0(w)-\frac{4\left(d+1\right)}{\ell^2}\,w_{\mu_1\mu_2\mu_3}+\frac{2\left(d+1\right)}{\ell^2}\,g_{(\mu_1\mu_2}\,w^\lambda{}_{\mu_3)\lambda}=0 \, , 
\eeq
where ${\cal K}_0(w)$ is  the spin-three Lichnerowicz operator: 
\beqa
{\cal K}_0(w)&=&\nabla^\nu\,\nabla_\nu\, w_{\mu_1\mu_2\mu_3}
-\nabla^\nu\, \nabla_{(\mu_1} \,w_{\mu_2\mu_3)\nu}
+\nabla_{(\mu_1}\,\nabla_{\mu_2}\,w_{\mu_3)\lambda}{}^\lambda\\
&+&\nabla^\nu\,\nabla^\lambda \,g_{(\mu_1\mu_2}\,w_{\mu_3) \nu\lambda}
-\nabla^\nu\,\nabla_\nu\, g_{(\mu_1\mu_2}\,w^\lambda{}_{\mu_3) \lambda}
-\frac{1}{2} \,g_{(\mu_1\mu_2}\,\nabla_{\mu_3)}\,\nabla^\nu\, w^\lambda{}_{\lambda\nu}\,.
\eeqa
The $(d+2)$-dimensional spin-three field, $\overline{w}_{M_1M_2M_3}$, decomposes to $(d+1)$-dimensional component fields of different spins:
\beq
w_{\mu\nu\rho}= \overline{w}_{\mu\nu\rho}
+\frac{1}{d+1}\,g_{(\mu\nu}\,\overline{w}_{\rho)\theta\theta}, \quad 
h_{\mu\nu}=\overline{w}_{\mu\nu\theta}
+\frac{1}{d-1}\,g_{\mu\nu}\,\overline{w}_{\theta\theta\theta}\,,\quad 
A_\mu=\overline{w}_{\mu\theta\theta}\,,\quad
\phi=\overline{w}_{\theta\theta\theta}\,.\label{spin3 line}
\eeq
We defined component fields as linear combinations of different polarizations, which is the counterpart of spin-two situation Eq.\eqref{lin spin2}, to remove cross terms in field equations between parity-odd spin-three and spin-one, respectively, parity-even spin-two and spin-zero. For each component fields, the field equations  read
\beqa 
&& {\cal K}_0(w)-4\left(d+1\right)w_{\mu_1\mu_2\mu_3}
	+2\left(d+1\right)g_{(\mu_1\mu_2}w^\lambda{}_{\mu_3)\lambda}
	+\L_{d-2}\,\L_{-4}\left(w_{\mu_1\mu_2\mu_3}-
	g_{(\mu_1\mu_2}\,w^\lambda{}_{\mu_3)\lambda}\right)\nn\\
&& \hskip3cm			-3\,\nabla_{(\mu_1}\, {h}_{\mu_2\mu_3)}
				+3\,\L_{d-2}\left(\nabla_{(\mu_1}\, {h}
				-4\,\nabla^\lambda\, {h}_{\lambda(\mu_1}\right)\,g_{\mu_2\mu_3)}\nn\\
&& \hskip6cm		+\frac{3\,(d+2)}{d+1}\,\L_{d-2}\,\L_{d-3}\,g_{(\mu_1\mu_2}\, {A}_{\mu_3)}=0\,,\label{eqspin33}\\
&&{\cal K}_0( {h})-4\left(d+1\right){h}_{\mu_1\mu_2}+(3\,d+4)\,g_{\mu_1\mu_2}\,  {h}  -\frac{3}{2}\,\L_{-4}\,\L_{d-2}\, g_{\mu_1\mu_2}\,  {h}\nn\\
&& \hskip1cm -\frac{1}{2}\,\L_{-4}\left(g_{\mu_1\mu_2}\,\nabla^\lambda\, {w}_{\lambda\nu}{}^\nu+2\,\nabla^\lambda\, {w}_{\lambda\mu_1\mu_2}-\nabla_{(\mu_1}\, {w}_{\mu_2)\lambda}{}^\lambda\right)\nonumber\\
&& \hskip2cm -\frac{d+2}{d+1}\,\L_{d-3}\left(\nabla_\mu\, {A}_\nu+\nabla_\nu\, {A}_\mu-2 \,g_{\mu\nu}\,\nabla^\rho\, {A}_\rho\right)
		+\frac{d+2}{d-1}\,\L_{d-3}\,\L_{d-4}\, \phi=0\,,\label{eqspin32}\\
&& \nabla^\rho\, {F}_{\rho\mu}-\frac{d+3}{d+1}\,\L_{-3}\,\L_{d-3}\, {A}_\mu-3\left(d+1\right) {A}_\mu+\L_{-3}\,\L_{-4}\, {w}^\rho{}_{\rho\mu}
\nonumber\\
&& \hskip4cm -2\,\L_{-3}\left(\nabla^\rho \,{h}_{\rho\mu}-\nabla_\mu  \,{h}\right)
-\frac{d+1}{d-1}\,\L_{d-4}\,\nabla_\mu\, \phi=0\,,\label{eqspin31}\\
 &&\square\, \phi-\frac{2\,(d+2)}{d-1}\,\L_{-2}\,\L_{d-4}\, \phi-2\,(d+2)\, \phi\nonumber\\
&& \hskip6.7cm  +3\,\L_{-2}\,\L_{-3}\, {h}-3\,\L_{-2}\,\nabla^{\lambda}\, {A}_\lambda=0\label{eqspin30}\,.
\eeqa

Repeating the spectral analysis as in the lower spin counterparts, from the structure of the above field equations, we find that mode functions are related one another by the following spin-three spectrum generating complex:
\vskip0.2cm
\begin{tcolorbox}
\vskip-0.2cm
\setlength\arraycolsep{2pt}{
\beqa
\begin{array}{rcll} \label{spin3}
 & \Theta^{3|3}_n & &\\
\L_{d-2} & \upharpoonleft \downharpoonright & \L_{-4}&: \qquad -M_{n,3|3}^2=-M_n^2 \\
 & \Theta^{2|3}_n && \\
\L_{d-3} & \upharpoonleft \downharpoonright & \L_{-3} &: \qquad -M_{n,2|3}^2=-\left(M_n^2 +d+1\right) \\
 & \Theta^{1|3}_n && \\
\L_{d-4} & \upharpoonleft \downharpoonright & \L_{-2} & : \qquad -M_{n,1|3}^2=-\left(M_n^2+2\,d\right)\\
&\Theta^{0|3}_n &&
\end{array}
\eeqa}
\end{tcolorbox}
\noindent

We now show that the complex Eq.\eqref{spin3} can also be derived from the Kaluza-Klein compactification of spin-three gauge transformations. After the compactification, we organize the component gauge parameters as
\beq
\xi_{\mu\nu} \equiv \bar\zeta_{\mu\nu}+\frac1{d+1}\,g_{\mu\nu}\,\bar\zeta_{\theta\theta}\,,\qquad
\xi_{\mu} \equiv \frac32\,\bar\zeta_{\mu\theta}\,,\qquad
\xi \equiv 3\,\bar\zeta_{\theta\theta}\,.
\eeq
such that the spin-two gauge parameter is traceless. The numerical factors are chosen for later simplicity. The gauge transformations of component fields in Eq.\eqref{spin3 line} then read
\beqa
&& \delta\, w_{\mu\nu\rho}=\partial_{(\mu}\,\xi_{\nu\rho)}
	+ 3\,g_{(\mu\nu}\, \L_{d-2}\,\xi_{\rho)}\,,  \nonumber \\
&& \delta\, h_{\mu\nu}=\partial_{(\mu}\,\xi_{\nu)}
	+\frac13 \,\L_{-4}\,\xi_{\mu\nu}
	+\frac{2\,(d+2)}{(d^2-1)}\,g_{\mu\nu}\, \L_{d-3}\,\xi \,,\nonumber\\
&& \delta\, A_{\mu}=\partial_{\mu}\,\xi
	+\L_{-3}\,\xi_\mu\nonumber \\
&& \delta\,\phi=3\,\L_{-2}\,\xi\,.
\eeqa
Mode expanding and equating terms of the same mode functions, we readily extract relations among the mode functions as 
\beqa
\Theta^{3|3}  \sim \L_{d-2}\,\Theta^{2|3}\,,&\qquad&
\Theta^{2|3}  \sim \L_{-4}\,\Theta^{3|3} \sim\L_{d-3}\,\Theta^{1|3}\,,\\
\Theta^{1|3}  \sim \L_{-3}\,\Theta^{2|3}\,,&\qquad&
\Theta^{0|3}  \sim \L_{-2}\,\Theta^{1|3}\, . 
\eeqa
We then see that these relations give rise precisely to the spin-three spectrum generating complex, Eq.\eqref{spin3}.
Moreover, after the mode expansion, these gauge transformations can be identified with spin-three Stueckelberg gauge symmetry \cite{Zino1}.
The complex in turn provides for all information needed for extracting mass spectrum after the compactification. For instance, the eigenvalue of Sturm-Liouville operator $\L_{d-2}\,\L_{-4}$ of index $(d-6)$ can be identified with minus the mass-squared of spin-three field in AdS$_{d+1}$ space.

In identifying possible boundary conditions, another new feature shows up for the spin starting from three: higher-derivative boundary conditions (HDBC) are unavoidable for any choice of boundary conditions. 
Recall that, for the spin-two situation discussed in section \ref{sec spin2 BC}, there was one choice of boundary condition, {\bf B.C.2}, which involved just the  one derivative and hence standard Robin boundary condition. 
Once the Dirichlet boundary condition is imposed to any of the component fields, the structure of spin-three spectrum generating complex Eq.\eqref{spin3} automatically imposes boundary conditions involving two or three derivatives to some other component fields, just like boundary conditions {\bf B.C.1} and {\bf B.C.3} for spin-two field involved two derivatives to some other component fields. For example, if we impose Dirichlet boundary condition for the field associated with $\Theta^{k|3}|_{\theta=\pm\alpha}=0$ for $k=3,2,1,0$, we automatically get boundary conditions for other component fields from Eq.\eqref{spin3}:
{\setlength\arraycolsep{0.2pt}
 \beqa
\begin{array}{ccccccccccc}
&\{& \Theta^{3|3}| =0&\,,
\quad   &\L_{d-2}\,\Theta^{2|3}| =0&\,,
\quad   &\L_{d-2}\,\L_{d-3}\,\Theta^{1|3}| =0&\,,
\quad   &\L_{d-2}\,\L_{d-3}\,\L_{d-4}\,\Theta^{0|3}| =0&\}&\\
&\{& \L_{-4}\,\Theta^{3|3}| =0&\,,
\quad   &\Theta^{2|3}| =0&\,,
\quad   &\L_{d-3}\,\Theta^{1|3}| =0&\,,
\quad   &\L_{d-3}\,\L_{d-4}\,\Theta^{1|3}| =0&\}&\\
&\{& \L_{-3}\,\L_{-4}\,\Theta^{3|3}| =0&\,,
\quad   &\L_{-3}\,\Theta^{2|3}| =0&\,,
\quad   &\Theta^{1|3}| =0&\,,
\quad   &\L_{d-4}\,\Theta^{0|3}| =0&\}&\\
&\{&\L_{-2}\, \L_{-3}\,\L_{-4}\,\Theta^{3|3}| =0&\,,
\quad   &\L_{-3}\,\L_{-4}\,\Theta^{2|3}| =0&\,,
\quad   &\L_{-2}\,\Theta^{1|3}| =0&\,,
\quad   &\Theta^{0|3}| =0&\}&
 \end{array}\,\nonumber
\eeqa
}
\noindent
Here, $\Theta|$ implies the boundary value: $\Theta|_{\theta=\pm\alpha}$.
In all cases above, one of the four component fields is subject to higher-derivative boundary condition (HD BC). We claim that this is a general pattern but relegate reasons and intuitive understandings exclusively to Section \ref{sec7}.

\begin{figure}[!h] 
\centering
{\includegraphics[scale=0.53]{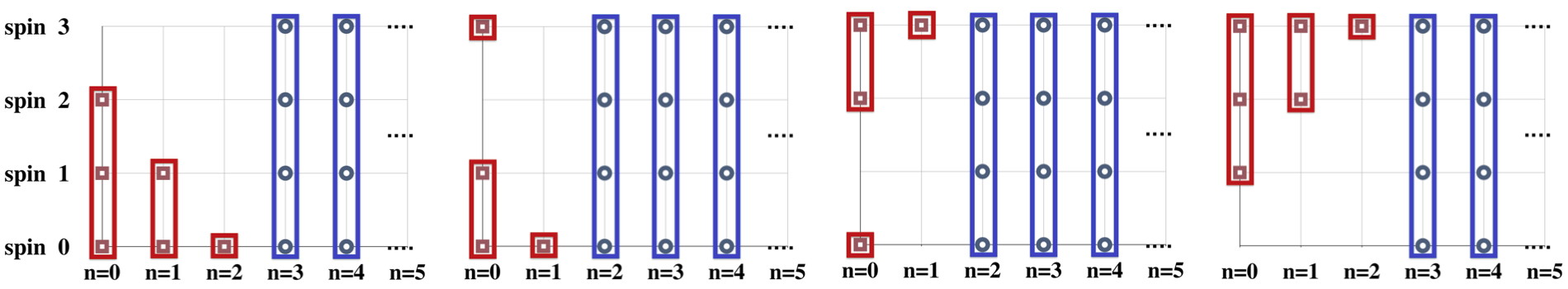}} \quad \quad \quad \quad
\caption{\sl
Mass spectra for spin-three higher-spin field.
The horizontal line labels the mode $n$ and the vertical line labels the spin $s$. 
Each point represents a single mode function.  Points in the same rectangle form the Stueckelberg system.
}
\label{spectra3}
\end{figure}

The pattern of mass spectra is similar to the spin-two situation. 
All Kaluza-Klein modes form spin-three Stueckelberg system. The ground modes for each of the four possible boundary conditions comprise of the irreducible representations of reducible spin-three Verma $\mathfrak{so}(d,2)$-module. As $s=3$, there are three Verma modules:
\beqa
\mathcal{V}\left(d+1,\,3\right) &=& 
\mathcal{D}\left(d+1,\,3\right)\oplus
\mathcal{D}\left(d+2,\, 2 \right) \nonumber \\
\mathcal{V}\left(d ,\,3\right) \,\, &=&  \,\, 
\mathcal{D}\left(d,\,3\right)\oplus
\mathcal{D}\left(d+2,\, 1 \right) \nonumber \\
\mathcal{V}\left(d- 1,\,3\right) &=& 
\mathcal{D}\left(d- 1,\,3\right)\oplus
\mathcal{D}\left(d+2,\, 0 \right) 
\label{breaking verma3}
\eeqa
From the relation between mass-squared and conformal dimension, 
we identify possible type of fields that are present as ground modes as follows:
\vskip1cm
\begin{table}[h]
\centering
\begin{tabular}{|c|c|c|c|}
  \hline
\textbf{type}	&	$\mathcal{D}(\Delta,s)_{\rm \mathfrak{s0}(d,2)}$	& Field types	& mass-squared \\
  \hline
\hline
\textbf{type I}	&	$\mathcal{D}(d+2,2)$	& Stueckelberg spin-two system	& $m^2=2\left(d+2\right)/\ell^2$ \\
\hline
\textbf{type II}	&	$\mathcal{D}(d+2,1)$	& Stueckelberg spin-one system	& $m^2=3\left(d+1\right)/\ell^2$ \\
\hline
\textbf{type III}	&	$\mathcal{D}(d+2,0)$	& massive spin-zero	&	$m^2=2\left(d+2\right)/\ell^2$ \\
\hline
\textbf{type IV}	&	$\mathcal{D}(d+1,3)$	& massless spin-three	& $m^2=0$ \\
\hline
\textbf{type V}	&	$\mathcal{D}(d,3)$	& Stueckelberg PM spin-three of depth-one system	& $m^2=-\left(d+1\right)/\ell^2$ \\
\hline
\textbf{type VI}	&	$\mathcal{D}(d-1,3)$	& Stueckelberg PM spin-three of depth-two  system	& $m^2=-2\,d/\ell^2$ \\
\hline
\end{tabular}
\vskip0.5cm
\caption{
\sl The type of irreducible representation fields upon inverse Higgs mechanism of spin-three Stueckelberg systems.}
 \label{spin3 actors}
\end{table}
\noindent 
The first three types are Goldstone modes, while the latter three types are massless or partially massless spin-three gauge fields. As the spin is three, there are two possible classes of partially massless fields. If the Stueckelberg system extends to spin-two, it is the depth-one gauge field. If the Stueckelberg system extends to spin-one, it is the depth-two gauge field.  
\vskip0.5cm
The patterns of mass spectra for each possible boundary conditions are depicted in Fig. \ref{spectra3} and in  Table. \ref{sp3}
\newline
 \\
\begin{table}[h]
\centering
\begin{tabular}{|c|c|}
  \hline
Boundary Condition	&  \textbf{Ground modes}\\
\hline
	$\Theta^{3|3}|_{\theta=\pm\alpha} =0$	&	\textbf{type I},\quad\textbf{type II} \text{  and  } \textbf{type III}\\
\hline
	$\Theta^{2|3}|_{\theta=\pm\alpha} =0$	&	\textbf{type II},\quad\textbf{type III} \text{  and  } \textbf{type IV}\\
\hline
	$\Theta^{1|3}|_{\theta=\pm\alpha} =0$	&	\textbf{type III},\quad\textbf{type IV} \text{  and  } \textbf{type V}\\
\hline
	$\Theta^{0|3}|_{\theta=\pm\alpha} =0$	&	\textbf{type IV},\quad\textbf{type V} \text{  and  } \textbf{type VI}\\
\hline
\end{tabular}
\caption{
\sl The contents of ground modes for each of the four possible boundary conditions. }
\label{sp3}
\end{table}
\vskip0.5cm
\noindent
We see that the pattern already observed for the spin-two field repeats in spin-three field. For the Kaluza-Klein modes, all boundary conditions yield a universal pattern, yielding a tower of spin-three Stueckelberg system. For the ground modes, the component field with Dirichlet boundary condition is absent from the spectrum while other component fields fill up massive fields and massless or partially massless spin-three gauge fields. 

\section{Higher-Derivative Boundary Condition}\label{sec7}
In the previous section, we discovered an emerging pattern that the Janus waveguide boundary conditions for higher-spin fields unavoidably involve boundary conditions containing derivatives of higher-order (higher than first order, Robin-type). From the viewpoint of Sturm-Liouville problem, such higher-derivative boundary condition (HD BC) is not only non-standard but also potentially problematic. The Sturm-Liouville differential operator is second-order in derivatives, so the operator in general fails to be self-adjoint in the Hilbert space of square-integrable functions if HD BCs are imposed. In this case, the set of Sturm-Liouville eigenfunctions are neither orthogonal nor complete. If so, how do we make a sense of the Kaluza-Klein compactification with HD BC? Stated in the approach of previous sections, how are the mode functions $\Theta_n^{s|S}$ defined and, lacking orthogonality and completeness, how are higher-spin fields in AdS$_{d+2}$ space decomposed into component fields of various spins in AdS$_{d+1}$ space? 

There is one more issue regarding the HD BC. We observed that, for HD BC's, some of the component fields in AdS$_{d+1}$ space become partially massless and have kinetic terms of wrong sign. This ties well with the fact that partially massless higher-spin fields in AdS space belong to non-unitary representations of the $\mathfrak{so}({d,2})$ module (though, in dS space,  they belong to unitary representations of $\mathfrak{so}(d+1, 1)$). Is it possible to trace the origin of this non-unitarity from the field equation and HD BC? In general, boundary conditions are tradable with boundary interactions such that variation of the interactions yield the boundary conditions. Is it then possible to treat the origin of non-unitarity of partially massless fields from non-unitarity of the boundary action of modes localized at the boundaries?

In this section, we answer these questions affirmatively positive by providing mathematical foundation and elementary physical setup from which we can build intuitive understandings.  The idea is this. 
In Sturm-Liouville problem with HD BC (and closely related eigenvalue-dependent boundary conditions), we can always ensure the self-adjointness by extending the inner product defining Hilbert space from ${\cal L}^2(D)$ class of square-integrable functions over domain $D$ to ${\cal L}^2(D) \oplus \mathbb{R}^N$ class of square-integrable functions over $D$ adjoined with $N$-dimensional vectors of finite norm:
\begin{tcolorbox}
\vskip-0.3cm
\beqa
\psi (D) \, \in \, {\cal H}[{\cal L}^2(D) ] \, \, \quad &\longrightarrow& \quad
\Psi(D) = \psi (D) \otimes {\bf q}(\partial D) \, \in \, {\cal H}[{\cal L}^2(D) \oplus \mathbb{R}^N] 
\nonumber \\
\langle \, \psi_1 , \, \psi_2 \rangle := \int_D \psi_1 \psi_2 \quad &\longrightarrow& \qquad \qquad
\llangle \Psi_1 , \Psi_2 \rrangle  := \langle \, \psi_1, \psi_2 \, \rangle + {\bf q}^T_1 \mathbb{G} {\bf q}_2.
\label{extendedinnerproduct}
\eeqa
\end{tcolorbox}
\noindent
Here, $\mathbb{G}$ specifies the metric on the $N$-dimensional vector space. 
We shall refer to the newly introduced inner product $\llangle \, , \, \rrangle$ as ``extended inner product".  The point is that failure of self-adjointness of the Sturm-Liouville operator all arise from contribution at $\partial D$ and so the self-adjointness can be restored by appropriate modification of the contribution coming from $\partial D$. Thus, defining appropriately chosen extended inner product, one can show~\cite{Eigenvalue BC, Eigenvalue BC2} that the Sturm-Liouville problem with HD BC can always be made self-adjoint. Intuitively, the finite-dimensional vector space $\mathbb{R}^N$ we introduced for the extension corresponds to adding nontrivial degrees of freedom localized at the boundary $\partial D$.

Utilizing this idea, we shall show below how partially massless higher-spin fields can be spectrally decomposed in a unique manner and also how to determine the boundary action which render the spectral problem well-posed. We shall first illustrate this idea by an elementary classical field theory consisting of a string with two-derivative boundary condition. We will then apply the understanding to the spin-two fields in AdS space with two-derivative boundary conditions {\bf B.C.1} and {\bf B.C.3},  which are direct counterpart of the above mechanical system.

\subsection{Case 1: Open string in harmonic potential}
The first case we study is the classical field theory of an open string attached to non-relativistic massive particles at each ends.\footnote{This example was considered in detail at \cite{TOS}.}
From the viewpoint of the open string, its motion is subject to boundary conditions. It is intuitively clear that the endpoint particles exert boundary conditions that interpolate between Neumann and Dirichlet types. If the masses are infinite, the string endpoints are pinned to a fixed position. If the masses are zero, the string endpoints move freely. What is less obvious and also less known, however, is that endpoint particles with a finite mass put the open string to higher-derivative boundary conditions \footnote{What is lesser known is that this is completely equivalent to the statement that the particle dynamics involves higher derivatives in time. This is carefully discussed for both flat and curved spacetime backgrounds in \cite{Bak:1999iq}.}. Here, we will study this system in three difference ways
and draw physical interpretations in each case.  We will then construct a refined inner product and a procedure for constructing boundary action from HD BC and vice versa. 

\begin{figure}[!h]  
\centering
{\includegraphics[scale=0.45]{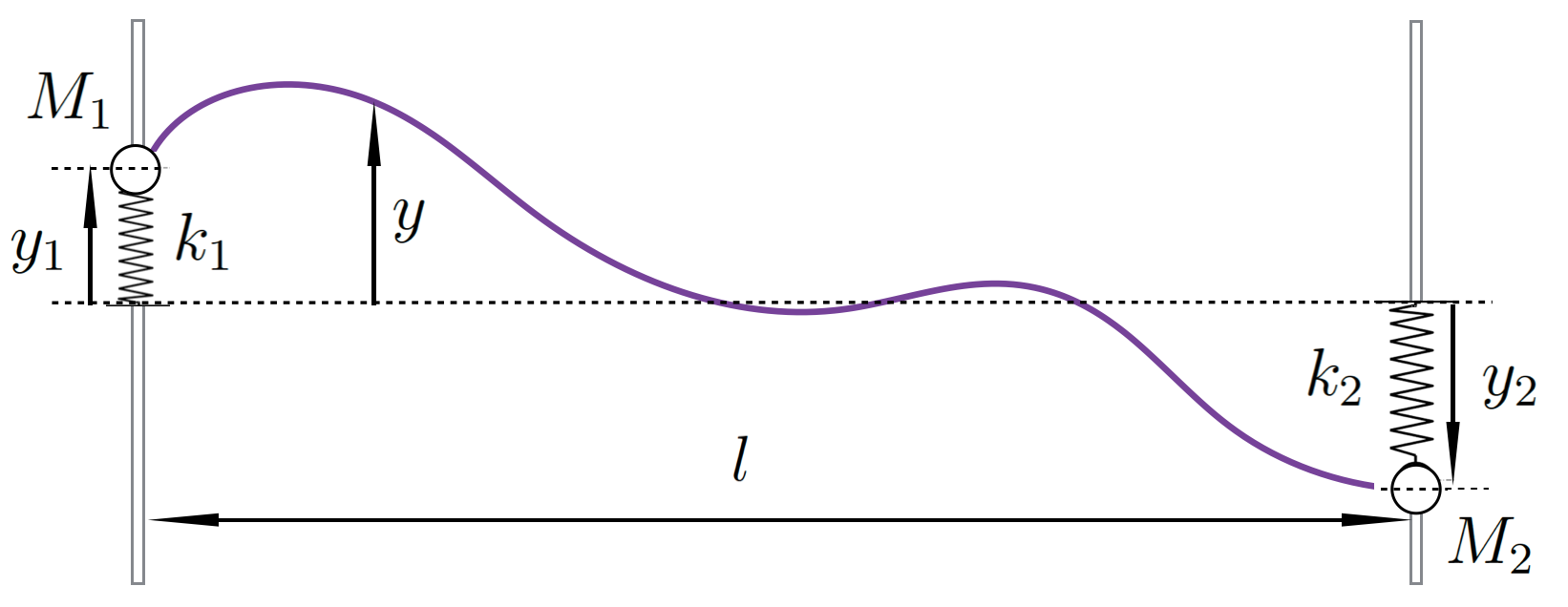}} \quad \quad \quad \quad
\caption{ \sl Open string connected to massive particles in harmonic potential. }\label{String System}
\end{figure}

As the first approach, we start with boundary degrees of freedom, integrate them out, and convert their dynamics to HD BC for the open string. Consider a relativistic open string of tension $T$, stretched along $x$-direction $0 \le x \le \ell $ and vibrating with vertical amplitude $y(x, t)$.  String's end points are attached to harmonic oscillator particles at $x = 0, \ell$ whose masses, vertical positions and Hooke's constants are $M_1, y_1(t), k_1$ and $M_2, y_2 (t), k_2$, respectively. See Fig. \ref{String System}. 
The system is described by the action
\beq
I = \int \textrm{d} t \big( L_{\rm{string}} + L_{\rm{particle},\, 1} + L_{\rm{particle},\, 2} \big) \, , 
\label{action}
\eeq
where the Lagrangians of open string and massive particles are  
\begin{align}
L_{\rm{string}}=\frac{T}{2} \int_0^l \mathrm{d}x\,\left( (\partial_t y)^2-(\partial_x y)^2\right) 
\qquad \mbox{and} \qquad
L_{\rm{particle,\,a}}=\frac{1}{2}\left(M_a\,\dot{y}_a^2-k_a\,y_a{}^2\right)\, \quad (a=1,2). \label{string action}
\end{align}
The system is closed, so the action determines dynamics of the variables $(y(x, t), y_1(t), y_2(t))$ completely without specifying any boundary conditions.  As the string is attached to the particles, the string amplitude is related to particle positions by
\begin{align} \label{strconst}
y(x,t) \Big\vert_{x=0}  = y_1 (t) \qquad \mbox{and} \qquad y(x,t) \Big\vert_{x=\ell}  = y_2 (t). 
\end{align}
Thus, one should be able to describe the system in terms of the string amplitude $y(x, t)$. This is achieved by eliminating the particle variables $y_1(t), y_2(t)$ and express them in terms of $y(x, t)$. In doing so, the constraints Eq.(\ref{strconst}) and particle actions turn into some boundary conditions to the string amplitude $y(x, t)$ at $x =0, \ell$.  Our goal is to derive these boundary conditions, starting from boundary actions $\int {\rm d} t L_{\rm{particle},\, 1, 2}$  that are provided by the endpoint particle actions. This procedure is nothing but the classical counterpart of Born-Oppenheimer approximation. 

To extract the boundary condition,  we derive the field equation of the string from the action Eq.(\ref{action}):
\begin{align}
\delta I = \int \mathrm{d}t \left( -T \int \mathrm{d}x \, \delta y \left[ \partial_t^2 y - \partial_x^2 y \right] + T \, \left[ \delta y \, \partial_x y \right]_{0}^{l} -\sum_{a=1,2} \delta y_a (M_a \ddot{y}_a + k_a y_a ) \right)
\end{align}
Imposing the constraints Eq.(\ref{strconst}), $\delta y(0,t) =\delta y_1 (t)$ and $\delta y(l,t) =\delta y_2 (t)$, we obtain the string field equation of motion 
\beq
\Big(\partial_t^2  - \partial_x^2 \Big) y (x, t) = 0 \qquad (0 \le x \le \ell)
\label{stringeom}
\eeq
and equations of motion for each particles
\begin{align}
M_1\,\ddot{y_1}+k_1\,y_1- T \, \partial_x y \Big\vert_{x=0} = 0 \qquad \mbox{and} \qquad 
M_2\,\ddot{y_2}+k_2\,y_2+ T \, \partial_x y \Big\vert_{x=\ell } = 0 \, .\label{point particle equation}
\end{align}
Integrating out the endpoint particles amount to relating $y_1(t), y_2(t)$ to the endpoints of string amplitude by combining Eq.(\ref{strconst}) with Eq.(\ref{stringeom}). We obtain the sought-for boundary conditions
\begin{align} \label{strHDBC}
M_1\, \partial_x^2 y - T \, \partial_x y +k_1\,y  \Big\vert_{x = 0} = 0\qquad \mbox{and} \quad \qquad M_2\,\partial_x^2 y+T\, \partial_x y +k_2\,y \Big\vert_{x = \ell} = 0\, . 
\end{align}
We see that, for finite $M_1$ and $M_2$,  the boundary conditions are second order in normal derivatives, so they are indeed HD BCs. Were if $M_1, M_2$ zero, the boundary conditions are the most general Robin boundary conditions. The Robin boundary condition is further reduced to Neumann and Dirichlet boundary conditions in the limit $k_{1,2}$ are zero and infinite, respectively.  Were if $M_1, M_2$ infinite, regularity of boundary conditions require that $\partial_x^2 y$ vanishes at the boundaries. In turn, $\partial_x y$ is constant-valued at the boundaries, and so the boundary conditions are again reduced to Dirichlet boundary conditions.    

Conversely, we can always reinterpret HD BCs on open string as attaching massive particles at the endpoints.  Start with an open string whose field equation Eq.(\ref{stringeom}) is subject to HD BCs Eq.(\ref{strHDBC}). This is the same situation as we have for the higher-spin field in the AdS waveguide.  Solving the open string field equation subject to the boundary conditions is the same as extremizing a modified action $\widetilde I$ whose variation is given by
\begin{align} \label{strVar}
\delta \widetilde I &= \int \mathrm{d}t \left( - T \int_{0}^{\ell} \mathrm{d}x \; \delta y \big[ \partial_t^2 y  - \partial_x^2 y \big] \right) \nonumber \\
&- \int_{x=0} \mathrm{d}t \lambda_1(t) \, \delta y \, \big[ M_1 \partial_x^2 y - T \partial_x y + k_1 y \big] 
- \int_{x= \ell} \mathrm{d}t \lambda_2(t) \, \delta y \, \big[ M_2 \partial_x^2 y + T \partial_x y + k_2 y \big] 
\nonumber \\
&- M_1 \int_{x=0} {\rm d} t \,  \delta y \, \big[ \partial_t^2 y - \partial_x^2 y \big] 
- M_2 \int_{x=\ell} {\rm d} t \, \delta y \, \big[ \partial_t^2 y - \partial_x^2 y \big]\, . 
\end{align}
Here, $\lambda_{1,2}(t)$ are Lagrange multipliers that imposes the HD BCs. The last line is redundant, since they vanish automatically when the open string field equation from the first line is obeyed. By reparametrization of time $t$ at both boundaries, it is always possible to put them to constant values, which we set to unity.  To reconstruct the action $\widetilde I$,  we combine derivative terms that depend on string tension $T$:
\begin{align}
T \int \mathrm{d}t \! \int_0^l \mathrm{d}x \, \delta y \, \partial_x^2 y 
-T \int \mathrm{d}t \, \big( \delta y \, \partial_x y \big)_{0}^{\ell} 
= - \delta \left( \frac{T}{2}\int \mathrm{d}t \! \int_0^\ell \mathrm{d}x \, (\partial_x y)^2 \right) ,
\end{align}
and also combine derivative terms that depend on the mass parameters $M_1, M_2$:
\begin{align}
& - M_1 \int_{x=0}  \mathrm{d}t \, \delta y \, \partial_x^2 y - M_1 \int_{x=0} \mathrm{d}t \, \, \delta y \big[\partial_t^2 {y} - \partial_x^2 y \big] = - M_1 \int_{x=0} \mathrm{d}t \, \delta y \, \partial_t^2 y = \delta \left( \frac{M_1}{2} \int \mathrm{d}t \, \dot{y}^2 \right) \nonumber \\
& - M_2 \int_{x=\ell}  \mathrm{d}t \, \delta y \, \partial_x^2 y - M_2 \int_{x=\ell} \mathrm{d}t \, \, \delta y \big[\partial_t^2 {y} - \partial_x^2 y \big] = - M_2 \int_{x=\ell} \mathrm{d}t \, \delta y \, \partial_t^2 y = \delta \left( \frac{M_2}{2} \int \mathrm{d}t \, \dot{y}^2 \right). 
\end{align}
Combining with other terms in the variation, we get 
\begin{align}
\widetilde I = \frac{T}{2} \int \mathrm{d}t \int_0^\ell \mathrm{d}x \, \left[ (\partial_t y)^2 - (\partial_x y)^{2} \right] + \frac{1}{2}\int \mathrm{d}t \, (M_1 \,  \dot{y}_1^2 - k_1 y_1^2) + \frac{1}{2}\int \mathrm{d}t \, (M_2 \, \dot{y}_2^2 - k_2 y_2^2) \, .  
\end{align}
By renaming the endpoint positions as in Eq.(\ref{strconst}), we find that the action $\widetilde I$ is precisely the action for an open string coupled to dynamical harmonic oscillator particles at each ends, Eq.(\ref{string action}).

We still need to understand how the Sturm-Liouville operator $-\partial_x^2$ of open string can be made self-adjoint for HD BC. It is useful to recall implication of self-adjointness for the Robin boundary condition. In this case, we can rewrite the open string action in terms of inner product for square-integrable functions
\begin{align}
I_{\rm{string}} = \frac{T}{2} \int \mathrm{d}t \, \Big( \langle \partial_t {y} , \partial_t {y} \rangle - \langle y , (-\partial_x^2) y \rangle \Big) \quad \mbox{where} \quad 
\langle f , g\rangle \equiv \int_0^\ell  \mathrm{d}x \, f(x) \, g(x) \, . 
\end{align}
Denote the square-integrable normal mode functions of $(-\partial_x^2)$ as $X_n$ $(n=0, 1, 2, 3, \cdots)$, viz. $(-\partial_x^2) X_n =\lambda_n X_n$. As the Sturm-Liouville operator $(-\partial_x^2)$ is self-adjoint for the Robin boundary condition, the normal mode functions can be made orthonormal and form a complete set of basis of the Hilbert space of square-integrable functions. So, we can decompose the string amplitude $y(x, t)$ as 
\beq
y(x, t) = \sum_n T_n(t) \, X_n(x) 
\eeq
and the open string action $I_{\rm string}$ as 
\begin{align} \label{strmodeaction}
I_{\rm{string}} = \sum_{n} \frac{T}{2} \int \mathrm{d}t \left( \dot{T}_n^2 - \lambda_n T_n^2  \right).
\end{align}

Motivated by this line of reasonings, we ask if the combined action of open string with HD BCs can be written in terms of some inner product $\llangle \, , \, \rrangle$: 
\begin{align} \label{straction2}
I_{\llangle, \rrangle} = \frac{T}{2} \int \mathrm{d}t \, \Big( \llangle \partial_t {y} , \partial_t {y} \rrangle - \llangle y, (- \partial_x^2) y \rrangle \Big) .
\end{align}
We now prove that the inner product $\llangle \, , \, \rrangle$ that renders the Sturm-Liouville operator $(-\partial_x^2)$ self-adjoint under the HD BC Eq.(\ref{strHDBC}) is precisely the extended inner product Eq.(\ref{extendedinnerproduct}). In the present case, the additional vector space is provided by the positions of two massive particles attached at the string endpoints. Therefore, it spans $\mathbb{R} \oplus \mathbb{R}$. The metric of this two-dimensional vector space is given by masses (measured in unit of the string tension).
For a function space ${\cal L}^2 \oplus \mathbb{R}^2$, a general element and its inner product with respect to HD BC Eq.(\ref{strHDBC}) would take the form
\begin{align}
{\bf f} = \left( \begin{array}{c} f(x) \\ f_1 \\ f_2 \end{array} \right) \in {\cal L}^2 \oplus \mathbb{R}^2, \qquad  {\bf f} \cdot {\bf g} = \int_0^\ell  \mathrm{d}x \, f(x)\, g(x) + G_{11} f_1 g_1 + G_{22} f_2 g_2 .
\end{align}
Roughly speaking, two new real numbers $f_{1,2}$ correspond to boundary values of $f(x)$ which are left undetermined by the Sturm-Liouville differential equation and the HD BC. The boundary conditions on element of ${\cal L}^2 \oplus \mathbb{R}^2$ are HD BC in Eq.(\ref{strHDBC}) for $f(x)$, together with $f_1 = f(0)$ and $f_2 = f(l)$. With these boundary conditions, we now define the "extended inner product" $\llangle \; ,  \; \rrangle$ for the open string with HD BC as 
\begin{align} \label{exN}
\llangle f , g \rrangle \equiv \int_0^\ell \mathrm{d}x \, f(x)\, g(x) +\frac{M_1}{T}f(0)g(0) + \frac{M_2}{T}f(l)g(l),
\end{align}
where the metric of $\mathbb{R}^2$ is chosen by the parameters in the HD BCs, Eq.(\ref{strHDBC}). With respect to this extended inner product, we now find that the Sturm-Liouville operator $(-\partial_x^2)$ of open string is indeed self-adjoint:
\begin{align}
& \llangle f, (-\partial_x^2) g \rrangle - \llangle (-\partial_x^2) f, g \rrangle \nonumber \\
=&- {1 \over T} f(M_1 \partial_x^2 g - T \partial_x g + k_1 g)\Big\vert_{x=0} + \frac{1}{T} (M_1 \partial_x^2 f - T \partial_x f +k_1 f)g \Big\vert_{x=0} \nonumber \\
&+\frac{1}{T}f(M_2 \partial_x^2 g + T \partial_x g + k_2 g) \Big\vert_{x=\ell}-\frac{1}{T} (M_2 \partial_x^2 f + T \partial_x f +k_2 f)g \Big\vert_{x=\ell} \nonumber \\
&=0 \, , 
\end{align}
where we arranged the harmonic force term (zero derivative terms in the boundary condition) and  the HD BC Eq.(\ref{strHDBC}) for $f$ and $g$. With the extended inner product, we shall expand the proposed action Eq.(\ref{straction2}) in terms of the original inner product over ${\cal L}^2$-space and additional inner product over $\mathbb{R}^2$ space. We observe that, after renaming the boundary values of $y(x, t)$ as Eq.(\ref{strconst}), the proposed action $I_{\llangle, \rrangle}$ in Eq.(\ref{straction2}) is precisely the action of open string attached to endpoint particles, $I = I_{\rm{string}} + I_{\rm{boundary}}$. We reiterate the key point here is that extended inner product, HD BCs, and boundary actions are bear the same information and dictate their structures one another. 

The "extended inner product" we introduced poses a new issue originating from the HD BC, equivalently, the endpoint particle dynamics. For some choices of the HD BCs, the extended Hilbert space can be indefinite, viz. the norm $\llangle y, y \rrangle$ can become negative. This happen precisely when the metric components ${M_{1,2}}/{T}$ have negative signs. Take, for instance, $M_1=M_2=M$ and $k_1=k_2=0$. There always exists at least one mode
\begin{align}
X_0(x) = N_0 \, \sinh\left[ m_0\, \left(x-\frac \ell 2\right)\right] \quad \text{with}\quad 
\frac{1}{m_0}=-\frac{M}{T}\tanh\frac{m_0\,\ell}{2} \, , 
\label{St Un}
\end{align}
whose extended norm is negative for negative value of $M$
\begin{align}
\llangle  X_0 ,  X_0 \rrangle = N_0{}^2\left[-\frac{\ell }{2}+\frac{M}{T}\,\sinh^2\frac{m_0\, \ell}{2}\right] <0\,.
\end{align}
This mode is problematic as, upon mode expansion, the corresponding component in the action Eq. (\ref{strmodeaction}) has the kinetic term with wrong sign,
\begin{align}
(-)\frac{T}{2} \int \mathrm{d}t \, \left( \dot{T}^2_0 - \lambda_0 T^2_0 \right) \, . 
\end{align}
This causes negative energy of the open string at classical level and negative probability (and hence lack of unitarity) at quantum level. Moreover, the mode eigenvalue $\lambda_0 = -  m^2_0$ is negative definite (which is again a consequence of negative value of $M$, as seen from Eq.(\ref{St Un})) and so the variable $T_0(t)$ develops an instability to grow exponentially large.  

{
There is another example demonstrating the utility of boundary degrees of freedom viewpoint. Consider $k_1 = k_2 = k < 0$, $M_1 = M_2 = M >0$ and $T > 0$ case. In this case, the extended inner product Eq.(\ref{exN}) ensures positivity of the norm. However, there are some modes with negative eigenvalue. Generic even (with respect to $x = \frac{\ell}{2}$) mode function with negative eigenvalue is $X_e (x) = \cosh\left[ \lambda \, \left( x - \frac{\ell}{2} \right) \right]$, $(-\partial^2_x ) X_e = -\lambda^2 X_e$. HD BC implies
\begin{align}
M \lambda^2 + T \, \lambda \, \tanh (\frac{\ell}{2}\lambda) = -k
\end{align}
and this equation always has solutions because for $\lambda \ge 0$, the left-hand side is starting from 0 and monotonically increasing. Also the HD BC of generic odd function $X_o (x) = \sinh\left[ \lambda \, \left( x - \frac{\ell}{2} \right) \right]$ implies
\begin{align}
M \, \lambda^2 \, \tanh(\frac{\ell}{2}\lambda) + T \, \lambda = -k \, \tanh(\frac{\ell}{2}\lambda)
\end{align}
and this equation has a solution for $T < -\frac{\ell}{2}\, k$~\footnote{In terms of boundary degrees of freedom, this inequality means that repulsive force from spring is bigger than string tension.}. Again, these negative eigenvalue modes are indications of instability of the system. In terms of HD BC, it is hard to see the origin of this instability. However, in terms of boundary degrees of freedom, it is immediate that the origin of instability is the negative spring constant. 
}

So, by relating HD BCs to boundary action of extra degrees of freedom, we gain better understanding of underlying physics. For $M$ negative, it is hard to recognize the above instability or non-unitarily at the level of equation of motion and boundary conditions. In contrast, the boundary action clearly displays the origin of instability or non-unitarity and it is simply a consequence of negative mass of the endpoint particles. 

\subsection{Spin-two waveguide with higher-derivative boundary conditions}\label{HD BC higher spin}
We now apply our understanding of the HD BC in the previous subsection to the spin-two field in AdS waveguide studied in section \ref{sec spin2 BC}. Recall that spin-two field is the first situation that HD BCs start to appear and, among three possible Dirichlet boundary conditions, \textbf{B.C. 1} and \textbf{B.C. 3} contain two-derivative boundary conditions to some of the component fields. In this subsection, we construct the extended inner product for these boundary conditions and, from that, intuitively explain the origin of non-unitarity for partially massless representations in AdS$_{d+1}$.

We first construct the extended inner product for spin-two fields in AdS space. 
The Sturm-Liouville problems with HD BCs that we will consider have the following form:
\begin{align}
\L_{b}\,\L_{a}\, \Theta_n = -\lambda_n\, \Theta_n \label{2DEVE}\,\qquad \mbox{where} \qquad
\L_{c}\,\L_{a}\, \Theta_n \big|_{\theta = \pm \alpha} = 0\, 
\end{align}
for some weights $a, b, c$. Note that the Sturm-Liouville equation and the boundary condition share the same operator $\L_a$.
From free action of the spin-two field, we get an ${\cal L}^2$ inner product
\begin{align}
\langle\Theta_m \, , \Theta_n\rangle =
\int_{-\alpha}^{\alpha}\mathrm{d}\theta \, (\hbox{sec}\theta)^{d-4} \, \Theta_m (\theta) \, \Theta_n (\theta)\,, \label{L2N}
\end{align}
where the weight factor in the integration measure originates from the conformal factor of the Janus metric Eq.(\ref{metricre}). As we deal with spin-two, $s=2$ and so $a+b = d-2s = d-4$. We thus take the weight factor as $(\hbox{sec}\theta)^{a+b}$. For any conformal factor $(\hbox{sec}\theta)^{c}$ with arbitrary weight $c$, we integrate by part 
\begin{align}
\int_{-\alpha}^{\alpha}\mathrm{d}\theta \, (\hbox{sec}\theta)^{c} \, \Theta_m (\L_{a} \Theta_n )= -\int_{-\alpha}^{\alpha}\mathrm{d}\theta \, (\hbox{sec}\theta)^{c} \, (\L_{c-a} \Theta_m )\, \Theta_n + (\hbox{sec}\alpha)^c \big[ \Theta_m \, \Theta_n \big]^{+\alpha}_{-\alpha}\,.\label{IbP}
\end{align}
Using this, one finds that the differential operator $\L_b\,\L_a$ is not self-adjoint on ${\cal L}^2$ functional space,
\begin{align}
\langle\Theta_m, (\L_b\,\L_a\,\Theta_n) \rangle - \langle (\L_b\,\L_a\,\Theta_m), \Theta_n \rangle
= (\hbox{sec}\alpha)^{a+b} \big[ \Theta_m (\L_{a} \Theta_n ) - (\L_{a}\Theta_m )\Theta_n \big]^{+\alpha}_{-\alpha}\neq0\,.\label{2DBC}
\end{align}
By inspection, however, we find an extended inner product which renders the Sturm-Liouville operator $\L_b \L_a$ self-adjoint. It is 
\begin{align}
\llangle \Theta_m ,  \Theta_n \rrangle \equiv 
\langle {\Theta}_m , {\Theta}_n \rangle
+ \sum_{\sigma = \pm}
\sigma \mathcal{N}_\sigma \Theta_m (\sigma \alpha) \Theta_n (\sigma \alpha) \, ,  
\label{ExN}
\end{align}
where 
\beqa
\mathcal{N}_{+} = \textcolor{red}{-}\mathcal{N}_{-} = (c-b)^{-1} \cot \alpha (\sec\,\alpha)^{a+b}.
\eeqa
We can confirm that $\L_b \L_a$ is indeed self-adjoint with respect to the extended inner product:
\begin{align}
&\llangle \Theta_m, ( \L_b \, \L_a \Theta_n) \rrangle
-\llangle (\L_b \, \L_a \Theta_m),  \Theta_n \rrangle \nonumber \\
= & (\hbox{sec}\alpha)^{a+b} \sum_{\sigma = \pm}  \sigma 
\Big( \Theta_m (\L_{a} \Theta_n ) - (\L_{a}\Theta_m )\Theta_n \Big) (\sigma \alpha) 
+ \sum_{\sigma = \pm} \sigma \mathcal{N}_{\sigma} \Big(\Theta_m \L_{b}\L_{a} \Theta_n - (\L_{b} \L_{a} \Theta_m )\, \Theta_n \Big)(\sigma \alpha) \nonumber \\
= & (\hbox{sec}\alpha)^{a+b} 
\sum_{\sigma = \pm} \sigma\,{{\cal N}_\sigma} \Big( \Theta_m (\L_{c}\L_{a} \Theta_n ) (\sigma \alpha){- (\L_{c}\L_{a} \Theta_m )\Theta_n (\sigma \alpha) }\Big)\, . 
\end{align}
The last expression vanishes by the HD BCs in Eq.\eqref{2DEVE}.

We apply the extended inner product to the ground modes for the HD BCs,  \textbf{B.C. 1} and \textbf{B.C. 3} in section \ref{sec spin2 BC}. In the last subsection, whether a given HD BC lead to non-unitarity or not depends on parameters specifying the boundary conditions. The extended norm-squared is positive definite if unitary, while it is negative definite if non-unitary. 
For \textbf{B.C. 1}, the HD BCs are imposed on spin-zero mode with $a = d-3$, $b = -1$ and $c = d-2$. We see that the normalization constants $\mathcal{N}_{\pm}$ in Eq.\eqref{ExN} are positive-definite, so the norm-squared is positive-definite.
In contrast, for \textbf{B.C. 3}, HD BCs are imposed on spin-two mode with $a = -2$, $b = d-2$, $c = -1$ and the normalization constants $\mathcal{N}_{\pm}$ are negative-definite. More explicitly, the ground modes of \textbf{B.C. 3} are
\beqa
\left\{ \begin{array}{ll}
\Theta_{1}^{2|2}=N_1\,\sec\theta\,\tan\theta\,, \quad
\Theta_{1}^{1|2}=N_2\,\sec\theta\,&\text{\textbf{type IV} in Table \ref{spin2 actors}}\,,\\
\Theta_{0}^{2|2}=N_3\,\sec^2\theta\,&\text{\textbf{type II} \, in Table \ref{spin2 actors}}\,,
\end{array}\right.
\eeqa
which correspond to the PM spin-two and massless spin-two fields, respectively.
Boundary condition of spin-one mode function is one-derivative boundary condition and its norm is positive-definite. In contrast, the norms of spin-two modes are
\beqa
&& \llangle \Theta^{2|2}_1 , \Theta^{2|2}_1 \rrangle = N_1{}^{2}  \left( \int^\alpha_{-\alpha}\, 
d\theta\,
\sec^{d-2}\theta\,\tan^2\theta
-\frac{2}{d-1}\,\sec^{d-2}\alpha\,\tan\alpha
\right) \,, \label{norm22}\\
&& \llangle \Theta^{2|2}_0 , \Theta^{2|2}_0 \rrangle = N_3{}^{2} \left( \int^\alpha_{-\alpha}\, 
d\theta\,
\sec^{d}\theta
-\frac{2\,\sec^{d}\alpha}{(d-1)\,\tan\alpha}\right)\,. \label{norm222}
\eeqa
It can be shown that the norm Eq.(\ref{norm22}) which corresponds to the PM mode, is always negative\footnote{
The other norm Eq.(\ref{norm222}) is negative for $\alpha\sim0$ and positive for $\alpha\sim\pi/2$. When one of the Kaluza-Klein mass hits zero mass, this norm vanishes. In this specific value of $\alpha$, there is no massless spin-two field, \textbf{type III}, in the spectrum and \textbf{type II} appears instead.
} 
by the following estimate:
\begin{align}
\llangle \Theta^{2|2}_1 , \Theta^{2|2}_1 \rrangle &= N_1{}^{2}  \left(2 \int^\alpha_{0}\, 
d\theta\,
\sec^{d}\theta\,\sin^2\theta
-\frac{2}{d-1}\,\sec^{d-1}\alpha\,\sin\alpha
\right) \nonumber \\
&< N_1{}^{2}  \left(2 \sin \alpha \int^\alpha_{0}\, 
d\theta\,
\sec^{d}\theta\,\sin\theta
-\frac{2}{d-1}\,\sec^{d-1}\alpha\,\sin\alpha
\right) \\
&= -N_1{}^{2} \frac{2}{d-1} \nonumber
\end{align}
Here, the inequality holds because $\sin \theta < \sin \alpha$ for $0 \le \theta <\alpha<\frac{\pi}{2}$. This negative norm implies that the kinetic term of PM mode has the wrong sign.

With the extended inner product, we can construct the boundary action which reveals physical properties of the imposed HD BCs. The action of free massless spin-two field $\bar{h}_{MN}$ on AdS$_{d+2}$ background is
\begin{align} \label{hds2a}
I_{\rm spin-two} = &\int\sqrt{\bar g} \, \mathrm{d}^{d+2}x \, \mathcal{L}_{2}\left(\bar{h}_{MN};\bar{g}_{MN},d+2\right) \nonumber \\
=&\int \sqrt{\bar g} \, \mathrm{d}^{d+2}x \, \Big[ -\frac{1}{2}\, \bar{\nabla}^{L}\, \bar{h}^{M N}\, \bar{\nabla}_{L}\, \bar{h}_{M N} + \bar{\nabla}^{M} \,\bar{h}^{N L}\, \bar{\nabla}_{N}\, \bar{h}_{M L} 
- \bar{\nabla}^{M}\, \bar{h}_{M N}\, \bar{\nabla}^{N}\, \bar{h} \nonumber \\
& \hskip2.5cm + \frac{1}{2}\, \bar{\nabla}^{L}\, \bar{h}\, \bar{\nabla}_{L}\, \bar{h} 
-\left(d+1\right)\big(\bar{h}^{M N}\, \bar{h}_{M N}-\frac{1}{2}\,\bar{h}^2 \big) \Big]
\end{align}
After the compactification on the AdS waveguide, each term of Eq.($\ref{hds2a}$) is decomposed into quadratic terms of component fields, $h_{\mu \nu}$, $A_{\mu}$ and $\phi$, which can be expressed as ${\cal L}^2$ inner product Eq.\eqref{L2N} with $a+b=d-2s = d- 4$. For example, 
\begin{align}
\int \mathrm{d}^{d+1}x \, \sqrt{-g}\, \int_{-\alpha}^{\alpha} \mathrm{d}\theta \, (\textrm{sec}\theta)^{d-4} \, \nabla^{\rho}\, h^{\mu \nu} \,\nabla_{\rho}\, h_{\mu \nu} = \int \mathrm{d}^{d+1}x\, \sqrt{-g} \, \langle\, \nabla^{\rho}\, h^{\mu \nu} ,\nabla_{\rho}\, h_{\mu \nu}\, \rangle \,. \nonumber
\end{align}
As in the open string case, we require that each term of the quadratic action to be expressed by appropriate inner product which ensures orthogonality and completeness of the mode functions. We now know that, depending on the nature of boundary conditions, some of these terms need to be the extended inner product which contain the contribution of boundary action. 
The situation is more involved as there are three component fields each of which obeys different boundary condition. From the spectrum generating complex, we have three kinds of boundary conditions:
\begin{itemize}
\item spin-two Dirichlet, expanded by $\Theta^{2|2}_n$ : \,\, $\phantom{\L_{-1}\L_{-2}} h_{\mu \nu}$, $\phantom{\L_{-2}}\L_{d-2}A_{\mu}$, \quad$\L_{d-2}\L_{d-3}\phi \quad$ 
\item spin-one Dirichlet, expanded by $\Theta^{1|2}_n$ : \,\, $\phantom{\L_{-1}}\L_{-2}h_{\mu \nu}$, $\phantom{\L_{-2}\L_{d-2}}A_{\mu}$, \quad$\phantom{\L_{d-2}}\L_{d-3}\phi \quad$ 
\item spin-zero Dirichlet, expanded by $\Theta^{0|2}_n$ : \,\, $\L_{-1}\L_{-2}h_{\mu \nu}$, $\phantom{\L_{d-2}}\L_{-1}A_{\mu}$, \quad$\phantom{\L_{d-2}\L_{d-3}}\phi \quad$.
\end{itemize} 
By a straightforward computation, we find that the action is decomposed as
\begin{align}
I &=  \int \mathrm{d}\theta \, (\textrm{sec}\theta)^{d-4} \, \mathcal{L}_2 \left(h_{\mu \nu};g_{\mu \nu},d+1\right) \nonumber \\
& + \Big[  -\frac{1}{2}\, \langle F^{\mu \nu}, F_{\mu \nu} \rangle 
- 2d\,\langle A^{\mu}, A_{\mu} \rangle  +\langle \L_{-2}h^{\mu \nu} , \nabla_{\mu}A_{\nu}+\nabla_{\nu}A_{\mu}-2g_{\mu \nu}\nabla^{\rho}A_{\rho} \rangle \nonumber \\
&-\frac{1}{2} \langle \L_{-2}\, h^{\mu \nu},  \L_{-2}\, h_{\mu \nu} \rangle 
+\frac{1}{2} \langle \L_{-2}\, h,  \L_{-2}\, h \rangle 
+ \frac{d(d+1)}{(d-1)^2} \langle \L_{d-4}\phi , \L_{d-4}\phi \rangle 
 - {d \over d-1}\, \langle \L_{-2}h , \L_{d-4}\phi \rangle \Big]
\nonumber \\
& + \Big[ - {d \over d-1} \big( \frac{1}{2}\,\langle \nabla^{\mu}\,\phi , \nabla_{\mu}\,\phi \rangle 
+\frac{d+1}{2}\, \langle \phi , \phi \rangle \big) 
+ {2d \over d-1} \,\langle \L_{-1}A^{\mu},  \nabla_{\mu}\phi \rangle  \Big] \, . 
 \label{expanded}
\end{align}

\noindent The first line, the second bracket and the third bracket are spin-$s$ component of modes: $\langle \Theta^{s|2}, \Theta^{s|2} \rangle$ for $s=2,1,0$, respectively
\footnote{\label{ambi} The classification appears somewhat arbitrary. For instance, $\langle \L_{d-3} \phi,  \L_{d-3}\phi \rangle$ belongs to $\langle \Theta^{1|2}, \Theta^{1|2} \rangle$, but its another form $\langle \phi, \L_{-1} \L_{d-3}\phi \rangle$ obtained by integration by parts belongs to $\langle \Theta^{0|2} , \Theta^{0|2} \rangle$.  We will show that the total action is nevertheless the same provided we keep track of boundary terms.}. 

Consider first \textbf{B.C. 1}. In this case,  the spin-zero component field 
obeys the HD BC: 
\beqa
\L_{-1}\L_{d-3}\Theta^{0|2} = -\lambda \Theta^{0|2} \qquad \mbox{and} \qquad
\L_{d-2}\L_{d-3}\Theta^{0|2}\big|_{\theta = \pm \alpha} = 0.
\eeqa
So, we need to adopt the extended inner product for terms involving the mode function $\Theta^{0|2} $. They are the terms in the third bracket of Eq.(\ref{expanded}). 
Using the extended inner product Eq.(\ref{ExN}) with $a = d-3$, $b=-1$ and $c=d-2$, we obtain the corresponding boundary action from the difference between extended inner product and original inner product:
 \begin{align}
\int \mathrm{d}^{d+1} x \sqrt{-g} \; \frac{d}{(d-1)^2}\frac{(\text{sec}\alpha)^{d-4}}{\text{tan}\alpha}
\sum_{\sigma = \pm} \Big[ -\Big(\frac{1}{2}\nabla^{\mu}\phi \nabla_{\mu}\phi + \frac{d+1}{2} \phi^2 \Big)
+ \Big( \L_{-1}A^{\mu} \nabla_{\mu} \phi \Big) 
\Big]_{\theta = \sigma \alpha}  \,.
\end{align}
Redefining the boundary values of spin-zero field as
\begin{align}
\phi^{\sigma} = \left(\frac{d}{(d-1)^2}\frac{(\text{sec}\alpha)^{d-4}}{\text{tan}\alpha} \right)^{1/2} \phi \Big\vert_{\theta = \sigma \alpha} \,, \qquad (\sigma = \pm), 
\end{align}
we get the boundary action as 
\begin{align}
I_{\rm boundary, BC1} = \sum_{\sigma = \pm} \int \mathrm{d}^{d+1} x \sqrt{-g} \; \Big[ -\Big(\frac{1}{2}\nabla^{\mu}\phi^{\sigma} \nabla_{\mu}\phi^{\sigma} + \frac{d+1}{2} (\phi^{\sigma})^2 \Big) 
+ C\Big( \L_{-1}A^{\mu}\big|_{\theta = \sigma \alpha} \nabla_{\mu} \phi^{\sigma} \Big) 
\Big] \,,
\end{align}
where $C=\big(\frac{d}{(d-1)^2} \cot \alpha {(\text{sec}\alpha)^{d-4}}\big)^{1/2}$.
Note that the sign of kinetic term for boundary spin-zero fields $\phi^\pm$ is standard in our convention. This matches precisely with the result of section 5.2 that the waveguide compactification with \textbf{B.C. 1} only yields unitary spectrum. For the second term of boundary action, we may interpret it two alternative ways. We can interpret that the bulk field $A_{\mu}$ is sourced by the boundary field $\phi^{\pm}$, equivalently, the boundary value of bulk field $A_\mu$ turns on the boundary field $\phi^\pm$. Alternatively, we can eliminate this term by writing the cross term $\langle \L_{-1} A^\mu , \nabla_\mu \phi \rangle$ as $\langle A^{\mu}, \L_{d-3} \nabla_{\mu}\phi \rangle$. This is related to the freedom which is explained in the footnote \ref{ambi}. We will revisit this issue at the end of this section.

Consider finally \textbf{B.C. 3}. In this case, the spin-two component field is subject to the HD BC: 
\beqa
\L_{d-2}\L_{-2}\Theta^{0|2} = -\lambda \Theta^{2|2} \qquad \mbox{ and} \qquad  \L_{-1}\L_{-2}\Theta^{2|2}\big|_{\theta = \pm \alpha} = 0. 
\eeqa
We thus need to adopt the extended inner product for terms involving the mode function  $\Theta^{2|2}$. 
It is the first term in Eq.(\ref{expanded}) that contains the kinetic and mass-like terms of spin-two field $h_{\mu \nu}$. 
Using the extended inner product Eq.(\ref{ExN})  with $a = -2$, $b=d-2$, $c=-1$, we get the boundary action as 
\begin{align}
I_{\rm boundary, BC3} = - \sum_{\sigma = \pm} \int \mathrm{d}^{d+1} x \sqrt{-g} \; 
\mathcal{L}_2 \left(h^\sigma_{\mu \nu};g_{\mu \nu},d+1\right)  
\label{boundary spin-2 action}
\end{align}
where we renamed the boundary value of the bulk spin-two field by 
\beq
h^\sigma_{\mu \nu} = \left(\frac{1}{d-1}\frac{(\text{sec}\alpha)^{d-4}}{\text{tan}\alpha} \right)^{1/2} h_{\mu \nu} \Big\vert|_{\theta = \sigma \alpha}, \qquad (\sigma = \pm).
\eeq
Most significantly, with extra minus sign in front of the boundary action Eq.(\ref{boundary spin-2 action}), the boundary spin-two field $h^\pm_{\mu \nu}$ has kinetic terms of wrong sign. 
Again, this fits nicely with the result of section 5.2 that the waveguide compactification with {\bf B.C. 3} yields non-unitary spectrum for the partially massless spin-two fields. As stressed already, we hardly anticipated that this is an obvious result just from the HD BCs. With the extended inner product, we now have a firm understanding for the origin of the non-unitarity of partially massless spin-two field without ever invoking $\mathfrak{so}(d,2)$ representation theory. 

Summarizing,
\vskip0.2cm
\begin{tcolorbox}
\begin{itemize}
\item
For the HD BC,
we need to extend the functional space from ${\cal L}^2$ to ${\cal L}^2\oplus \mathbb{R}^N$ to ensure the Sturm-Liouville operator self-adjoint. We showed that this extension can be physically understood as adding $N$ many boundary degrees of freedom. 
\item 
From the extended inner product, we constructed the boundary action for a given HD BC.
The boundary action enabled to directly trace the origin of (non)unitarity of waveguide spectrum. 
\item
For \textbf{B.C. 3} in section \ref{sec spin2 BC}, the boundary action of boundary spin-two fields have kinetic term of wrong sign. This explained why the partially massless spin-two field is non-unitary in AdS space.
\item 
For \textbf{B.C. 1} in section \ref{sec spin2 BC}, the boundary action of boundary spin-zero fields have kinetic term of conventional sign. This explain why the massive spin-zero field is unitary. 
\end{itemize}
\vskip-0.4cm
\end{tcolorbox}
\noindent

Before concluding this subsection, let us revisit the ambiguity mentioned in the footnote {\ref{ambi}}. Consider the \textbf{B.C. 1} and the term $\langle \L_{d-3} \phi ,  \L_{d-3} \phi \rangle$. Such term was classified as originating from $\langle \Theta^{1|2} ,  \Theta^{1|2} \rangle$, so appears not to be refined. On the other hand, using Eq.(\ref{IbP}), this term can also be rewritten as $-\langle \phi , \L_{-1} \L_{d-3} \phi \rangle$ with surface term $\big[(\text{sec}\alpha)^{d-4} \, \phi \, \L_{d-3}\phi \big] \big|_{-\alpha}^{+\alpha}$. This term belongs to $\langle \Theta^{0|2}, \Theta^{0|2} \rangle$, so needs to be refined. Though this appears to pose  an ambiguity, it actually is not. From
\begin{align}
\llangle \, \phi ,  \L_{-1} \L_{d-3} \phi \, \rrangle &= \langle \phi ,  \L_{-1} \L_{d-3} \phi \rangle + \sum_{\sigma = \pm} \mathcal{N}_\sigma \Big( \phi \, \L_{-1} \L_{d-3} \phi 
\Big)_{\theta = \sigma \alpha} \nonumber \\
&= -\langle \L_{d-3} \phi , \L_{d-3} \phi \rangle + \sum_{\sigma = \pm} \mathcal{N}_\sigma \Big( \phi \, \L_{d-2} \L_{d-3} \phi 
\Big)_{\theta = \pm \alpha} , 
\end{align}
we see that $\llangle \phi ,  \L_{-1} \L_{d-3} \phi \rrangle$ and $-\langle \L_{d-3} \phi,  \L_{d-3} \phi \rangle$ are the same, up to boundary conditions.
One can start with any bulk action, but the extended action is always the same. There is no ambiguity.

\section{Waveguide Spectrum of Spin-$s$ Field}\label{sec8}

From lower-spin field cases, we found that the linearized field equations of the waveguide compactification are determined uniquely by the linearized gauge symmetries. In this section, we shall take the same route and extend the results to arbitrary higher-spin fields.

For spins greater than three, however, yet another issue crops out.  
In the spin-three case, we had to sort out the issue that 
the gauge parameter is traceless. 
Now, in the Fronsdal formulation of higher-spin fields, not only gauge parameters are traceless but also higher-spin fields are doubly traceless.
As in the spin-three case, naive component decomposition does not yield the Stueckelberg fields after compactification because of the (doubly) traceless conditions for gauge fields or gauge parameters.  
In this section, we will explicitly construct the correct linear combination of higher-spin fields and gauge parameters. Fortuitously, analogous to the lower spin cases, we are able to explicitly construct the Stueckelberg spin-$s$ field and relations among mode functions. We shall compare our results with previous work~\cite{Zino1}, as we can determine the equation of motions and the mass spectrum for all possible boundary conditions, including partially massless higher-spin fields.

\subsection{Stueckelberg gauge transformations}\label{sec: lin}
In this subsection, we show that the gauge variation of AdS$_{d+1}$ higher-spin field $\phi^{(k)}$ for $0 \le k \le s$ in AdS$_{d+1}$ space, which is a linear combination of AdS$_{d+2}$ spin-$s$ field  $\bar{\phi}_{M_1,\,M_2,\cdots,\,M_s}$ after the AdS waveguide compactification, is given by
\begin{align} \label{sgt}
\delta \phi^{(k)}_{\mu_1 \cdots \mu_k} = &\frac{k}{s}\, \nabla^{\phantom{a}}_{(\mu_1}\, \xi^{(k)}_{\mu_2 \cdots \mu_k )} +  a_1\,\L_{-(s + k-1)}\, \xi^{(k+1)}_{\mu_1 \cdots \mu_k} 
+ a_2\, \L_{d-(s -k)-2}\, g^{\phantom{a}}_{(\mu_1 \mu_2}\,\xi^{(k-1)}_{\mu_3 \cdots \mu_k )}\,,
\end{align}
where the coefficients are 
\beq
a_1=\frac{s -k}{s} \qquad \mbox{and} \qquad a_2=\frac{k\,(k-1)\,(d+s +k-3)}{s \,(d+2k-5)\,(d+2k-3)}\, . \nonumber
\eeq
Extending the pattern of lower-spin fields in the previous sections, we expect that $\phi^{(k)}$'s are the candidates of higher-spin Stueckelberg field in AdS$_{d+1}$ space. These fields were first considered in ~\cite{Zino1}. We now show that the gauge transformations Eq.\eqref{sgt} exactly matches with the higher-spin Stueckelberg gauge transformations, which were previously derived directly in~\cite{Zino1}.

We first construct the correct linear combinations that render the corresponding higher-spin field doubly traceless. The AdS$_{d+2}$ spin-$s$ field $\bar{\phi}^{(s)}_{M_1 \cdots M_s}$ is totally symmetric and doubly traceless, while its gauge transformation parameter $\bar{\xi}^{(s)}_{M_2 \cdots M_s }$ is totally symmetric and traceless. Upon the AdS waveguide compactification, we define the AdS$_{d+1}$ higher-spin fields and gauge transformation parameters in terms of the combinations,
\beqa
\psi^{(s-m)}_{\mu_1 \cdots \mu_{s-m}} \, \, & \equiv & \, \, \bar{\phi}^{(s)}_{\mu_1 \cdots \mu_{s-m} \theta \cdots \theta} 
\nonumber \\
\zeta^{(s-m)}_{\mu_1 \cdots \mu_{s-m-1}} & \equiv & \, \bar{\xi}^{(s)}_{\mu_1 \cdots \mu_{s-m-1} \theta \cdots \theta},
\eeqa 
where $m$ indices are taken along the waveguide $\theta$-direction. A complication is that, while $\bar{\xi}$ ($\bar{\phi}$) obey the (double) traceless conditions, $\zeta$ ($\psi$) are not. 
To get the double traceless AdS$_{d+1}$ fields $\phi^{(k)}$, we thus need to consider linear combination of $\psi^{(\ell)}$'s. So, our ansatz for  double traceless higher-spin field is
\begin{equation}
\phi^{(k)}_{\mu_1 \cdots \mu_k}=\sum_{m=0}^{[k/2]}\, c_{m,k}\, \psi^{(m,k)}_{\mu_1 \cdots \mu_k}
\qquad \mbox{where} \qquad
\psi^{(m,k)}_{\mu_1 \cdots \mu_k} =g^{\phantom{aa}}_{(\mu_1 \mu_2}\cdots g^{\phantom{aa}}_{\mu_{2m-1} \mu_{2m}}\, \psi^{(k-2m)}_{\mu_{2m+1}\mu_{2m+2}\cdots \mu_k )}\,.
\label{spins linear}
\end{equation}
The trace part of $\psi$-fields is not included in this linear combination as it would contain the divergence term $\nabla^{\mu} \xi_{\mu \nu \cdots}$ in the gauge transformation.
Requiring the double traceless condition to $\phi^{(k)}$ field leads to recursion relations to the coefficients $c_{m, k}$.  Taking the normalization $c_{0,k} = 1$, the solution is
\begin{eqnarray}
c_{m,k}
&=& \frac{1}{4^m\, m!}\frac{\Gamma(k+1)\,\Gamma\left(k+(d+1)/2-2-m\right)}{\Gamma(k-2m+1)\,\Gamma\left(k+(d+1)/2-2\right)}\,
\qquad\text{for}\qquad4 \le k \le s.
\end{eqnarray}
By a similar method, we get the traceless linear combination of gauge transformation parameters:
\beq
\xi^{(k)}_{\mu_1 \cdots \mu_{k-1}} \equiv \sum_{m=0}^{[k/2]}\,\frac{k-2m}{k}\,c_{m,k}\, \zeta^{(m,k-1)}_{\mu_1 \cdots \mu_{k-1}}\,,
\eeq
where $\zeta^{(m,k-1)}$ is defined similarly to Eq.\eqref{spins linear}.
With these higher-spin fields and higher-spin gauge transformation 
parameters, the AdS$_{d+2}$ gauge transformations are reduced to the gauge transformations Eq.(\ref{sgt}).

\subsection{Kaluza-Klein modes and ground modes}\label{mode s}
Having identified the correct gauge transformations Eq.(\ref{sgt}) of irreducible decompositions, we now derive the relations between expansion modes $\Theta_n^{k|s}$s and their differential relations. Requiring each term in the gauge transformations Eq.(\ref{sgt}) expanded by the same mode functions, we get the complexes
\vskip0.2cm
\begin{tcolorbox}
\vskip-0.5cm
\begin{eqnarray}
&& \begin{pmatrix} 0 & \L_{d-(s-k)-2}  \\ \L_{-(s+k-2)}  & 0 \end{pmatrix} 
\begin{pmatrix} \Theta^{k|s}_n \,\,\,\\
\,\,\, \Theta^{k-1|s}_n \end{pmatrix}
=
\begin{pmatrix} 
c^{k-1|k}_n\, \Theta^{k|s}_n\,\, \\
\,\,\, c^{k|k-1}_n\, \Theta^{k-1|s}_n \end{pmatrix}\,,
\qquad k=1, \cdots, s. \qquad 
\label{lad}
\end{eqnarray}
\end{tcolorbox}
\noindent
These relations determine the Sturm-Liouville differential equations of $\Theta^{\ell|s}_{n}$'s for all $\ell=0, \cdots, s$:
\begin{align}
\L_{d-(s-k)-2}\, \L_{-(s+k-2)}\, \Theta^{k|s}_{n}
&=c^{k|k-1}\,c^{k-1|k}\, \Theta^{k|s}_{n} \label{eve1}\,, \\
\L_{-(s+k-1)}\, \L_{d-(s-k)-1}\, \Theta^{k|s}_{n} 
&=c^{k|k+1}\,c^{k+1|k}\, \Theta^{k|s}_{n}\,.\label{eve2}
\end{align}
 Here, $-M^2_{n,k|s}$ is used to represent the $n$-th characteristic value of the Sturm-Liouville problems Eqs.(\ref{eve1},\ref{eve2}).  One can show that Eq.\eqref{eve1} and Eq.\eqref{eve2} are equivalent, as the identity 
 \beq
 \L_m\,\L_n - \L_{n-1}\, \L_{m+1}=(n-m-1)
 \nonumber
 \eeq
relates the Sturm-Liouville operators each other, and the eigenvalues each other by
\beq
M_{n,k|s}^2=
M_{n,k+1|s}^2+d+2k-3.
\eeq
All relations are summarized by the spin-$s$ spectrum generating complex :
\vskip0.3cm
\begin{tcolorbox}
\vskip-0.2cm
\setlength\arraycolsep{2pt}{
\beqa
\begin{array}{rcll} \label{Thr2}
 & \Theta^{s|s}_n & &\\
\L_{d-2} & \upharpoonleft \downharpoonright & \L_{-(2s-2)}& -M_{n,s|s}^2=-M_n^2 \\
&\vdots&&\qquad \qquad \vdots \\
\L_{d-(s-k-1)-2} & \upharpoonleft \downharpoonright & \L_{-(s+(k+1)-2)} &-M_{n,k+1|s}^2= -\left(M_n^2 +(s-k-1)\,(d+s+k-3)\right)\\
 & \Theta^{k|s}_n && \\
\L_{d-(s-k)-2} & \upharpoonleft \downharpoonright & \L_{-(s+k-2)} & -M_{n,k|s}^2=-\left(M_n^2 +(s-k)\,(d+s+k-4)\right) \\
&\vdots& &\qquad \qquad \vdots \\
\L_{d-s-1} & \upharpoonleft \downharpoonright & \L_{-(s-1)} & -M_{n,1|s}^2=-\left(M_n^2+(s-1)\,(d+s-3)\right)\\
&\Theta^{0|s}_n &&
\end{array}
\eeqa}
\end{tcolorbox}
\noindent
Here, $M^2_{n,s|s}$ is the mass-squared of $n$-th mode of spin-$s$ field. They in turn determine mass-squared of lower spin fields, $k=s-1, s-2, \cdots, 1$. This spectrum-generating complex enables us to interpret the gauge transformations Eq.(\ref{sgt}) as the Stueckelberg gauge transformations. Let us choose, for convenience, the relative normalizations in Eq.\eqref{Thr2} as
\beq
\L_{-(s+k-2)}\,\Theta^{k|s}_n=-a_{k|s}\,M_{n,k|s}\,\Theta^{k-1|s}_{n}\quad \mbox{and} \quad \L_{d-(s-k)-2}\,\Theta^{k-1|s}_n=\frac{M_{n,k|s}}{a_{k|s}}\,\Theta^{k|s}_n\label{Thr}
\eeq
where factors independent of mode index $n$ are put together to
\beq
a^2_{k|s}=\frac{k\,(d+s+k-3)}{(s-k+1)\,(2k+d-3)}. 
\nonumber
\eeq
Then, the gauge transformation Eq.(\ref{sgt}) precisely gives rise to the Stueckelberg spin-$s$ gauge transformation in AdS$_{d+1}$ space previously derived in~\cite{Zino1}:
\begin{align}
&\delta \phi^{(k)}_{\mu_1 \cdots \mu_k} = \frac{k}{s}\,\nabla^{\phantom{aa}}_{(\mu_1}\,\xi^{(k)}_{\mu_2 \cdots \mu_k )}
+\alpha_k\, \xi^{(k+1)}_{\mu_1 \cdots \mu_k} 
+ \beta_k\, g^{\phantom{aa}}_{(\mu_1 \mu_2}\xi^{(k-1)}_{\mu_3 \cdots \mu_k )}\,,\label{Zc}
\end{align}
where
\begin{align}
&\alpha^2_k = \frac{(k+1)\,(s-k)\,(d+s+k-2)}{s^2\,(d+2k-1)}\left( M^2 +(s-k-1)\,(d+s+k-3) \right) \label{ZR}\,, \\
&\beta_k = -\frac{(k-1)}{(d+2k-5)}\,\alpha_{k-1}. \nonumber
\end{align}
Here, the dependence on mode label $n$ enters only through the mass-squared $M^2 := M_n^2$. Apart from this, all modes of spin-$k$ fields have the same structure of gauge transformations.  
Therefore, spin-$k$ gauge transformation of $n$-th mode is simply the Stueckelberg gauge transformation of spin-$k$ field with mass $M_n$.
In turn, these gauge transformations completely fix the equations of motion for each spin $k=0, 1, \cdots, s$ and for each mode $n$. They constitute the Kaluza-Klein modes. 

As in the lower-spin counterparts, were if $M_n^2$ is tuned to special negative values, it can happen that $\alpha_k=0$. These special values are the values at which $M_{n,k+1|s}=0$ as well. In this case, the Stueckelberg system of spin-$s$ field decomposes into two subsystems: the partially massless spin-$s$ system of depth $t = (s- k-1)$ and the Stueckelberg spin-$k$ field. Importantly, the massless spin-$s$ field is also part of the spectrum, since it is nothing but partially massless spin-$s$ field of depth-0~\footnote{Note that our conventions of the mass-squared of higher-spin field is such that it is zero when the higher-spin fields have gauge symmetries. So, it differs form the mass-squared that appears in the AdS Pauli-Fierz equation,  $\left(\nabla^2+\kappa^2_s\right)\phi_{\mu_1\,\mu_2\,\cdot\mu_s}=0$. }. 
Together, they constitute the ground modes: \label{subsystem}
\vskip0.2cm
\begin{tcolorbox}
\vskip-0.2cm
\begin{itemize}
\item 
The upper subsystem consists of $\left(\phi^{(s)},\,\phi^{(s-1)},\,\cdots\,\phi^{(k+1)}\right)$ and forms the Stueckelberg system of partially massless field with depth $t=(s-k-1)$. Their mass spectra are given by
\begin{align} \label{mass1}
M^2=-t\left(d+2s-t-4\right)/\ell^2.
\end{align}
\item 
The lower subsystem consists of $\left(\phi^{(k)},\,\phi^{(k-1)},\,\cdots\,\phi^{(0)}\right)$ and forms the Stueckelberg spin-$k$ field. Their mass spectra are given by
\begin{align} \label{mass2}
M^2=(s-k+1)(d+s+k-3)/\ell^2.
\end{align}
\end{itemize}
\end{tcolorbox}
\noindent 

Group theoretically, the decomposition pattern of the ground modes can be understood in terms of the Verma $\mathfrak{so}(d,2)$-modules. At generic conformal weight $\Delta$, the Verma module $\mathcal{V}\left(\Delta,\,s\right)$ is irreducible. At special values of $\Delta=d+k-1$, however, $\mathcal{V}(\Delta, s)$ decomposes into $\mathfrak{so}(d,2)$ irreducible representations \cite{Xi1, Shaynkman:2004vu}:
\beq
\mathcal{V}\left(d+k-1,\,s\right)= 
\mathcal{D}\left(d+k-1,\,s\right)\oplus
\mathcal{D}\left(d+s-1,\,k\right)\,.\label{breaking verma}
\eeq
In Eq.\eqref{breaking verma}, the irreducible representation ${\cal D}(d+k-1, s)$ represents the partially massless spin-$s$ field, while $\mathcal{D}\left(d+s-1,\,k\right)$ represents the massive spin-$(k+1)$ field whose mass-squared is set by the conformal weight $\Delta$:
\beq
m^2\,\ell^2=\Delta\left(\Delta-d\right)-\left(s-2\right)\left(d+s-2\right)\, . \label{mass conformal}
\eeq

\subsection{Waveguide boundary conditions}
\begin{figure}[!h] 
 \label{fig4}
\centering
{\includegraphics[scale=0.43]{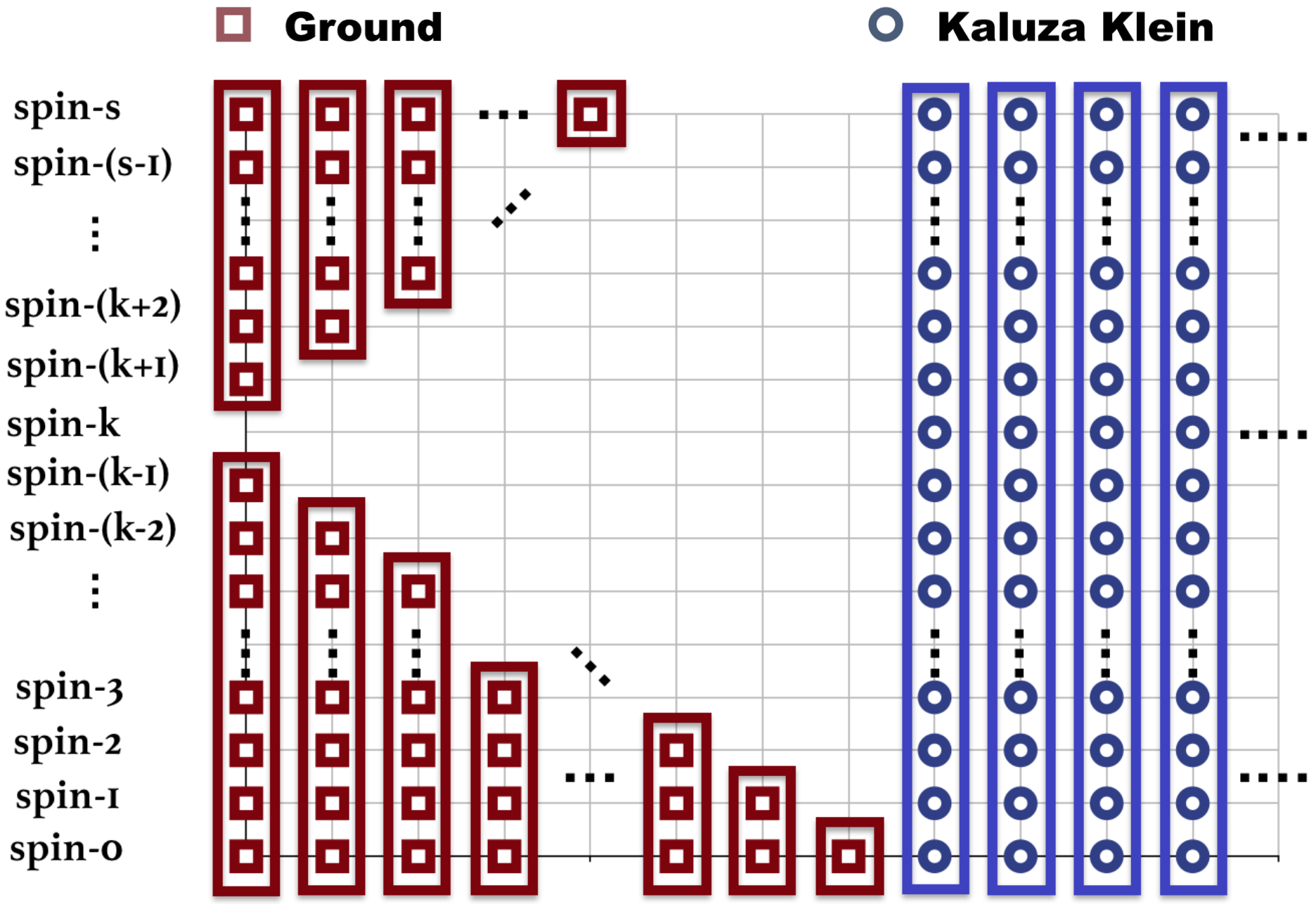}}
\caption{\sl
Mass spectra for all possible boundary conditions characterized by Dirichlet conditions on $\Theta^{k|s}$.
Each point represents a mode function. Points in the same rectangle form the Stueckelberg system with the highest spin. The the upper triangle consists of the Stueckelberg system of partially massless field,
while the lower triangle consists of the Stueckelberg spin $\ell=0,\,1,\cdots k-1$, 
as described in Eq.\eqref{Spins Spect}.
}
\label{spectrumspins}
\end{figure}

We next classify all possible AdS waveguide boundary conditions and determine the mass spectra. As before, we shall only consider boundary conditions derived from the Dirichlet conditions on $\Theta^{k|s}|_{\theta=\pm\alpha}=0$ for some $k$. For $0\le k \le s$, there are $(s+1)$ possible Dirichlet conditions.  The relations Eq.(\ref{Thr2}) then fix the boundary conditions for all other fields originating from the same mode functions: 
\beqa
&& \L_{-(s+k-1)}\,\cdots \, \cdots \, \cdots \L_{s+\ell -2}\,\Theta^{\ell |s}\Big\vert_{\theta=\pm\alpha}=0 \qquad \qquad (\ell = k, k+1, \cdots, s) \nonumber \\
&&  \L_{d-(s-k)-2}\,\cdots\L_{d-(s-\ell-1)-2}\,\Theta^{\ell k|s} \Big\vert_{\theta=\pm\alpha}=0 \qquad \qquad (\ell = k, k+1, \cdots, s)
\,.\label{spins BC}
\eeqa
Below, we show that the pattern of mass spectrum takes the form of Fig.\ref{spectrumspins}. 
First, to counter cluttering indices, we define simplifying notations as $\Theta^{\ell}:=\Theta^{k+\ell |s}$, $M_{n,\ell}^2:=M_{n,\,k+\ell |s}^2$, $U_\ell := \L_{d-s-2+\ell+k}$ and $D_\ell=\L_{-s+2- \ell -k}$. Then, the sub-complex of Eq.\eqref{Thr2} can be written in the form
\beq
\begin{array}{rcll}
 & \Theta^{\ell}_n & &\\
U_{\ell} & \upharpoonleft \downharpoonright & D_{\ell}&   : \qquad -M_{n,\ell}^2=-\left(M_n^2 +(s-k-\ell)\,(d+s+k+\ell -4)\right) \\
 & \Theta^{\ell -1}_n & &\\
\end{array} \,  . 
\eeq
By this complex, there is one-to-one map between $\Theta^{\ell}_n$ and $\Theta^{\ell-1}_n$ for $M_{n, \ell}^2 \neq0$. If $M_{n,\ell}^2 =0$, there exists one additional mode $\Theta^\ell_0$ ($\Theta^{\ell -1}_0$) when $\ell$ is positive (negative). 
This additional mode satisfies $D_l\, \Theta^l_0=0$ for positive $l$, $U_l\,\Theta^{l-1}$ for negative $l$.
After inductively applying this relation from $\ell =0$, one can show that the structure of mode function is given by
\beq
\{\Theta^{k+ \ell |s}\}=\{K^\ell_i,\, G^{l}_{a=1,\,2,\,\cdots\,\ell}\}\,,\qquad \mbox{and} \qquad
\{\Theta^{k-\ell |s}\}=\{K^{-\ell}_i,\, G^{-\ell}_{a=1,\,2,\,\cdots\,\ell}\}\,.
\eeq
Here, $K^\ell_i$'s are the Kaluza-Klein modes, and $G^{\ell }_a$'s are the ground modes which satisfy the equations
\beq
\left\{ \begin{array}{lll}
D_a\,D_{a+1}\cdots D_{\ell}\, G_a^\ell =0 &\quad \text{with}\;\; D_k \,D_{k+1}\cdots D_\ell \, G_a^\ell \neq 0 \quad &\text{for all}\;\; a<k\\
U_{-a+1}\,U_{-a}\cdots U_{-\ell +1}\, G_a^{-\ell}=0 &\quad \text{with} \;\; U_k \, U_{k+1}\cdots U_{-\ell +1}\, G_a^{-\ell }\neq 0 \;\; &\text{for all}\quad a<-k+1\\
\end{array}\right. \, .
\eeq
The ground modes $G_i^\ell$ with the same subscript $i$ have the same eigenvalues. Their eigenvalues can be obtained by the first-order differential equation $D_\ell \,G^\ell _\ell =0$ and $U_{-\ell +1}G^{-\ell }_\ell =0$ for positive $\ell$.
Finally, fields corresponding to $G_a^\ell$, $\ell= s-k,\,\cdots, a$ form  the Stueckelberg system of partially massless spin-$s$ field with depth-$(s-k-a+1)$. See our earlier expositions on this in Eq.\eqref{ZR} and thereafter. 
Fields corresponding to $G_a^\ell$, $\ell =-a,\,\cdots, -k$ form the massive spin-$(k-a)$  Stueckelberg field with mass-squared $M^2=(s-k+a+1)(d+s+k-a-3)/\ell^2$. These spectra are depicted in Fig. \ref{spectrumspins}.

Summarizing,
\vskip0.3cm
\begin{tcolorbox}
\vskip0.1cm
\begin{itemize}
\item   
Mass spectrum for the boundary condition characterized by Dirichlet condition at $\Theta^{k|s}$ consists of three parts. The first part is the set of massive spin-$s$ Kaluza-Klein tower, whose mass-squared is given by the eigenvalue of the Sturm-Liouville problem, Eq. \eqref{eve1} with $k=s$.
The second part is the set of partially massless spin-$s$ field with depth-$(0,\,1,\,2,\cdots,\, s-k-1)$.
The third part consists of the set of massive Stueckelberg spin $\ell =0,\,1,\cdots,\, k-1$ with mass-squared, $M^2=(s-\ell+1)(d+s+\ell-3)/\ell^2$. \label{Spectra}
\item
The $\mathfrak{so}(d,2)$ representations of ground modes are
\beq
\left\{ \begin{array}{lll}
\mathcal{D}\left(d+s-t-2,\,s\right)&\text{for}&t=0,\,1,\cdots,\, s-k-1\\
\mathcal{D}\left(d+s-1,\,\ell \right)&\text{for}& \ell =0,\,1,\cdots,\, k-1
\end{array}\right.\label{Spins Spect}
\eeq
whose masses are set in terms of conformal weights by Eq.\eqref{mass conformal}.
\end{itemize}
\end{tcolorbox}

\subsection{More on Boundary Conditions}
So far, we studied a class of AdS waveguide boundary conditions for spin-$s$ field that are characterized by Dirichlet condition on a mode function. Quite satisfactorily, the resulting $(s+1)$ possible boundary conditions led to the mass spectra encompassing all of massive, partially massless and massless higher-spin fields upon the compactification. Here, we dwell further on variations of this idea. 

\subsubsection*{Higher-derivative boundary conditions and boundary action}
Two remarks are in order regarding the boundary conditions and boundary action.
We showed in section \ref{sec7} that, for spin-two system, HD BC imply the presence of  boundary degrees of freedom with nontrivial dynamics described by their boundary action.
Mainly due to algebraic complexity, extension of the analysis to higher-spin is not performed yet with full generality. However, the factorization property of Sturm-Liouville differential equations renders the analysis simplified. Based on the intuition gained so far, we here mention the expected patterns for arbitrary spin-$s$.
\begin{itemize}
\item 
For the boundary conditions derived from the Dirichlet conditions $\Theta^{k|s}|_{\theta=\pm\alpha} = 0$ for $k\le s-2$, the spectrum contains partially massless higher-spin fields. Their non-unitarity can be traced to the non-unitarity of boundary action. One can show that the boundary action has some of the kinetic terms with wrong sign, leading to indefinite norm in the extended Hilbert space. These boundary degrees of freedom is precisely associated with the partially massless fields.
In contrast, for the cases with $k\ge s-1$, one can also show that all kinetic terms in the boundary action have positive sign. Correspondingly, they are all unitary.
\item
Explicit analysis for spin-three and spin-four fields suggests that the masses of the boundary degree of freedom is proportional to the masses in Eq.(\ref{mass1}) and Eq.(\ref{mass2}).
For instance, in spin-three system, we verified that the boundary action of \textbf{type II} in Table \ref{spin3 actors} is associated with the first spectrum in Fig. \ref{spectra3}
\end{itemize}

\subsubsection*{Extended Dirichlet conditions}\label{extension BC}
More generally, we can also consider extended form of the boundary conditions beyond those characterized by the Dirichlet conditions \footnote{In section \ref{sec10}, we will discuss other class of extended boundary conditions.
}. We can see from the pattern of raising and lowering operators between adjacent spins in the spectrum generating complex Eq.\eqref{Thr2} that a new class of boundary conditions are possible. To this end, let us extend the mode functions $\Theta^{k|s}|_{\theta=\pm\alpha}=0$  to the regime, $k>s$ and $k<0$. Though there is no corresponding physical excitations, formal analytic extension is possible;
\beqa
\begin{array}{rcll}
 & \Theta^{s+1|s}_n & &\\
\L_{d-2} & \upharpoonleft \downharpoonright & \L_{-2s+1}& -M_{n,s+1|s}^2=-\left(M_n^2-2\,s-d+3 \right)\\
 & \Theta^{s|s}_n &&
 \end{array}\,.
\eeqa
By analysis identical to the ones in the previous cases, one can derive the spectrum of the ground modes and relate them to $\mathfrak{so}(d,2)$-modules. For $k>s$, the ground modes consist of two parts: the ground modes of $k=s$ and additional towers
\beq
M^2=\left(l-s\right)\left(d+s+l-4\right)/\ell^2\,,\quad
{\mathcal D}(d+l-2,s)\quad\text{for  }\,\,l=s+1,\,s+2\,,\cdots\,,k \, .
\eeq 
For $k<0$, the ground modes also consist of two parts: the ground modes of $k=0$ and additional towers,
\beq
M^2=\left(l-s\right)\left(d+s+l-4\right)/\ell^2\,,\quad
{\mathcal D}(d+l-2,s)\quad\text{for  }\,\,l=-|k|+1, -|k|+2, \cdots, -1, 0 \,.
\eeq 
We see that the  $k<0$ case gives rise to non-unitary representations of $\mathfrak so(d,2)$. 

\subsection{Decompactification limit} \label{alpha two pi limit} 
In our setup, the AdS waveguide ranges over the Janus wedge $[- \alpha, \alpha]$. If the wedge $\alpha$ approaches $\pi/2$,  the waveguide decompactifies to the entire AdS$_{d+2}$.
In other words, the $\alpha=\pm \pi/2$ hyperplanes correspond to the AdS$_{d+2}$ boundary. As such, one might anticipate that the spectra of AdS$_{d+1}$  waveguide asymptotes to the spectra of AdS$_{d+2}$. This seems to be in tension with our result, as the mass spectra of spin-$s$ field in AdS$_{d+1}$ space arises only for special set of boundary conditions. Here, we discuss subtleties involved in the decompactification limit.

Consider the AdS$_{d+2}$  massless spin-$s$ spectrum from the viewpoint of AdS$_{d+1}$ space. This is just like the $L \rightarrow \infty$ limit of flat space waveguide we studied in section 2. The ${\cal L}^2$ square-integrable modes of massless spin-$s$ field form the $\mathfrak{so}(d+1,2)$-module: $\mathcal{D}(d+s-1,s)_{\rm \mathfrak{so}(d+1,2)}$. Representation theoretically, we can decompose this module into $\mathfrak{so}(d,2)$-modules, a procedure referred to as the "dimensional degression" in \cite{Dimensional Degression}:
\beq
\mathcal{D}(d+s-1,s)_{\rm \mathfrak{so}(d+1,2)}
=
\bigoplus_{n=0}^\infty\, {\mathcal D}(d+n+s-1,s)_{\rm \mathfrak{so}(d,2)}
\oplus
\bigoplus_{l=0}^{s-1}\,{\mathcal D}(d+s-1,l)_{\rm \mathfrak{so}(d,2)}\,.\label{spins degression}
\eeq
We can relate these modules with states that arise from the compactification of higher-spin field in the AdS waveguide, as the foliation of Fig.\ref{fol} precisely matches with the above dimensional degression.
There are two kinds of ${\mathfrak{so}(d,2)}$-modules in the right hand side of Eq.\eqref{spins degression}:
the first set of modules have the same spin, spin-$s$ but different conformal dimensions, while the second set of modules have the same conformal dimension but different spins ranging over 0 to $s-1$. 
We see that the second set of modules in Eq.\eqref{spins degression} coincide with the set of ground modes for $k=s$ ($\Theta^{s|s}|_{\theta=\pm \alpha} = 0$) in Eq.\eqref{Spins Spect}.
In order to reconstruct the $\mathfrak so(d+1,2)$ module in the left side of Eq.(\ref{spins degression}), we would then need the Kaluza-Klein modes from $k=s$ to match with the the first set of modules. Below, we demonstrate this affirmatively.

In the $k=s$ case, the mass spectra of spin-$s$ field are determined by the Sturm-Liouville equation Eq.\eqref{eve1} and the Dirichlet condition for mode functions $\Theta^{s|s}$:
\beq
\L_{d-2}\,\L_{-2(s-1)}\,\Theta_n^{s|s}=-M_n^2\, \Theta_n^{s|s}\,,\qquad \mbox{where} \qquad \Theta^{s|s}|_{\theta=\pm\alpha}=0\,.
\eeq
The solution is given by
\beq
\Theta_n^{s|s}=\left(\cos\theta\right)^{\mu}\,\left(c_1\,P_\nu^\mu(\sin\theta)+c_2\,Q_\nu^\mu(\sin\theta)\right)\,,
\eeq
where $P_\nu^{\mu}$ and $Q_\nu^{\mu}$ are the associated Legendre functions with arguments,
$\mu=\frac12\left(d+2\,s-3\right)$ and $\nu\left(\nu+1\right)=M_n^2-\frac14\left(1-\left(d+2s-4\right)^2\right)$\,. 
In the decompactification limit, the boundary conditions $\Theta^{s|s}|_{\theta=\pm\alpha} = 0$ take the form
\beqa
0&=&-\frac\pi2\,\sin A\Big((P_\nu^\mu)^2-\frac{4}{\pi^2}\,(Q_\nu^\mu)^2\Big)
-2 \cos A\, P_\nu^\mu\,Q_\nu^\mu\\
&\simeq&\left\{ \begin{array}{lll}
-\frac1{2\,\pi}\,\sin A
\left(\cos(\mu\,\pi)\,\Gamma(\mu)\right)^2\left(\frac{2}{\epsilon}\right)^{\mu}\quad&\text{for even $d$}\\
\\
-\frac\pi2\,\sin A
\left(\frac{1}{\Gamma(1-\mu)}\right)^2\left(\frac{2}{\epsilon}\right)^{\mu}\quad&\text{for \, odd $d$}
\end{array}\right.
\eeqa
where $A=\pi\left(\mu+\nu\right)$ and $1\gg\epsilon=1-\sin\alpha>0$.
Therefore, it must be that $\mu+\nu$ are integer-valued in the decompactification limit.
From the relation Eq.\eqref{mass conformal}, it immediately follows that the modules that correspond to the Kaluza-Klein modes are precisely $\bigoplus_{n=0}^\infty\, {\mathcal D}(d+n+s-1,s)$.
\vskip0.3cm
\begin{tcolorbox}
Spectrum for the cases of $(k=s)$ goes to the spectrum of ``dimensional degression \cite{Dimensional Degression}" in the decompactification limit(i.e. $\alpha\rightarrow \pi/2$).
\end{tcolorbox}

All are well so far, so one might anticipate that the spectral match with the dimensional degression continues to hold for $k\neq s$. This however is no longer true.  The point is that some of the ground modes in Eq.\eqref{Spins Spect} contain the modules which are not in the massless spin-$s$ modules of AdS$_{d+2}$ space, $\mathcal{D}(d+s-1,s)_{\rm \mathfrak{so}(d+1,2)}$ in Eq.\eqref{spins degression}.
The mode functions that would potentially match with are actually singular (equivalently, the normalization factor goes to zero) in the decompactification limit. In particular, massless spin-$s$ field in the AdS$_{d+2}$ space belongs to one of these singular modes. For spin-two case, this was already shown in Eq.\eqref{bc2spin2}. Conversely, this explains transparently why the dimensional degression \cite{Dimensional Degression} of AdS$_{d+2}$ space does not generate ``massless" spin-$s$ fields in AdS$_{d+1}$ space.

\section{Holographic Dual: Isotropic Lifshitz Interface}
In this section, we discuss an interface conformal field theory whose holographic dual would exhibit the AdS waveguide and higher-spin field theory on it. For concrete discussion, we shall study this in the context of free $O(N)$ vector model in $ 2< d < 4$.  However, the setup is completely general, and can easily be extended to other critical models and higher dimensions. The CFT action is schematically structured as
\beqa
I_{\rm CFT} = 
{1 \over g^2_{\rm B}} \int_{B} {\rm d}^{d+1} x \, (\nabla {\boldsymbol \phi})^2 + \sum_{z=1}^\infty {1 \over g_{z, \rm I}^2} \int_I {\rm d}^{d} x \,   (\partial^z {\boldsymbol \phi})^2 \,  , 
\label{icft}
\eeqa
where $\nabla, \partial$ are $(d+1)$-dimensional and $d$-dimensional derivatives, respectively. 
This CFT system is a juxtaposition of $(d+1)$-dimensional Heisenberg model in the bulk and $d$-dimensional isotropic Lifshitz Heisenberg model in the interface. 

In the action, $g_B$ refers to the coupling parameter in the bulk, while $g_I$ refers to the coupling parameters in the interface. By normalization, one can always set one of the coupling parameters, say, $g_B$ to unity, such that $g_I$'s relative to $g_B$ are physically meaningful. However, for the sake of classification, we will keep the normalizations as above. 

The behavior of the system is extremely rich, controlled by the relative strength of the coupling parameters. 
We can classify them as follows. 
\begin{itemize}
\item $g_B^2 \gg g_{k, I}$: The $(d+1)$-dimensional bulk is strongly interacting, so the dynamics is described by $(d+1)$-dimensional $O(N)$ Heisenberg magnet. This system exhibits conserved (even) higher-spin currents starting from $s=2$. The holographic dual would be higher-spin gauge theory on AdS$_{d+2}$ space. As the bulk is strongly interacting, the energy-momentum of the interface is not separately conserved. Only the $(d+1)$-dimensional energy-momentum tensor is conserved. The same goes for higher-spin tensor currents. In terms of AdS waveguide, this is the decompactification limit.  
\item $g_{1, \rm I}^2 \gg g_{z \ne 1, \rm I}^2, g_B^2$: The $d$-dimensional interface is strongly interacting with Lifshitz exponent $z=1$. The dynamics is described by $d$-dimensional $O(N)$ Heisenberg magnet. This system exhibits conserved (even) higher-spin current starting from $s=2$. The holographic dual would be higher-spin gauge theory on AdS$_{d+1}$ space. In terms of AdS waveguide, this corresponds to the Dirichlet condition to each higher-spin field. The interface CFT is $d$-dimensional and cannot interpolate to $(d+1)$ dimensions. In terms of holographic dual AdS waveguide, the decompactification limit is singular. 
\item $g_{k, \rm I}^2 \gg g_{z \ne k, \rm I}^2, g_B^2$:  The $d$-dimensional interface is strongly interacting with the leading Lifshitz exponent $z = k >1$. The dynamics is described by a $d$-dimensional Lifshitz $O(N)_z$ Heisenberg magnet. As the bulk has the Lifshitz exponent $z=1$, this critical behavior is achieved only if the interface is fine-tuned. This is possible only when the interface maintains nontrivial interactions (even though weak) with the bulk. Thus, the interface is an open system interacting with the reservoir bulk system) and hence in general non-unitary. 
\end{itemize}

The critical behavior classified as above fits perfectly with the AdS waveguide spectra we analyzed in the previous sections. In the gravity dual, the AdS wedge angle $\alpha$ measures the extent the extra dimension extends. Thus, we expect that it is related to the coupling parameters of the above interface Lifshitz $O(N)$ Heisenberg magnet as
\beqa
\tan \alpha \simeq {g_{z, \rm I}^2 \over g_{\rm B}^2}.
\eeqa

\begin{figure}[!h] 
 \label{fig-icft}
\centering
\vskip-2cm
\hskip-3cm
{\includegraphics[scale=0.7]{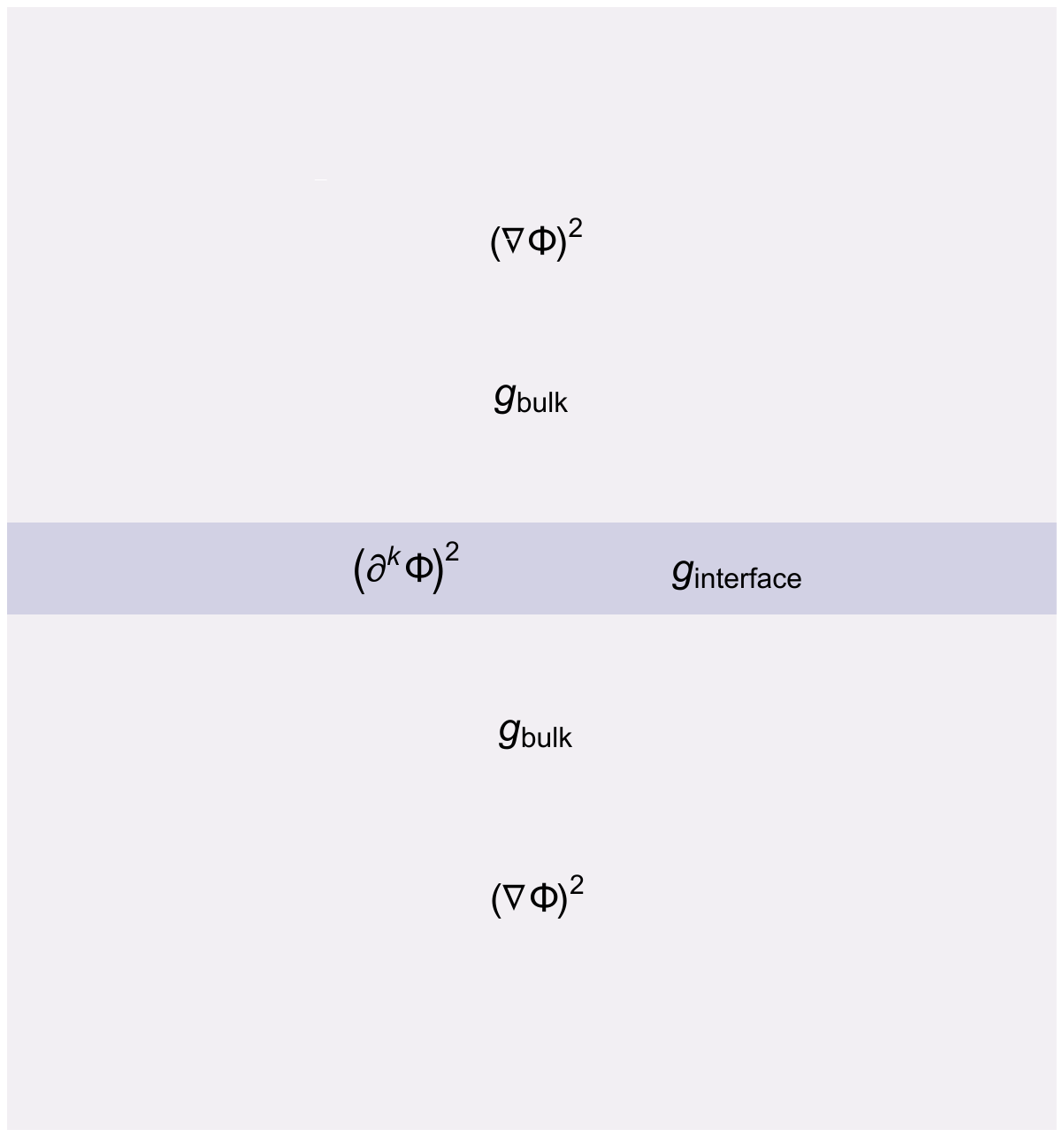}} 
\vskip-7cm
\caption{ {\sl The interface configuration of Lifshitz $O(N)$ Heisenberg magnet. The bulk of the CFT is $(d+1)$-dimensional system. The boundary of the CFT is $d$-dimensional isotropic Lischitz system of exponent $z > 1$}. }
\end{figure}

The critical behavior as above is quite rich and intricate. Nevertheless, it can be easily achieved within the Lifshitz-Janus system. The idea is that the coupling parameters $g_{\rm B}, g_{z, \rm I}$ can be arranged by a nontrivial kink profile of the dilaton field, which is $O(N)$ singlet. The interface is a hypersurface of $(d+1)$ dimensions in which the bulk Heisenberg magnet live in. Denote the local coordinate normal to the interface as $y$ and the dilaton field as $\Phi(y)$. The anisotropic, Lifshitz $O(N)$ Heisenberg magnet controlled by the dilaton field is described by the action
\beqa
I_{\rm CFT}  &=& \int {\rm d}^d x \int {\rm d} y \, (\nabla \boldsymbol \phi)^2 + \sum_{z \ge 2}^\infty \int {\rm d}^d x \int {\rm d} y \, ({\partial_y^{z-1} \Phi(y)}) (\partial^z {\boldsymbol \phi})^2 \, . 
\eeqa
If the dilaton field is fine-tuned to behave across the interface at $y = 0$ as $\Phi(y) \sim g_{k, \rm I}^2 \, y^{k-1} \delta (y)$, the second term gives rise precisely to the type of interactions of the form Eq.(\ref{icft}). 
This argument then provides the existence proof for the CFT dual of the higher-spin system compactified on the Janus waveguide. 

\section{Discussions and Outlooks}\label{sec10}
In this paper, we developed a new approach for realizing (inverse) Higgs mechanism of all known massive and partially massless higher-spin fields, at the linearized level. Our approach is geometric and utilizes the Kaluza-Klein compactification. We pointed out that the Janus wedge provides the AdS waveguide with which the Kaluza-Klein compactification of higher-spin fields can be performed.  We showed that, upon compactification of AdS$_{d+2}$ down to AdS$_{d+1}$, two classes of higher-spin fields appear. The first-class, which we referred to as the Kaluza-Klein modes, comprises of infinite tower of massive higher-spin fields. Their masses depend on the waveguide wedge size $\alpha$, and become infinitely heavy as $\alpha$ goes to zero. As in flat space, these Kaluza-Klein masses are generated by the AdS$_{d+1}$ Stueckelberg couplings, whose origin is the higher-spin gauge symmetry in AdS$_{d+2}$ space.
The second-class, which we referred to as ground modes, consists of higher-spin fields of all possible types: massive, partially massless and massless. Their masses, however, are independent of the wedge angle $\alpha$ and are just set by the AdS$_{d+1}$ radius. Again, their equations of motion are organized by the AdS$_{d+1}$ Stueckelberg couplings coming from the higher-spin gauge symmetry in AdS$_{d+2}$ space. The mass spectrum of ground modes fits to the breaking pattern of Verma $\mathfrak{so}(d,2)$-modules.
The ground modes can be massless, partially massless and massive with specific masses as described in Eq.\eqref{Spins Spect}.

In this paper, we mainly concentrated on the compactification of massless spin-$s$ from AdS$_{d+2}$ to AdS$_{d+1}$ for the set of Dirichlet boundary conditions  $\Theta^{k|s}|_{\theta = \pm \alpha} = 0$. We can extend the analysis along the following directions. 
\begin{itemize}
\item 
We could consider more general boundary conditions, ${\cal M}\,\Theta^{k|s}|_{\theta=\pm\alpha}$ where ${\cal M}$ is an arbitrary differential operator.
This would lead to a set of new boundary conditions
\beq
{\cal M}\,\L_{-(s+k-1)}\,\cdots\L_{s+l-2}\,\Theta^{l\geq k|s}|_{\theta=\pm\alpha}=0
\,,\quad
{\cal M}\,\L_{d-(s-k)-2}\,\cdots\L_{d-(s-l-1)-2}\,\Theta^{l\geq k|s}|_{\theta=\pm\alpha}=0
\label{general BC}
\eeq
The spectrum is more complicated but, by construction, the new ground modes necessarily include the ground modes we discussed in section \ref{sec8} for the Dirichlet boundary condition.
\item
We assumed that the waveguide wedge is parity-symmetric, $-\alpha<\theta<\alpha$. We can generalize it to a general domain $\alpha_1<\theta<\alpha_2$  and impose different boundary conditions at each boundary. 
\item
We can also compactify more than one directions. We worked out details of the dimensional reduction method from AdS$_{d+k}$ to AdS$_d$ in Appendix \ref{DR}. 
It is the AdS generalization of the non-abelian Kaluza-Klein compactification \cite{Cho:1975sw}. This generalization also provides a geometric framework for the colored higher spin theory proposed in \cite{Color Vasiliev, Gwak:2015vfb}.
\item
We can also consider the fermionic higher-spin fields as well as super symmetric higher-spin fields. The latter is particularly interesting from the viewpoint of boundary conditions, as supersymmetry is generally reduced by the boundary conditions.
\item
Although far less interesting, we can also start from partially massless fields in AdS$_{d+2}$ space. Partially massless fields and their Stueckelberg couplings retain partial gauge symmetries. At the linearized level, these partial gauge symmetries are sufficient enough to fix the action, so it should be possible to classify boundary conditions that retain massless fields and understand physical mechanism of the new gauge symmetries.
\end{itemize}

Further development of the Higgs mechanism we studied from the dual conformal field theory viewpoint are within the reach and would be very interesting. 
\begin{itemize}
\item
We argue that isotropic Lifshitz interface of $O(N)$ Heisenberg magnet in the spintronics and the Gross-Neveu fermion system in graphene stacks at Dirac point are ideal candidates for the concrete realization of dual conformal field theory. Both interface and Lifshitz behaviors are realizable in Type IIB string theory from the Janus configuration. 
\item
Further interesting direction of research is the emergence of partially massless fields in the infrared and subsequent renormalization group flow to $(d-1)$-dimensional conformal field theories. For partially massless fields, dual CFTs were conjectured in \cite{Dolan:2001ih, Xi1,higherSin}. It would also be interesting to investigate the reduction of a partially massless field on AdS$_{d+2}$ on codimension-two defects. 
\item
In the context of AdS/CFT correspondence, the free energy of both side of the correspondence should match and, especially, one-loop free energy of AdS side correspond to 1/N correction of free energy of CFT side. The Casimir free energy calculation \cite{CasimirE} will provide clues to the quantum aspects of dual CFTs \footnote{For similar calculation on HS/CFT duality, see \cite{Giombi:2013gk}.}. 
\end{itemize}

We are currently pursuing these issues and will report our results in future publications.

\section*{Acknowledgement}
We are grateful for useful discussions to Karapet Mkrtchyian, Tasso Petkou, Sasha Polyakov, Augusto Sagnotti, Kostas Skenderis, Misha Vasiliev and Yuri Zinoviev. SHG and JWK acknowledges the APCTP Focus Program "Liouville, Integrability and Branes (11)" for providing stimulating scientific environment. SJR acknowledges warm hospitality and stimulating atmosphere of Tessalloniki Workshop "Aspects of Conformal Field Theory" (September  21-25, 2015), Singapore Workshop "Higher Spin Gauge Theories" (November 4 - 6, 2015), Munich IAPP Workshop "Higher Spin Theory and Duality" (May 2 - 27, 2016), where various parts of this work were presented. This work was supported in part by the National Research Foundation Grants 2005-0093843, 2010-220-C00003 and 2012-K2A1A9055280.

\appendix
\section{AdS Space}\label{adsC}
Conventions are demonstrated in $d$ dimension. Space-time coordinate indices $\mu$, $\nu$ and tangent indices $a$, $b$ run from $0$ to $d$. We will use most positive sign for metric and $\eta_{ab}=\text{diag}(-1,1,\cdots,1)$. We will call ${e_\mu}^a$ and (sometimes  its inverse ${e_a}^\mu$ or $e^a={e_\mu}^a\,dx^\mu$) as vielbein if it satisfies ${e_a}^\mu\,\eta_{ab}\,{e_b}^\nu=g^{\mu\nu}$.
Covariant derivatives are defined as follow.  $\omega^{ab}={\omega_\mu}^{ab}\, dx^\mu$ is spin connection.
\beqa
\nabla_\mu A^\rho=\pd_\mu A^\nu + \Gamma^\nu_{\rho\mu}A^\rho,   &\quad& \nabla_\mu A_\nu=\pd_\mu A_\nu -\Gamma^\rho_{\mu\nu}A_\rho\\
\nabla_\mu A^a=\pd_\mu A^a+{{\omega_\mu}^a}_b A^b, &\quad& \nabla_\mu A_a=\pd_\mu A_a-{{\omega_\mu}^b}_a A_b.
\eeqa

(A)dS$_{d+1}$ metric and  Riemann tensor are given as following.
\beqa
ds^2&=&g_{\mu\nu}dx^\mu dx^\nu=-\frac{\ell^2}{z^2}(dt^2-d\vec{x}_{d-1}^2-dz^2),\\
\mathfrak{R}_{\mu\nu\rho\sigma}&=&\frac{\sigma}{\ell^2}\left(-g_{\mu\rho}\,g_{\nu\sigma}+g_{\mu\sigma}\,g_{\nu\rho}\right)
=\frac{2\,\Lambda}{d\,\left(d-1\right)}\left(g_{\mu\rho}\,g_{\nu\sigma}-g_{\mu\sigma}\,g_{\nu\rho}\right)\,,
\eeqa
where $\ell$ is $(A)dS$ radius and $\sigma$ is $(+1)$ for AdS and $(-1)$ for dS.
$\Lambda=-\frac{\sigma}{2\,\ell^2}\,d\left(d-1\right)$ is the cosmological constant.

\section{Verma module and partially massless(PM) field}\label{Verma M}
Here we recall the definition of the Verma $\mathfrak{so}(d,2)$-module. Consider a finite dimensional module $\mathcal{Y}(\Delta,\, Y)$ of sub-algebra $\mathfrak{so}(2)\oplus\mathfrak{\mathfrak{so}}(d)$.
We use $\Delta$ to denote conformal dimension and $Y$ to denote Young diagram of $\mathfrak{so}(d)$.
For the analysis of symmetric higher spin, we limit ourself to the Young diagram of a single row of length $s$. The Verma $\mathfrak{so}(d,2)$-module $\mathcal {V}(\Delta,\,s)$
is the space generated by action of the raising operators to the module $\mathcal{Y}(\Delta,\, Y)$. 
We will also denote $\mathcal{D}(\Delta,s)$ for the irreducible quotient of Verma module $\mathcal {V}(\Delta,\,s)$.
For generic value, Verma module $\mathcal {V}(\Delta,\,s)$ is irreducible and therefore coincides with $\mathcal{D}(\Delta,s)$.
However, for specific values, it becomes reducible with a non-trivial submodule.
For instance, $\Delta=d+k-1$ with an integer $0\le k\le s-1$, there is a submodule $\mathcal{D}\left(d+s-1,\,k\right)$.
Therefore, $\mathcal{D}\left(d+k-1,\,s\right)$ is not equal to Verma module but is to the quotient of Verma module:
\beq
\mathcal{V}\left(d+k-1,\,s\right)\simeq
\mathcal{D}\left(d+k-1,\,s\right)\oplus
\mathcal{D}\left(d+s-1,\,k\right)\,,\quad
\mathcal{D}\left(d+k-1,\,s\right)\simeq
\frac{\mathcal{V}\left(d+k-1,\,s\right)}{\mathcal{D}\left(d+s-1,\,k\right)}\,.
\eeq
For $k = s-1$, $\mathcal{D}(d+(s-1)-1,s)$ is unitary and its field theoretical realization is the massless spin-$s$ field propagating in AdS$_{d+1}$. 
For $0\le k<s-1$, $\mathcal{D}(d+k-1,s)$ is non-unitary and their field theoretical realizations are partially massless(PM) fields\footnote{
Extrapolating our convention, the massless higher-spin field is the partially-massless higher-spin field with depth-zero.} with depth $t = (s-k-1)$. 
(For more general cases, see \cite{Xi1, Shaynkman:2004vu}.)
The action for PM field has the PM gauge symmetry which contains covariant derivatives up to order $t-1$.
This can be derived by Stueckelberg form of PM field. 
\beq
\delta\,\phi_{\mu_1\mu_2\cdots\mu_s}
=\nabla_{(\mu_1}\,\cdots\nabla_{\mu_{t+1}}\,\xi_{\mu_{t+2}\cdots\mu_{s})}+\cdots
\eeq
See  paragraph below Eq.\eqref{spin2PMgauge} for PM spin-two case. 
The followings are properties of PM field:
\begin{table}[h]
\centering
\begin{tabular}{|c|c|c|l|}
  \hline
Field type	& $\Delta_+$	& $m^2$ & Gauge variation: $\delta\,\phi_{\mu_1\mu_2\cdots\mu_s}$\\
  \hline
depth-$t$ PM field		& $d+s-t-2$	& $-\frac{\sigma}{\ell^2}\,t\left(d+2s-t-4\right)$ 
& $
\nabla_{(\mu_1}\,\cdots\nabla_{\mu_{t+1}}\,\xi_{\mu_{t+2}\cdots\mu_{s})}+\cdots$\\
  \hline
\end{tabular}
\caption{
\footnotesize{Partially massless(PM) field}
 \label{PM property}}
\end{table}
\newline
In Table \ref{PM property}, $m$ is defined by the following convention. By the mass of a field, we refer to the mass in flat limit.
Therefore, it is zero when the higher spin gauge symmetry exist.
\label{mass square}
In this convention, the relation between mass-squared and conformal dimension is given by
\beq
m^2\,\ell^2=\Delta\left(\Delta-d\right)-\left(s-2\right)\left(d+s-2\right)\,.
\eeq
Note that this is different from the mass-squared which appears in Fierz-Pauli equation in AdS \cite{Camporesi:1993mz}: $\left(\nabla^2+\kappa^2\right)\phi_{\mu_1\,\mu_2\,\cdot\mu_s}=0$ which is given as $\kappa^2\,\ell^2=\Delta\left(\Delta-d\right)-s$.

Finally, $\mathfrak{so}(d+1,2)$-module for massless spin-$s$ can be decomposed 
into $\mathfrak{so}(d,2)$-modules by the following branching rules \cite{Dimensional Degression}:
\beq
\mathcal{D}(d+s-1,s)_{\rm \mathfrak{so}(d+1,2)}
=
\bigoplus_{n=0}^\infty\, {\mathcal D}(d+n+s-1,s)_{\rm \mathfrak{so}(d,2)}
\oplus
\bigoplus_{l=0}^{s-1}\,{\mathcal D}(d+s-1,l)_{\rm \mathfrak{so}(d,2)}
\eeq
In main text we open omit subscripts $_{\rm \mathfrak{so}(d,2)}$ for brevity.


\section{From (A)dS$_{d+k}$ to (A)dS$_{d+1}$}\label{DR}
In Section \ref{sec3}, we described the AdS waveguide by using Poincar\'e coordinate of AdS$_{d+2}$. 
However, it can be generalized to other coordinate system 
and one can show that the results which are given in the main text do not change.
In this Appendix, we analyze the more general AdS waveguide which can be built from (A)dS$_{d+k}$. 

The main consistency condition for the waveguide is the covariance condition: 
covariant derivatives of tensors in AdS$_{d+k}$ dimension are also tensors in AdS$_{d+1}$. 
To see more explicit form of covariant conditions, 
let us write down a AdS$_{d+k }$ covariant derivative in term of AdS$_{d+1}$ language, 
\beqa
\bar\nabla_\mu\, \bar{B}_\nu{}^{a} &=& 
\nabla_\mu B_\nu{}^{a}
-\bar\Gamma^{\theta_i}{}_{\mu\nu}\, B_{\theta_i}{}^{a}
+\bar{\Omega}^a{}_{\mu\,m}\, A_{\nu}{}^{{m}}.
\eeqa

Here, $\mu, \nu$ are for (A)dS$_{d+1}$ indices
 and $\theta_i$  internal space indices in lower dimensional space-time point of view. 
$A=(a,m)$ are defined in similar ways: $a$, $b$ are indices for (A)dS$_{d+1}$ local space-time indices, 
 $m$ and $n$ are for internal space. 
 ``Bar" are used to represent quantities which are tensors or covariant derivatives in $d+k$ dimension. 
 Finally, $\bar\Omega$ is spin connection in $d+k$ dimension and $\bar\Gamma$ is Christoffel symbol.
The covariance condition of $\bar\nabla_\mu\, \bar{B}_\nu{}^{a}$ implies that
$\bar{\Gamma}^{\theta_i}{}_{\mu\nu}$ and $\bar{\Omega}_\mu{}^{a\,m}$ must be tensor in the lower dimension.

To achieve these covariance conditions,
we shall take the ansatz for spin connection $\bar\Omega^{AB}$ and vielbein $\bar{E}^A$:
\beq
\bar{E}^a=f_0\,E^a,\quad \bar{E}^m={M^m}_n\, d\theta^n\,,
\quad \bar{\Omega}^{ab}= {\Omega}^{ab}\,, 
\quad \bar{\Omega}^{am}= N^m\, {E}^{a}\,,
\quad \bar{\Omega}^{mn}=L^{mn}\,.
\label{ansatz}
\eeq
Here, $f_0$, $N^m$ and $M^m{}_n$ are functions of $\theta^i$ and independent of $x^\mu$. 
The one-form tensor $L^{mn}$ is also independent of $x^\mu$.
$E^a$ and $\Omega^{ab}$ are the vielbein and spin-connection of AdS$_{d+1}$:
\beq
dE^a+\Omega^{ab}\wedge E_b=0,\quad d\,\Omega^{ab}+\Omega^{ac}\wedge{\Omega_c}^b+\sigma\,E^a\wedge E^b=0\,.
\eeq
where $\sigma$ is $+1$ for AdS and $-1$ for dS. 
The first equation is the torsionless condition and second is locally (A)dS conditions.
The similar conditions for $\bar\Omega^{AB}$ and $\bar{E}^{A}$ constrain unknowns quantities in the ansatz Eq.\eqref{ansatz}:
\beqa
\delta\, f_0 = N^m\,\bar{E}_m\,,\quad 
\delta\, N^m=N_n\,\bar{\Omega}^{nm}+\sigma'\,f_0\,\bar{E}^m\,,\quad 
\sigma\,f_0^2-N^m\,N_m=\sigma\,,\label{condi1}
\\
\delta\, \bar{E}^m+\bar{\Omega}^{mn}\wedge \bar{E}_n=0\,,
\quad 
\delta\,\bar{\Omega}^{mn}+\bar{\Omega}^{n}{}_{l}\wedge\bar{\Omega}^{ln}+\sigma'\,\bar{E}^m\wedge \bar{E}^n=0\,,\label{Hyper}
\eeqa
where $\delta$ represents the exterior derivatives of internal space $\theta^i$. i.e. $\delta f=\sum_{i=1}^k\frac{\pd f}{\pd \theta^i}d\theta^i$. 
The $\sigma'$ is introduced to represent the curvature sign of (A)dS$_{d+k}$.
The equations in Eq.\eqref{Hyper} imply that $\bar{E}^m$ and $\bar{\Omega}^{mn}$ are vielbein and spin connection 
for the Euclidean space with constant curvature which is equal to the curvature of AdS$_{d+k}$.  
After imposing this ansatz (or solving constraints Eqs.(\ref{condi1}, \ref{Hyper})), one can easily check the covariance conditions, 
as in the explicit examples for $k=1$ and $k=2$.

\paragraph{($k=1$ case)} 
In this case, $M^m{}_{n}\rightarrow M$ and $N^m\rightarrow N$ and Eq.\eqref{Hyper} is automatically satisfied.
The constraints Eq.\eqref{condi1}, Cristoffle symbols and (A)dS$_{d+2}$ metric are,
\beqa
&&\sigma'\,f_0{}^2-N^2=\sigma\,,\quad
\frac{d\,f_0}{d\,\theta}=N\,M\,,\quad
\frac{d\,N}{d\,\theta}=\sigma' f_0\, M\,,\label{condi sim}\\
&&\bar\Gamma^\theta{}_{\nu\lambda}=-\frac{f_0 \,N}{M}g_{\nu\lambda},
\quad\bar\Gamma^\lambda{}_{\nu\theta}=\frac{M\, N}{f_0}\delta^\lambda{}_\nu\,,\quad
\bar{\Gamma}^{\theta}{}_{\theta\theta}=\frac{1}{M}\frac{d\,M}{d\,\theta}\,,
\label{Christoffle}\\
&&d\,s^2_{d+2}=f_0^2\,d\,s_{d+1}^2+M^2\,d\theta^2\,. 
\eeqa
AdS waveguide from AdS$_{d+1}$ can be constructed by solving constraints \eqref{condi sim} with $\sigma=1$ and $\sigma'=1$.
It can be checked that $f_0=M=\sec\theta$ and $N= \tan\theta$ are solutions of constraints Eq.\eqref{condi sim}\footnote{
There are other cases, for examples $f_0=\cosh\theta$ , $M=1$ and  $N= \sinh\theta$ or $f_0=1/\tanh\theta$ , $M=-1/\sinh\theta$ and  $N= 1/\sinh\theta$. But they are just coordinate change of first one.}. 
In this case metric is given by $ds^2_{\text{AdS}_{d+2}}=\frac{1}{\cos^2\theta}(ds^2_{\text{AdS}_{d+1}}+\ell^2 d\theta^2)$ 
where $ds^2_{\text{AdS}_{d+1}}$ is an arbitrary locally AdS metric as we advertised. 

By solving conditions in Eq \eqref{condi sim}, we can construct various type of (A)dS waveguide.
One can obtain the dS$_d$ waveguide from dS$_{d+1}$ and AdS$_{d+1}$.
However, we cannot obtain the AdS waveguide  from dS$_{d+1}$ because of the first equation in Eq.\eqref{condi sim}.
For the other cases, see Table \ref{AdStoAdS}.
\begin{table}[h]
\centering
\begin{tabular}{|c|c|c|c|c|c|c|}
  \hline
Waveguide($\sigma'$/$\sigma$) & $f_0$ & $M$ & $N$ &  	$\bar{\Gamma}_1$	&$\bar{\Gamma}_2$	&$\bar{\Gamma}^\theta{}_{\theta\theta}$\\
  \hline
    \hline
AdS$_{d+2}$ to AdS$_{d+1}$ (1/1)& $\sec\theta$ & $\sec\theta$ & $\tan\theta$ 
&	$-\tan\theta$	&	$\tan\theta$ &	$\tan\theta$ 
 \\  \hline
dS$_{d+2}$ to dS$_{d+1}$ (-1/-1)& $\text{sech}\,\theta$ & $-\text{sech}\,\theta$ & $\tanh\theta$ 
&	$\tanh\theta$	&	$-\tanh\theta$ &	$-\tanh\theta$ 
 \\  \hline
AdS$_{d+2}$ to dS$_{d+1}$ (1/-1)& $\tan\theta$ & $\sec\theta$ & $\sec\theta$ 
&	$-\tan\theta$	&	$\sec^2\theta\,\cot\theta$ &	$\tan\theta$ 
 \\
  \hline
\end{tabular}
\caption{
\footnotesize{Various waveguide from
(A)dS$_{d+2}$ to (A)dS$_{d+1}$ and related factors.
$\bar{\Gamma}_1$ and $\bar{\Gamma}_2$ are the parts of $\bar{\Gamma}^\theta{}_{\mu\nu}$ and $\bar{\Gamma}^\mu{}_{\nu\theta}$:
$\bar{\Gamma}^\theta{}_{\mu\nu}=\bar{\Gamma}_1\,g_{\mu\nu}$\,,
$\bar{\Gamma}^\mu{}_{\nu\theta}=\bar{\Gamma}_2\,\delta^\mu{}_\nu$\,.}
}
  \label{AdStoAdS}
\end{table}

\paragraph{($k=2$ case)} Ansatz for vielbein and spin connection in $(d+3)$-dimension are,
\beq
\bar{E}^a=f_0E^a,\quad \bar{E}^i={M^i}_j d\theta^i,
\quad \bar{\Omega}^{ai}= N^i {E}^{a},
\eeq
where $i$ and $j $ represent  two compactifying indices.  
Constraints Eq.\eqref{condi1} and Eq.\eqref{Hyper} can be solved as
\beqa
\bar{E}^{i}=
\left(
\begin{array}{c}
 \sinh \rho\, du\\
 d\rho\\
\end{array}
\right)\,,
\quad \bar\Omega^{ij}=\cosh \rho
\left( \begin{array}{cc}
 0 & du\\
 -du & 0
\end{array}\right)\,,\quad
f_0=\cosh\rho\,,\quad N_1=0\,,\quad N_2=\sinh\rho\,.\nn
\eeqa
The metric in $(d+3)$-dimension can be written as
\beq
ds^2_{\text{AdS}_{d+3}}=\cosh^2\rho\, (ds^2_{\text{AdS}_{d+1}})+ d\,\rho^2+\sinh^2\rho\, du^2.
\eeq
One can easily generalize this for an arbitrary k.



\begin{thebibliography}{99}

\bibitem{Chow:1997sg}
  C.~K.~Chow and S.~J.~Rey,
 {\sl Chiral perturbation theory for tensor mesons}, 
  JHEP {\bf 9805} (1998) 010
  [hep-ph/9708355].
    
\bibitem{Camanho:2014apa}
  X.~O.~Camanho, J.~D.~Edelstein, J.~Maldacena and A.~Zhiboedov,
  {\sl Causality Constraints on Corrections to the Graviton Three-Point Coupling},
  JHEP {\bf 1602} (2016) 020
 [\href{http://arxiv.org/abs/1407.5597}{{hep-th/1407.5597}}].

\bibitem{Porrati:2001db}
  M.~Porrati,
  {\sl Higgs phenomenon for 4-D gravity in anti-de Sitter space},
  JHEP {\bf 0204} (2002) 058
  [hep-th/0112166].

\bibitem{Porrati:2003sa}
  M.~Porrati,
  {\sl Higgs phenomenon for the graviton in ADS space},
  Mod.\ Phys.\ Lett.\ A {\bf 18} (2003) 1793
  [hep-th/0306253].

\bibitem{Kaluza:1921 Klein:1926}
	T.~Kaluza, 
	{\sl {Zum	 Unit\"atsproblem der Physik}},
	Sitz.\ Preuss.\ Akad.\ Wiss.\ Phys. Berlin. (Math. Phys.): (1921) 966--972;\\
	 O.~Klein, 
	 {\sl{Quantentheorie und f\"unfdimensionale Relativit\"atstheorie}}, 
	 Zeits.\ Phys.\ {\bf 37} (1926) 895.

\bibitem{Metsaev} 
R.~R.~Metsaev,
{\sl Massive fields in AdS(3) and compactification in AdS space time},
  Nucl.\ Phys.\ Proc.\ Suppl.\  {\bf 102} (2001) 100
 [\href{http://arxiv.org/abs/hep-th/0103088}{{hep-th/0103088}}].

\bibitem{Dimensional Degression} 
A.~Y.~Artsukevich and M.~A.~Vasiliev,
{\sl On Dimensional Degression in AdS(d)},
  Phys.\ Rev.\ D {\bf 79}, 045007 (2009)
   [\href{http://arxiv.org/abs/0810.2065}{{hep-th/0810.2065}}]. 

\bibitem{DD Stueckel} 
C.~Aragone, S.~Deser and Z.~Yang,
{\sl Massive Higher Spin From Dimensional Reduction of Gauge Fields},
  Annals Phys.\  {\bf 179}, 76 (1987);\\  
S.~D.~Rindani and M.~Sivakumar,
{\sl Gauge - Invariant Description of Massive Higher - Spin Particles by Dimensional Reduction},
  Phys.\ Rev.\ D {\bf 32}, 3238 (1985).

\bibitem{Gover:2014vxa}
  A.~R.~Gover, E.~Latini and A.~Waldron,
  {\sl Metric projective geometry, BGG detour complexes and partially massless gauge theories},
  Commun.\ Math.\ Phys.\  {\bf 341} (2016) no.2,  667
  [arXiv:1409.6778 [hep-th]].

\bibitem{PaMa} 
S.~Deser and A.~Waldron,
{\sl Gauge invariances and phases of massive higher spins in (A)dS,}
  Phys.\ Rev.\ Lett.\  {\bf 87}, 031601 (2001)
 [\href{http://arxiv.org/abs/hep-th/0102166}{{hep-th/0102166}}];\\
{\sl Null propagation of partially massless higher spins in (A)dS and cosmological constant speculations,}
  Phys.\ Lett.\ B {\bf 513}, 137 (2001)
 [\href{http://arxiv.org/abs/hep-th/0105181}{{hep-th/0105181}}].
 
\bibitem{Stue} 
E. ~C. ~G. ~Stueckelberg, {\sl \"Die Wechselwirkungskr\"afte in der Elektrodynamik und in der Feldtheorie der Kr\"afte,}
 Helv.\ Phys.\ Acta {\bf 11}, 225 (1938).

\bibitem{Zino1} 
Y.~M.~Zinoviev,
{\sl On massive high spin particles in AdS,}
 [\href{http://arxiv.org/abs/hep-th/0108192}{{hep-th/0108192}}].

\bibitem{RaRe}
T.~Biswas and W.~Siegel,
{\sl Radial dimensional reduction: Anti-de Sitter theories from flat,}
  JHEP {\bf 0207} (2002) 005
 [\href{http://arxiv.org/abs/hep-th/0203115}{{hep-th/0203115}}];\\
K.~Hallowell and A.~Waldron,
{\sl Constant curvature algebras and higher spin action generating functions,}
  Nucl.\ Phys.\ B {\bf 724} (2005) 453
 [\href{http://arxiv.org/abs/hep-th/0505255}{{hep-th/0505255}}];\\
R.~Manvelyan, R.~Mkrtchyan and W.~Ruehl,
{\sl Radial Reduction and Cubic Interaction for Higher Spins in (A)dS space,}
  Nucl.\ Phys.\ B {\bf 872}, 265 (2013)
 [\href{http://arxiv.org/abs/1210.7227}{{hep-th/1210.7227}}].

\bibitem{Gaiotto:2008sa}
  D.~Gaiotto and E.~Witten,
  {\sl Supersymmetric Boundary Conditions in N=4 Super Yang-Mills Theory},
  J.\ Statist.\ Phys.\  {\bf 135} (2009) 789
  [arXiv:0804.2902 [hep-th]];\\
  D.~Gaiotto and E.~Witten,
  {\sl Janus Configurations, Chern-Simons Couplings, And The theta-Angle in N=4 Super Yang-Mills Theory},
  JHEP {\bf 1006} (2010) 097
  [arXiv:0804.2907 [hep-th]].

\bibitem{Bak:2003jk}
  D.~Bak, M.~Gutperle and S.~Hirano,
  {\sl Dilatonic deformation of AdS(5) and its field theory dual},
  JHEP {\bf 0305} (2003) 072
  [hep-th/0304129].
    
\bibitem{Eigenvalue BC}
C.~ T.~ Fulton, 
{\sl Two-point boundary value problems with eigenvalue parameter contained in the boundary conditions,} 
Proceedings of the Royal Society of Edinburgh: Section A Mathematics 77.3-4 (1977): 293-308.

\bibitem{Eigenvalue BC2}
P.~A.~ Binding, P.~J.~Browne, K.~Seddighi,  {\sl Sturm-Liouville problems with eigenparameter dependent boundary conditions,}
Proceedings of the Edinburgh Mathematicla Society (1993) 37, 57-72

\bibitem{TOS}
J.~ W.~ Strutt, B.~ Rayleigh, {\st The Theory of Sound (vol.1),}
London, Macmillan and co. (1877): p. 200-201

\bibitem{Bak:1999iq}
  D.~Bak and S.~J.~Rey,
  {\sl Holographic view of causality and locality via branes in AdS / CFT correspondence},
  Nucl.\ Phys.\ B {\bf 572} (2000) 151
  [hep-th/9902101].



\bibitem{Xi1} 
{ X.~Bekaert and M.~Grigoriev,
{\sl Notes on the ambient approach to boundary values of AdS gauge fields,}
  J.\ Phys.\ A {\bf 46} (2013) 214008
  [\href{http://arxiv.org/abs/1207.3439v2}{{hep-th/1207.3439}}]};\\
X.~Bekaert and M.~Grigoriev,
{\sl Higher order singletons, partially massless fields and their boundary values in the ambient approach,}
  Nucl.\ Phys.\ B {\bf 876}, 667 (2013)
 [\href{http://arxiv.org/abs/1305.0162}{{hep-th/1305.0162}}].

\bibitem{Shaynkman:2004vu} 
  O.~V.~Shaynkman, I.~Y.~Tipunin and M.~A.~Vasiliev,
{\sl Unfolded form of conformal equations in M dimensions and o(M + 2) modules,}
  Rev.\ Math.\ Phys.\  {\bf 18}, 823 (2006)
   [\href{http://arxiv.org/abs/hep-th/0401086}{{hep-th/0401086}}].

\bibitem{Cho:1975sw} 
  Y.~M.~Cho,
{\sl Higher - Dimensional Unifications of Gravitation and Gauge Theories,}
  J.\ Math.\ Phys.\  {\bf 16}, 2029 (1975).

\bibitem{Color Vasiliev}
  M.~A.~Vasiliev,
  {\sl Extended Higher Spin Superalgebras and Their Realizations in Terms of Quantum Operators},
  Fortsch.\ Phys.\  {\bf 36} (1988) 33;\\
  S.~E.~Konshtein and M.~A.~Vasiliev,
  {\sl Massless Representations and Admissibility Condition for Higher Spin Superalgebras},
  Nucl.\ Phys.\ B {\bf 312} (1989) 402;\\
    M.~A.~Vasiliev,
  {\sl Consistent Equations for Interacting Massless Fields of All Spins in the First Order in Curvatures},
  Annals Phys.\  {\bf 190} (1989) 59;\\
    S.~E.~Konstein and M.~A.~Vasiliev,
  {\sl Extended Higher Spin Superalgebras and Their Massless Representations},
  Nucl.\ Phys.\ B {\bf 331} (1990) 475;\\
    M.~A.~Vasiliev,
  {\sl Higher Spin Superalgebras in Any Dimension and Their Representations},
  JHEP {\bf 0412} (2004) 046
   [\href{http://arxiv.org/abs/hep-th/0404124}{{hep-th/0404124}}].

\bibitem{Gwak:2015vfb}
S.~Gwak, E.~Joung, K.~Mkrtchyan and S.~J.~Rey,
{\sl Rainbow Valley of Colored (Anti) de Sitter Gravity in Three Dimensions},
 [\href{http://arxiv.org/abs/1511.05220}{{hep-th/1511.05220}}];\\
{\sl Rainbow Vacua of Colored Higher Spin Gravity in Three Dimensions},
 [\href{http://arxiv.org/abs/1511.05975}{{hep-th/1511.05975}}].
 
\bibitem{Klebanov:2002ja} 
  I.~R.~Klebanov and A.~M.~Polyakov,
{\sl AdS dual of the critical O(N) vector model,}
  Phys.\ Lett.\ B {\bf 550}, 213 (2002)
 [\href{http://arxiv.org/abs/hep-th/0210114}{{hep-th/0210114}}].

\bibitem{Sezgin:2002rt} 
  E.~Sezgin and P.~Sundell,
{\sl Massless higher spins and holography,}
  Nucl.\ Phys.\ B {\bf 644}, 303 (2002)
[\href{http://arxiv.org/abs/hep-th/0205131}{{hep-th/0205131}}].

\bibitem{Bae:2014hia}
  J.~B.~Bae and S.~J.~Rey,
  {\sl Conformal Bootstrap Approach to O(N) Fixed Points in Five Dimensions},
  arXiv:1412.6549 [hep-th].

\bibitem{Dolan:2001ih}
  L.~Dolan, C.~R.~Nappi and E.~Witten,
{\it Conformal operators for partially massless states},
  JHEP {\bf 0110} (2001) 016
[\href{http://xxx.lanl.gov/abs/hep-th/0109096}{{hep-th/0109096}}].

\bibitem{higherSin} 
T.~Basile, X.~Bekaert and N.~Boulanger,
{\sl Flato-Fronsdal theorem for higher-order singletons,}
  JHEP {\bf 1411}, 131 (2014)
  [\href{http://arxiv.org/abs/1410.7668}{{hep-th/1410.7668}}].

\bibitem{CasimirE} J.~Kim, S.~Gwak, to appear.

\bibitem{Giombi:2013gk}
  Simone Giombi, Igor R. Klebanov, 
{\sl One Loop Tests of Higher Spin AdS/CFT},
  JHEP {\bf 1312} (2013) 68
[\href{http://arxiv.org/abs/1308.2337}{{hep-th/1308.2337}}];\\
  Simone Giombi, Igor R. Klebanov, Arkady A. Tseytlin,
{\sl Partition Functions and Casimir Energies in Higher Spin AdS$_{d+1}$/CFT$_{d}$},
  Phys.\ Rev.\ D {\bf 90} (2014) 2,  024048
[\href{http://arxiv.org/abs/1402.5396}{{hep-th/1402.5396}}];\\
  M. Beccaria, Arkady A. Tseytlin,
{\sl On Higher Spin Partition Functions},
  J.\ Phys.\ A {\bf 48} (2015) 27,  275401
[\href{http://arxiv.org/abs/1503.08143}{{hep-th/1503.08143}}].

\bibitem{Camporesi:1993mz}
  R.~Camporesi and A.~Higuchi,
{\sl Arbitrary spin effective potentials in anti-de Sitter space-time,}
  Phys.\ Rev.\ D {\bf 47} (1993) 3339;\\
{\sl Spectral functions and zeta functions in hyperbolic spaces,}
  J.\ Math.\ Phys.\  {\bf 35} (1994) 4217.


\bibitem{Rey:2010ry}
  S.~J.~Rey and T.~Suyama,
  {\sl Exact Results and Holography of Wilson Loops in N=2 Superconformal (Quiver) Gauge Theories},''
  JHEP {\bf 1101} (2011) 136
  doi:10.1007/JHEP01(2011)136
  [arXiv:1001.0016 [hep-th]].

\bibitem{Cho:1992rq}
  Y.~M.~Cho and S.~W.~Zoh,
  {\sl Explicit construction of massive spin two fields in Kaluza-Klein theory},
  Phys.\ Rev.\ D {\bf 46} (1992) 2290.

\bibitem{Aharony:2015zea}
  O.~Aharony, M.~Berkooz and S.~J.~Rey,
{\sl Rigid holography and six-dimensional $ \mathcal{N}=\left(2,0\right) $ theories on AdS$_{5} \times \mathbb{S}^{1}$},
JHEP {\bf 1503} (2015) 121
[\href{http://arxiv.org/abs/1501.02904}{{hep-th/1501.02904}}].

\bibitem{Fernandez-Melgarejo:2016xiv}
  J.~J.~Fernandez-Melgarejo, S.~J.~Rey and P.~Sur�wka,
  {\sl A New Approach to Non-Abelian Hydrodynamics},
  arXiv:1605.06080 [hep-th].
  
\bibitem{Bak:2016rpn}
  D.~Bak, A.~Gustavsson and S.~J.~Rey,
  {\sl Conformal Janus on Euclidean Sphere},
  arXiv:1605.00857 [hep-th].

\bibitem{Rey:1991ka}
  S.~J.~Rey,
  {\sl Enhanced mass spectrum of pseudoGoldstone bosons?},
  Phys.\ Lett.\ B {\bf 277} (1992) 141.

\bibitem{HS3Sym1}  
M.~Henneaux and S.~J.~Rey,
{\sl Nonlinear $W_{\infty}$ as Asymptotic Symmetry of Three-Dimensional Higher Spin Anti-de Sitter Gravity,}
  JHEP {\bf 1012}, 007 (2010)
 [\href{http://arxiv.org/abs/1008.4579}{{hep-th/1008.4579}}].

\bibitem{HS3Sym3} 
A.~Campoleoni, S.~Fredenhagen, S.~Pfenninger and S.~Theisen,
{\sl Asymptotic symmetries of three-dimensional gravity coupled to higher-spin fields,}
  JHEP {\bf 1011}, 007 (2010)
 [\href{http://arxiv.org/abs/1008.4744}{{hep-th/1008.4744}}].

\bibitem{HS3Sym2} 
M.~Henneaux, G.~Lucena G\"omez, J.~Park and S.~J.~Rey,
{\sl Super- W($\infty$) Asymptotic Symmetry of Higher-Spin $AdS_3$ Supergravity,}
  JHEP {\bf 1206}, 037 (2012)
 [\href{http://arxiv.org/abs/1203.5152}{{hep-th/1203.5152}}].

\bibitem{Rey:2008zz}
  S.~J.~Rey,
  {\sl String theory on thin semiconductors: Holographic realization of Fermi points and surfaces},
  Prog.\ Theor.\ Phys.\ Suppl.\  {\bf 177} (2009) 128
  [arXiv:0911.5295 [hep-th]].

\bibitem{Gover:2005mn}
  A.~R.~Gover,
{\sl Laplacian operators and Q-curvature on conformally Einstein manifolds},
[\href{http://arxiv.org/abs/math/0506037}{{math/0506037}}].

\bibitem{Juhl:2011ua}
  A.~Juhl,
{\sl Explicit formulas for GJMS-operators and $Q$-curvatures},
[\href{http://arxiv.org/abs/1108.0273}{{math.DG/1108.0273}}].

\bibitem{Joung:2012rv}
  E.~Joung, L.~Lopez and M.~Taronna,
  {\sl On the cubic interactions of massive and partially-massless higher spins in (A)dS},  JHEP {\bf 1207} (2012) 041
  [arXiv:1203.6578 [hep-th]].

\bibitem{Joung:2012hz}
  E.~Joung, L.~Lopez and M.~Taronna,
  {\sl Generating functions of (partially-)massless higher-spin cubic interactions},
  JHEP {\bf 1301} (2013) 168
  [arXiv:1211.5912 [hep-th]].
  
\bibitem{Joung:2014aba}
  E.~Joung, W.~Li and M.~Taronna,
  {\sl No-Go Theorems for Unitary and Interacting Partially Massless Spin-Two Fields},
  Phys.\ Rev.\ Lett.\  {\bf 113} (2014) 091101
  [arXiv:1406.2335 [hep-th]].

\bibitem{Deser:2013bs}
  S.~Deser, E.~Joung and A.~Waldron,
  {\sl Gravitational- and Self- Coupling of Partially Massless Spin 2},
  Phys.\ Rev.\ D {\bf 86} (2012) 104004
  [arXiv:1301.4181 [hep-th]].
  
\bibitem{Deser:2012qg}
  S.~Deser, E.~Joung and A.~Waldron,
  {\sl Partial Masslessness and Conformal Gravity},
  J.\ Phys.\ A {\bf 46} (2013) 214019
  [arXiv:1208.1307 [hep-th]].

\bibitem{Rey:1989ti}
  S.~J.~Rey,
  {\sl The Higgs Mechanism for Kalb-ramond Gauge Field},
  Phys.\ Rev.\ D {\bf 40} (1989) 3396.


\bibitem{Joung:2015jza}
  E.~Joung and K.~Mkrtchyan,
  {\sl Partially-massless higher-spin algebras and their finite-dimensional truncations},
  JHEP {\bf 1601} (2016) 003
  [arXiv:1508.07332 [hep-th]].


\end{thebibliography}
\end{document}